\documentclass[12pt]{article}
\pdfoutput=1
\usepackage{amssymb,amsmath}
\usepackage{epsfig}
\usepackage{epstopdf}
\usepackage{latexsym}
\usepackage{graphicx}
\usepackage{booktabs}
\usepackage{bbm}
\usepackage{rotating}
\usepackage{enumitem}
\usepackage[numbers,compress]{natbib}

\usepackage{url}

\usepackage{slashed}
\usepackage{multirow}

\usepackage[T1]{fontenc}

\usepackage[margin=20pt,small]{caption}
\usepackage{subcaption}

\usepackage[toc]{appendix}

\usepackage{color}
\usepackage{datetime}
\usepackage[
      colorlinks=false,
      linkcolor=darkblue,  
      urlcolor=blue,    
      filecolor=blue,     
      citecolor=red,
linktocpage=true,
      pdfstartview=FitV,
      bookmarksopen=true,
	  hidelinks
      ]{hyperref}

\ifpdf
\fi

\newcommand*{\boxedcolor}{red}
\makeatletter
\renewcommand{\boxed}[1]{\textcolor{\boxedcolor}{%
  \fbox{\normalcolor\m@th$\displaystyle#1$}}}
\makeatother

\definecolor{cardinal}{rgb}{0.6,0,0}
\definecolor{darkgreen}{rgb}{0,0.5,0}
\definecolor{golden}{rgb}{0.92, 0.7, 0}
\definecolor{midnight}{rgb}{0, 0, 0.5}
\definecolor{darkblue}{rgb}{0.2, 0, 0.8}

\def\Re{{\rm Re}} \def\Im{{\rm Im}}

\def\cO{{\cal O}}

\def\sl{\frak{sl}}

\newcommand{\abs}[1]{\left\lvert #1 \right\rvert}

\newcommand {\be} {\begin {equation}}
\newcommand {\ee} {\end {equation}}

\newcommand {\bes} {\begin {equation*}}
\newcommand {\ees} {\end {equation*}}

\newcommand{\es}[2] {\begin{equation} \label{#1} \begin{split} #2 \end{split} \end{equation}}

\newcommand{\Z}{\mathbb{Z}}

\newcommand{\R}{\mathbb{R}}

\def\<{\langle}
\def\>{\rangle}

\def\cO {{\cal O}}
\def\cG {{\cal G}}

\begin{document}  

\begin{titlepage}

 \begin{flushright}
{\tt PUPT-2544}
\end{flushright}

\medskip
\begin{center} 
{\Large \bf  Decoding a Three-Dimensional Conformal Manifold}

\bigskip
\bigskip
\bigskip

{\bf Marco Baggio,$^1$ Nikolay Bobev,${}^{1}$ Shai M. Chester,${}^{2}$ \\Edoardo Lauria,${}^{1,3}$ and Silviu S. Pufu${}^{2}$  \\ }
\bigskip
\bigskip

${}^{1}$Instituut voor Theoretische Fysica, KU Leuven, \\Celestijnenlaan 200D, B-3001 Leuven, Belgium
\vskip 5mm

${}^{2}$Joseph Henry Laboratories, Princeton University, \\Princeton, NJ 08544, USA

\vskip 5mm

${}^{3}$Perimeter Institute for Theoretical Physics \\
31 Caroline Street North, ON N2L 2Y5, Canada

\vskip 5mm

\texttt{marco.baggio@kuleuven.be,~nikolay.bobev@kuleuven.be,~schester@princeton.edu,\\~edo.lauria@kuleuven.be,~spufu@princeton.edu} \\
\end{center}

\bigskip
\bigskip

\begin{abstract}

\noindent  
\end{abstract}

\noindent 

We study the one-dimensional complex conformal manifold that controls the infrared dynamics of a three-dimensional $\mathcal{N}=2$ supersymmetric theory of three chiral superfields with a cubic superpotential. Two special points on this conformal manifold are the well-known XYZ model and three decoupled copies of the critical Wess-Zumino model. The conformal manifold enjoys a discrete duality group isomorphic to $S_4$ and can be thought of as an orbifold of $\mathbf{CP}^1$. We use the $4-\varepsilon$ expansion and the numerical conformal bootstrap to calculate the spectrum of conformal dimensions of low-lying operators and their OPE coefficients, and find a very good quantitative agreement between the two approaches.

\end{titlepage}


\setcounter{tocdepth}{2}
\tableofcontents

\section{Introduction}
\label{sec:intro}

Conformal manifolds are the manifolds parametrized by the exactly marginal coupling constants of a given conformal field theory (CFT)\@.  Of course, not every CFT can have exactly marginal couplings, and in fact their existence imposes non-trivial constraints on the CFT data (for recent discussions, see \cite{Behan:2017mwi,Bashmakov:2017rko,Hollands:2017chb,Sen:2017gfr} and references therein).  In two dimensions, exactly marginal couplings occur in many CFTs, including all string compactifications with moduli (see for example  \cite{Seiberg:1988pf,Kutasov:1988xb}), of which there are many examples.  In $d>2$ spacetime dimensions, conformal manifolds are less common, and all known examples are in superconformal field theories (SCFTs).  As shown in \cite{Cordova:2016xhm} using an abstract approach, the superconformal algebra allows for the existence of marginal couplings only in SCFTs with ${\cal N} = 1$ or $2$ supersymmetry in 3d and ${\cal N} = 1$, $2$, or $4$ supersymmetry in 4d.  There are no 5d or 6d SCFTs with exactly marginal couplings, and there are no interacting SCFTs in more than six dimensions \cite{Nahm:1977tg}.

In 4d, one way to construct SCFTs with exactly marginal deformations is to use the fact that, classically, the gauge coupling constant is marginal.  Choosing the charged matter content appropriately and relying on supersymmetry, one can then ensure that this coupling is exactly marginal, thus parameterizing a conformal manifold.   A lot has been learned about conformal manifolds of four-dimensional gauge theories employing important insights from S-duality, D- and M5-brane realizations as well as some exact non-perturbative calculations---see for instance \cite{Leigh:1995ep,Argyres:2007cn,Gaiotto:2009we} and references thereof for a sample of illustrative examples. 

Conformal manifolds of 3d SCFTs have been somewhat less studied, in part because the small amount of supersymmetry (${\cal N} \leq 2$) required for their existence does not lead to the powerful constraints present in 4d ${\cal N} = 2$ and ${\cal N} = 4$ CFTs.  On the flip side, unlike in 4d, in 3d it is possible to construct interacting (S)CFTs with UV Lagrangian descriptions that use only scalars and fermions, without any need for gauge interactions.  Our goal in this work is to study one of the simplest examples of an SCFT in more than two dimensions with an exactly marginal coupling:  a Wess-Zumino model with a cubic superpotential \cite{Strassler:1998iz} that we will describe shortly.  We will calculate various quantities along the conformal manifold, both perturbatively in the $4-\varepsilon$ expansion and exactly using the conformal bootstrap technique directly in 3d.  

Our interest in 3d conformal manifolds comes partly from the conformal bootstrap, which is a technique that can be used primarily to put bounds on various quantities in CFT (see \cite{Rychkov:2016iqz,Simmons-Duffin:2016gjk} for a review and further references).  In special cases, such as the 3d Ising model or critical $O(N)$ vector models, these bounds take the form of islands in theory space, and consequently this technique can be turned into a precision study of these CFTs \cite{Kos:2014bka,Kos:2015mba,Kos:2016ysd,ElShowk:2012ht,El-Showk:2014dwa}.  Other 3d examples in which one can perform precision studies are Gross-Neveu-Yukawa theories \cite{Iliesiu:2015qra,Iliesiu:2017nrv}, the ${\cal N} = 2$ super-Ising model \cite{Bobev:2015jxa,Bobev:2015vsa}, and ${\cal N} = 8$ SCFTs with holographic duals \cite{Chester:2014mea,Chester:2014fya,Agmon:2017xes}.  It would be very nice to expand the set of theories that can be solved exactly using the conformal bootstrap technique, and, as we will see, the current work provides another example.  Apart from the pure theoretical interest in 3d SCFTs with exactly marginal operators, it has been pointed out recently that similar $\mathcal{N}=2$ conformal models may find phenomenological applications in condensed matter physics \cite{Lee:2006if,Yu:2010zv,Ponte:2012ru,Grover:2013rc,Jian:2016zll,Li:2017dkj}.

Before we focus on our particular model, let us summarize some general results on 3d $\mathcal{N}=2$ conformal manifolds.  It can be shown along the lines of \cite{Asnin:2009xx} (see also \cite{Tachikawa:2005tq,deAlwis:2013jaa}) that 3d $\mathcal{N}=2$ conformal manifolds admit a K\"ahler metric. The dimension of the conformal manifold can be determined either by the well-known Leigh-Strassler method \cite{Leigh:1995ep} or using the results in \cite{Green:2010da,Kol:2002zt,Kol:2010ub}. The Leigh-Strassler approach is more explicit but is suitable only for theories with an explicit Lagrangian description since it relies on knowing the $\beta$-functions of the coupling constants in the theory. On the other hand one can generally show that locally the conformal manifold $\mathcal{M}_C$ equals the space of complex marginal couplings, $\{\lambda_a\}$, modded out by the complexification of the group, $G$, of continuous flavor symmetries \cite{Green:2010da,Kol:2002zt,Kol:2010ub}
\begin{equation}
\mathcal{M}_C = \{\lambda_a\}/G^{\mathbb{C}}\;.
\end{equation}
We emphasize that this result is local and fails to capture the global properties of the conformal manifold.  We will see an explicit example in the particular model of interest in this work.

The theory we study in detail here is that of three chiral multiplets $X_i$, $i=1, 2, 3$, with canonical K\"ahler potential and cubic superpotential interaction
\begin{equation}\label{superintro}
 W = h_1\, X_1X_2X_3+\frac{h_2}{6}\,(X_1^3+X_2^3+X_3^3)\;,
\end{equation}
where $h_1$ and $h_2$ are complex coupling constants.   The model in \eqref{superintro} has two special limits that have been well-studied (see for instance \cite{Aharony:1997bx}):   when $h_2=0$ one finds the superpotential of the well-known XYZ model;  while for $h_1=0$ we have three decoupled copies of the Wess-Zumino model describing the ${\cal N} = 2$ super-Ising model.  In 3d the two complex couplings $h_{1,2}$ in \eqref{superintro} are relevant and one can argue for the existence of a manifold of IR fixed points parametrized by the complex coupling $\tau = h_2/h_1$ taking values in ${\bf CP}^1$ \cite{Strassler:1998iz}.\footnote{The model \eqref{superintro} is related by 3d mirror symmetry \cite{Intriligator:1996ex,deBoer:1997kr,Aharony:1997bx} to ${\cal N} = 2$ supersymmetric QED with 1 flavor (SQED$_1$).  This SQED$_1$ theory should also exhibit a conformal manifold, with the marginal direction being a superpotential deformation by a chiral monopole operator (see Section 4.1 of \cite{Benini:2017dud} for a recent discussion).}

The conformal manifold parameterized by $\tau$ does not admit a weakly coupled region  and thus is hard to access quantitatively.  To understand it, we will employ various complementary strategies.   First, a careful study of the superpotential interaction~\eqref{superintro} reveals that this theory enjoys an order 54 discrete flavor symmetry group.  In addition, there exist field redefinitions that can be absorbed into a redefinition of the couplings $h_{1, 2}$ (acting on $\tau$ as certain fractional linear transformations). This action manifests itself as a duality of the IR conformal manifold.  As we will explain, the duality symmetry group is isomorphic to the symmetric group $S_4$.  It is akin to S-duality in four-dimensional gauge theories, but with the notable difference that here, it acts linearly on the local operators of the SCFT\@. We use the duality transformations in order to first understand, qualitatively, a general picture of how the CFT data must change as a function of $\tau$.  The duality group also serves as a stringent check on our more quantitative analyses.

To study the conformal manifold more quantitatively, we use two strategies.  The first approach is to continue this theory away from 3d.  In 2d, this theory is equivalent to a $\Z_3$ orbifold of an $\mathcal{N}=(2,2)$ chiral multiplet and was studied in \cite{Lerche:1989cs,Verlinde:1991ci} using the power of the Virasoro symmetry that is not available in $d>2$.\footnote{See \cite{Lin:2016gcl} for a curious appearance of this model in the context of the numerical bootstrap for two-dimensional $\mathcal{N}=(2,2)$ CFTs.}  In $d = 4 - \varepsilon$ dimensions, the RG flow triggered by the interaction \eqref{superintro} becomes ``short,'' so the conformal manifold is accessible in perturbation theory \cite{Wilson:1971dc,Wilson:1973jj}.  The four-loop $\beta$-function for the couplings $h_{1,2}$ can be extracted from results available in the literature, which then allow us to determine the scaling dimensions of all unprotected quadratic operators in $X_i$ to order $\varepsilon^4$. These perturbative results can be used to estimate the scaling dimensions of these operators in 3d.   In addition to scaling dimensions, we also compute some OPE coefficients as well as the Zamolodchikov metric in the $4-\varepsilon$ expansion up to two-loops. 

As already mentioned, the second approach we employ is the numerical conformal bootstrap \cite{Rychkov:2016iqz,Simmons-Duffin:2016gjk}. This strategy has been applied successfully to extract constraints on the spectrum of conformal dimensions and OPE coefficients in both the critical WZ model \cite{Bobev:2015jxa,Bobev:2015vsa} and the XYZ model \cite{Chester:2015lej}. Here we refine and generalize this analysis along the whole conformal manifold parametrized by $\tau$. The chiral ring relations that follow from \eqref{superintro} and the structure of the crossing equations in our model imposed by supersymmetry and the flavor symmetry allow us to extract numerical constraints on the spectrum and OPE coefficients as a function of $\tau$. To the best of our knowledge this is the first time the numerical conformal bootstrap program has been applied successfully as a function of a marginal coupling in $d>2$.\footnote{See \cite{Beem:2013qxa,Beem:2016wfs} for conformal bootstrap studies of 4d $\mathcal{N}=4$ SYM which also has a one-dimensional complex conformal manifold.} The results from the conformal bootstrap are non-perturbative in nature and are applicable directly to the strongly coupled theory in three dimensions. They confirm the general qualitative analysis based on symmetries and dualities and match the perturbative $4-\varepsilon$ expansion to remarkable precision.

The rest of this paper is organized as follows.  In Section~\ref{model}, we set the stage by presenting the properties of the model \eqref{superintro}, including the existence of a conformal manifold, global symmetry, and duality group. We continue in Section \ref{sec:epsilon} with a detailed study of the model of interest using the perturbative $4-\varepsilon$ expansion. In Section \ref{sec:bootstrap} we describe the constraints imposed by unitarity and crossing symmetry and apply the numerical conformal bootstrap technology to extract bounds on conformal dimensions and OPE coefficients. We conclude in Section~\ref{sec:discussion} with a discussion and a summary of some interesting open questions. Many technical details on the perturbative analysis, four-point function crossing equations, as well as the global symmetries of our model are delegated to the Appendices. We also summarize some results about the 2d analogue of the model \eqref{superintro} in Appendix \ref{2d}.

\section{The cubic model}
\label{model}

In this section we introduce the model \eqref{superintro} in more detail and study abstractly some of its properties. In particular, we first identify the flavor symmetry group, which for generic values of the couplings turns out to be a discrete group of order $54$.  We use it to argue that the model flows in the IR to a family of CFTs parametrized by the ratio of the two coupling constants $\tau = h_2/h_1$.  We then show that field redefinitions imply that theories at different points in the conformal manifold are dual to each other. This allows us to identify the conformal manifold with a 2-dimensional orbifold with three special points. Finally, we derive some non-perturbative consequences of the duality on the operator spectrum. We emphasize from the outset that our main interest is in studying this cubic model in 3d, however many of the results we find below are applicable for any (even non-integer) value of the dimension $2\leq d\leq 4$. Thus whenever possible we keep the dimension $d$ general.

\subsection{Global symmetries}
\label{symmetries}

In 3d, our model is an $\mathcal{N}=2$ theory that consists of three chiral superfields $X_1, X_2, X_3$, with the following K\"ahler potential $K$ and superpotential $W$:
\begin{align}
    K & = \sum_{i=1}^3 X_i \overline{X}^i\;, \\
    \label{eq:superpotential}
    W &= h_1\, X_1X_2X_3+\frac{h_2}{6}\,(X_1^3+X_2^3+X_3^3)~.
\end{align}
In the absence of the superpotential interaction, the theory of the free massless chiral superfields $X_i$ and canonical K\"ahler potential has a $U(3)$ flavor symmetry.  We use lower and upper indices for the $\mathbf{3}$ and $\bar{\mathbf{3}}$ representations of $U(3)$, respectively.  In the presence of the superpotential interaction, the complex couplings $h_1$ and $h_2$ are relevant and the model becomes strongly coupled in the infrared.\footnote{When one expands the superfields $X_i$ in components the resulting Lagrangian has quartic bosonic interactions and the usual Yukawa couplings between the scalars and fermions, see equation \eqref{Toines}.} 

The model at hand enjoys a $U(1)_R$ R-symmetry that acts with charge $2/3$ on the complex scalar fields $X_i$ in the three chiral superfields (we use $X_i$ to denote both the chiral superfields as well as their scalar components), ensuring that the superpotential in \eqref{eq:superpotential} has R-charge $2$.  Generically, the model \eqref{eq:superpotential} does not have any other Abelian symmetries, so if it flows to a superconformal fixed point in the IR, it must be that the $U(1)_R$ symmetry mentioned above is the one appearing in the ${\cal N} = 2$ superconformal algebra.\footnote{We assume that there are no accidental continuous symmetries emerging in the IR.} At a superconformal fixed point, the scaling dimension of a chiral primary operator, $\Delta_{\mathcal{O}}$, is fixed in terms of its superconformal R-charge, $q_{\mathcal{O}}$, through the relation
\begin{equation}\label{Deltaqrel}
\Delta_{\mathcal{O}} = \frac{d-1}{2} q_{\mathcal{O}}\;.
\end{equation}
Therefore we conclude that $\Delta_{X_i}=(d-1)/3$ at a superconformal fixed point.

At \emph{generic} values of the coupling constants $h_i$, the superpotential \eqref{eq:superpotential} is also invariant under an order $54$ discrete flavor symmetry group $G=(\mathbb{Z}_3\times \mathbb{Z}_3)\rtimes S_3$, generated by the three $U(3)$ matrices
\es{G}{
g_1=\begin{pmatrix}
0&1&0\\
1&0&0\\ 
0&0&1\\
\end{pmatrix}\,,\qquad
g_2=\begin{pmatrix}
0&0&1\\
1&0&0\\
0&1&0\\
\end{pmatrix}\,,\qquad
g_3=\begin{pmatrix}
1&0&0\\
0&\omega&0\\
0&0&\omega^2\\
\end{pmatrix}\,,
}
where $\omega=e^{2\pi {\rm i}/3}$ is a cubic root of unity.  The matrices $g_1$ and $g_2$ generate an $S_3$ subgroup of $G$ that simply permutes the three chiral superfields. More details on this discrete group, including the classification of irreducible representations and the character table, can be found in Appendix~\ref{sec:discretegroup}. It is important to notice that there are special values of $h_1$ and $h_2$ at which the symmetry group is enhanced. For example, it is well-known that the $h_2=0$ theory (also known as the XYZ model) enjoys a continuous flavor symmetry $U(1) \times U(1)$ (see for example \cite{Aharony:1997bx}). We postpone the classification of these special points to the next subsection, where we see that they correspond to orbifold singularities on the conformal manifold.

\subsubsection{Conformal manifold}

We now argue that for generic $h_i$, the theory flows to a family of strongly interacting CFTs with $\mathcal{N}=2$ superconformal symmetry, parametrized by the ratio of the coupling constants\footnote{To the best of our knowledge this was first pointed out in \cite{Strassler:1998iz}.}
\begin{equation}
    \label{eq:marginaltau}
    \tau = \frac{h_2}{h_1}~.
\end{equation}
Since the theory is supersymmetric, we can choose a scheme where only the K\"ahler potential is renormalized
\begin{equation}
    K = \sum_{i,j} {Z^i}_j X_i \overline{X}^j~.
\end{equation}
Invariance under $G$ implies that the matrix ${Z^i}_j$ is proportional to the identity matrix
\begin{equation}
    {Z^i}_j = Z\, {\delta^i}_j~.
\end{equation}
As a consequence, the three fields $X_i$ receive the same wave-function renormalization, and in turn the ``physical'' coupling constants $\tilde{h}_i \equiv Z^{-3/2} h_i$ are renormalized in the same way. This immediately implies that the ratio in equation \eqref{eq:marginaltau} is not renormalized and parametrizes a marginal direction. Indeed, $\tau$ can be viewed as a coordinate on the conformal manifold, taking values in $\mathbf{CP}^1$.  

An alternative way to argue in favor of the existence of a one-dimensional conformal manifold is offered by the method presented in \cite{Green:2010da}  (see also \cite{Kol:2002zt,Kol:2010ub}). At the XYZ point we have three complex cubic operators, $X_i^3$. These operators are chiral with R-charge $q=2$ and thus according to \eqref{Deltaqrel} have dimension $\Delta=2$. In the same superconformal multiplet there is a scalar descendant operator which is obtained by acting with two supercharges on the superconformal primary and thus has dimension $\Delta=3$. Therefore we conclude that we have three complex marginal couplings. However we also have a $U(1)\times U(1)$ global flavor symmetry at the XYZ point (see \eqref{XYZspecial} below). As argued in \cite{Green:2010da} the space of exactly marginal couplings is locally the quotient of the space of marginal couplings by the complexified global continuous symmetry group. Applying this to our setup with three marginal operators and a two-dimensional global symmetry group we conclude that there is one exactly marginal complex operator at the XYZ point.

\subsubsection{Spectrum of operators}
The spectrum of operators of the theory can be described in terms of irreducible representations of $G$. Since $G$ is a discrete group, there are only a finite number of irreps:
\begin{equation}
\mathbf{1}~,\quad \mathbf{1}'~, \quad \mathbf{2}~, \quad\mathbf{2}'~, \quad\mathbf{2}''~, \quad\mathbf{2}'''~, \quad\mathbf{3}~, \quad\mathbf{3}'~, \quad\overline{\mathbf{3}}~, \quad\overline{\mathbf{3}}'~.
\end{equation}
The scalar operators $X_i$ sit in the $\mathbf{3}$. It is straightforward to decompose generic operators built out of the $X_i$'s (and their complex conjugates) into irreducible representations using the character table in Appendix \ref{sec:discretegroup}. 

For instance, let us describe the scalar operators that are quadratic in the $X_i$ and/or $\overline{X}^j$.  The nine operators $X_i \overline{X}^j$ can be organized according to the decomposition
\begin{equation}
\label{eq:33bar}
    \bold3\otimes\bar{\bold3}=\bold1\oplus\bold2\oplus\bold2'\oplus\bold2''\oplus\bold2'''\;,
\end{equation}
as 
\es{bilinears}{
    &\cO_{\bold1,0}= \frac{1}{\sqrt{3}}X_i\overline X^i\,,\\
    &\cO_{\bold2,0}= \frac{1}{\sqrt{2}}\left(X_1\overline X^1-X_3\overline X^3,\quad  X_1\overline X^1-X_2\overline X^2\right)\,, \\
    &\cO_{\bold2',0}=\frac{1}{\sqrt{3}}\left(X_2\overline X^1+X_3\overline X^2+X_1\overline X^3,\quad  X_3\overline X^1+X_1\overline X^2+X_2\overline X^3\right)\,,\\
    &\cO_{\bold2'',0}= \frac{1}{\sqrt{3}}\left(\omega^2 X_2\overline X^1+ X_3\overline X^2+\omega X_1\overline X^3,\quad\omega X_1\overline X^2+ X_2\overline X^3+\omega^2 X_3\overline X^1\right)\,,\\
    &\cO_{\bold2''',0}=\frac{1}{\sqrt{3}}\left(\omega^2 X_1\overline X^2+ X_2\overline X^3+\omega X_3\overline X^1,\quad \omega X_2\overline X^1+ X_3\overline X^2+\omega^2 X_1\overline X^3\right)\,.
}
Each of these operators is the bottom component of a long superconformal multiplet. We see that all of the four inequivalent two-dimensional irreps of $G$ appear, and so there should be five distinct eigenvalues for the conformal dimensions in this sector.   In Section \ref{sec:epsilon} we explicitly compute the conformal dimensions of these operators using the $4-\varepsilon$ expansion to order $\varepsilon^4$ and verify the predicted degeneracy of the spectrum.

To examine the operators $X_i X_j$ (or their complex conjugates), consider the decomposition
\begin{equation}
\label{eq:33}
   \bold3\otimes{\bold3}=\bar{\bold3}_s\oplus\bar{\bold3}_s\oplus\bold3'_a\,,
\end{equation}
where $s/a$ denotes the symmetric/antisymmetric product. Since the $X_i$ are bosonic fields and commute with each other, the scalar operators quadratic in $X_i$ appear in the symmetric product of $\bold3\otimes{\bold3}$, namely in the $\bar{\bold3}$ irrep, which appears twice.  The linearly independent operators can be written as
\es{chiralBis}{
&\cO_{\bar{\bold3}_1,0}=  \left(X_1 X_1,\quad  X_2 X_2,\quad   X_3 X_3\right)\,,\\
&\cO_{\bar{\bold3}_2,0}=   \left(X_2 X_3,\quad  X_1 X_3,\quad  X_1 X_2\right) \,.
}
Due to the chiral ring relations, discussed in Section \ref{sec:chiralring} below, only half of these operators flow to chiral primaries of $\Delta = 2\frac{d-1}{3}$.  The others, namely, $\cO_{\bar{\bold3}_2,0} + \frac{\tau}{2} \cO_{\bar{\bold3}_1,0}$ flow to superconformal descendants of anti-chiral primaries and have dimension $\Delta = d-2\frac{d-1}{3}$ .

\subsection{The duality group}
\label{sec:dualities}
Since the K\"ahler potential is invariant under $U(3)$, we can use elements of this group to perform field redefinitions. A generic $U(3)$ element will transform the superpotential \eqref{eq:superpotential} into a generic cubic superpotential of the form $W \sim h^{ijk} X_i X_j X_k$. As we now explain, there is a discrete subgroup of $U(3)$ that leaves the form of the superpotential invariant and only changes the coupling constants $h_1$ and $h_2$. Theories whose coupling constants are related in such a way are then equivalent, and define the same CFT in the infrared.
 
This ``duality subgroup'' of $U(3)$ is generated by the following elements
\begin{align}
\label{dualMat}
u_1 & =\begin{pmatrix}
\omega^{2}&0&0\\
0&1&0\\
0&0&1\\
\end{pmatrix}\,,& u_2 & = \frac{1}{\sqrt{3}}
\begin{pmatrix}
1&1&\omega^2\\
1&\omega&\omega\\
\omega&1&\omega\\
\end{pmatrix}\,,
\end{align}
where as before $\omega = e^{2\pi {\rm i}/3}$ is a cubic root of unity. These field redefinitions lead to the following duality transformation on the coupling constants:
\begin{align}
\label{dualPot}
    d_1(\tau) & =\omega \tau\,,& d_2(\tau)&=\frac{\tau+2\omega^2}{\omega\tau-1}\,,
\end{align}
meaning that two CFTs characterized by distinct values of the marginal coupling $\tau$ and $\tau' = d_i(\tau)$ are equivalent. It is easy to check that the group generated by the transformations in \eqref{dualMat} and their compositions is the alternating group $A_4$, that is the group of even permutations on four objects. This result was derived for the same superpotential in two dimensions in \cite{Lerche:1989cs}.

Theories related by $h_i' = h_i^*$ are also equivalent under complex conjugation $X_i \to \overline{X}^i$, so we can enlarge the group of dualities by including
\begin{equation}\label{d3def}
    d_3(\tau) = \bar{\tau}~.
\end{equation}
The duality group generated by \eqref{dualMat} and \eqref{d3def} is then the symmetric group $S_4 = A_4 \rtimes \mathbb{Z}_2$, where $\mathbb{Z}_2$ is complex conjugation. This $S_4$ is precisely the outer automorphism group of the discrete symmetry group $G$, and it acts by permuting the four inequivalent two-dimensional irreps of the group. This property will be very important when we discuss the action of the duality group on the spectrum of operators.

\subsubsection{The global structure of the conformal manifold}
According to the preceding discussion, the conformal manifold $\mathcal{M}$ for our model is given by the quotient
\begin{equation}
    \mathcal{M} = \mathbf{CP}^1 \Big/ S_4~.
\end{equation}
It turns out that the action of $S_4$ is not free since there are fixed points under some elements of $S_4$, so the conformal manifold has the structure of a 2-orbifold.

\begin{figure}[t!]
\begin{center}
   \includegraphics[width=0.49\textwidth]{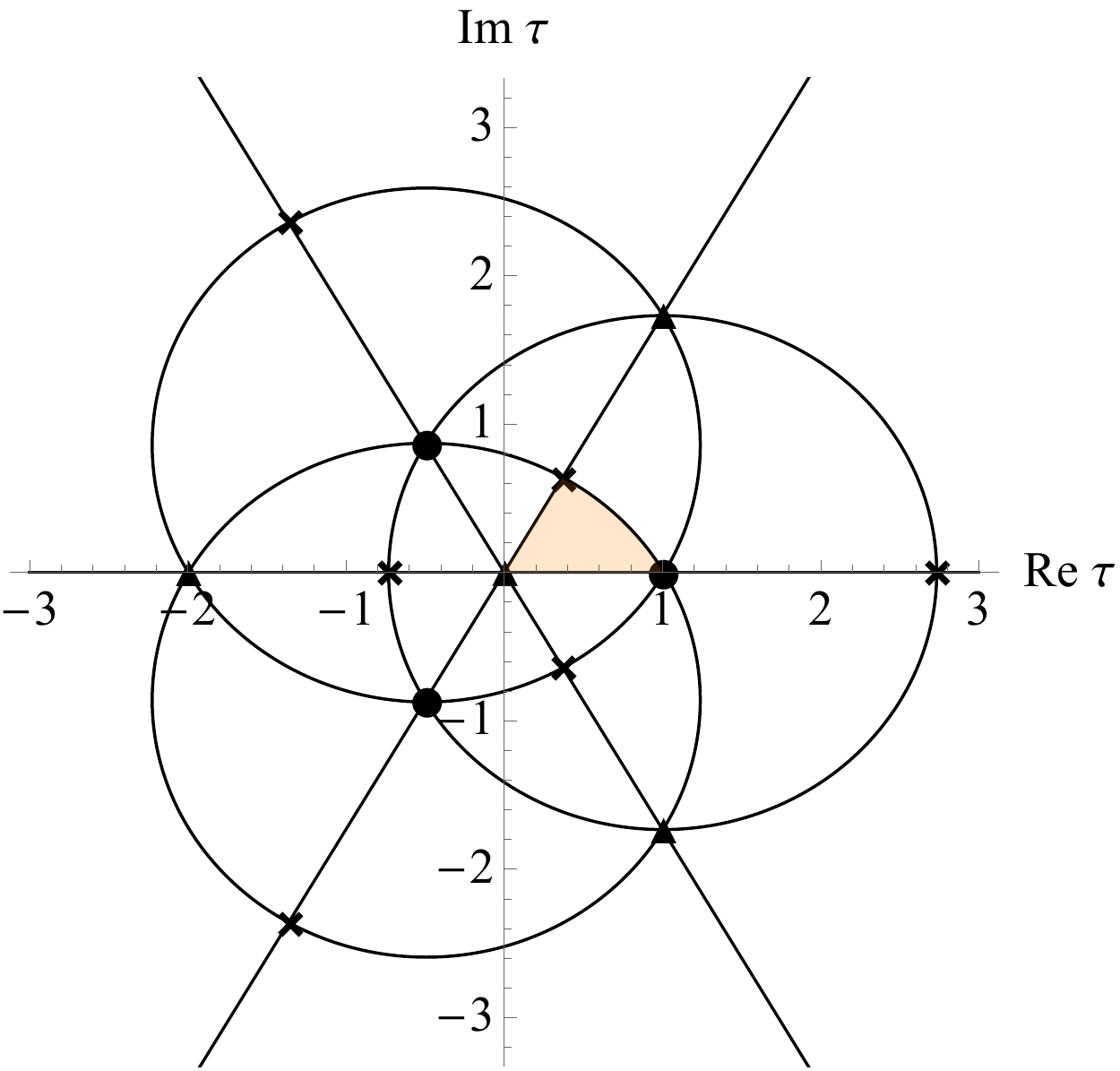}   \includegraphics[width=0.5\textwidth]{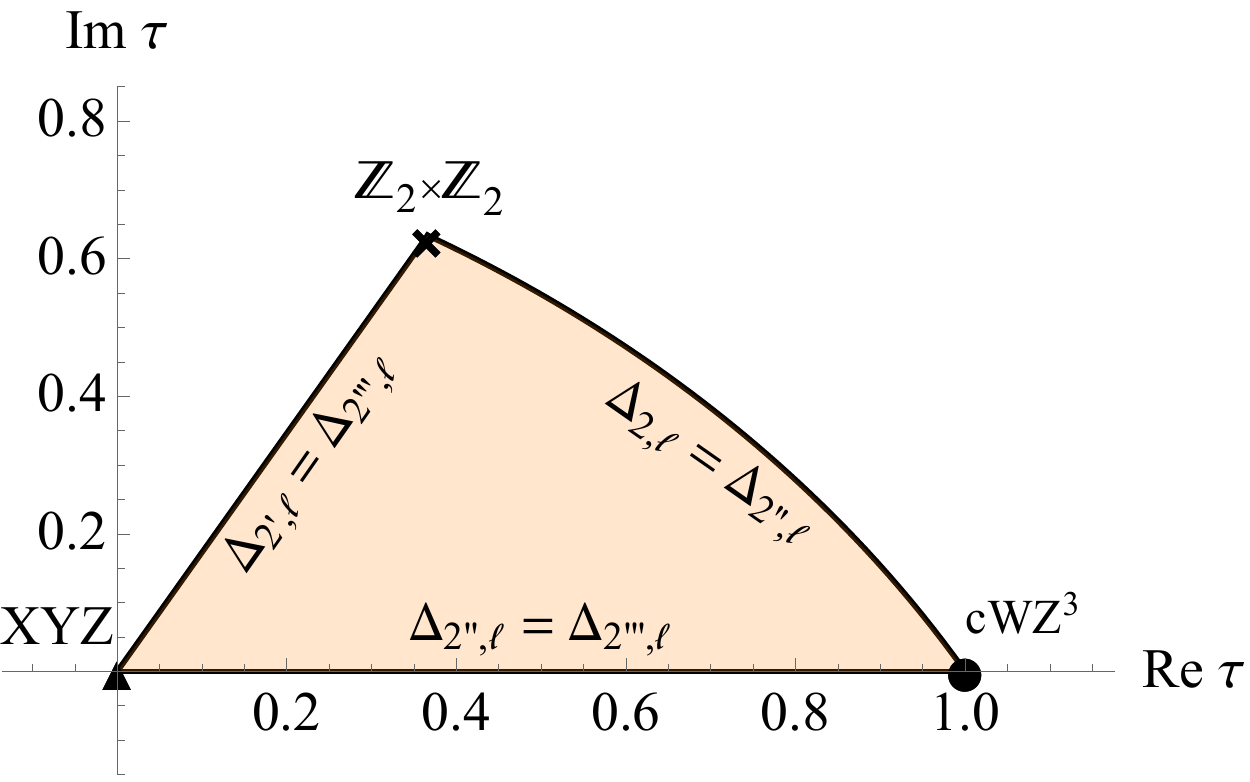}
 \caption{The black lines denote the images of the real line $\Im\,\tau = 0$ under the various dualities, and therefore each of them is invariant under an appropriate $\mathbb{Z}_2$ reflection subgroup of the duality group. Each cell defines a fundamental domain, and the orange shading is the domain we chose in \eqref{fun}. The triangles, circles, and crosses are dual to the XYZ, $\text{cWZ}^3$, and $\mathbb{Z}_2\times\mathbb{Z}_2$ theories with $\tau=0,1$ and $(1-\sqrt{3})\omega^2$, which are self-dual under the $S_3$, $S_3$, and $ \mathbb{Z}_2\times \mathbb{Z}_2$ subgroups of $S_4$ defined in \eqref{special}. There is an extra circle at $\tau=\infty$. The plot on the right is focused on a particular fundamental domain and shows additional degeneracies in the spectrum of quadratic operators \eqref{bilinears} along the boundaries of the fundamental domain.}
\label{fund}
\end{center}
\end{figure}

We can choose the fundamental domain $\mathbb{F}$ in the complex $\tau$ plane to be bounded by the curves
\es{fun}{
&\mathbb{L}_0:\quad\Im\,\tau=0\quad\text{for}\quad 0\leq\Re\,\tau\leq1\,,\\
&\mathbb{L}_1:\quad\Im\,\tau=\sqrt{3}\Re\,\tau\quad\text{for}\quad 0\leq\Re\,\tau\leq\frac{-1+\sqrt{3}}{2}\,,\\
&\mathbb{L}_2:\quad\left(\Re\,\tau+\frac12\right)^2+\left(\Im\,\tau+\frac{\sqrt{3}}{2}\right)^2=3\quad\text{for}\quad  \frac{-1+\sqrt{3}}{2}\leq\Re\,\tau\leq1\,,\\
}
as shown in Figure \ref{fund}. 
The three boundary curves $\mathbb{L}_0$, $\mathbb{L}_1$, and $\mathbb{L}_2$ are self-dual under the $\mathbb{Z}_2$ reflections $d_3$, $d_1 d_3$ and $d_2 d_1 d_3$ respectively.

The three vertices of the boundary are also fixed points under the action of the following subgroups of $S_4$:
\es{special}{
&\tau=1\quad \text{fixed by}\quad S_3 \quad \text{generated by}\quad \{d_3\,,d_2d_1\}\,,\\
&\tau=0\quad \text{fixed by}\quad  S_3 \quad \text{generated by}\quad \{d_3\,,d_1\}\,,\\
&\tau=(1-\sqrt{3})\omega^2\quad \text{fixed by}\quad  \mathbb{Z}_2\times\mathbb{Z}_2 \quad \text{generated by}\quad \{d_1d_3\,,d_2\}\,.
}
Two-dimensional orbifolds have been classified in \cite{thurston1978geometry} (see Chapter 13.3).\footnote{Reference \cite{thurston1978geometry} is available online at \url{http://library.msri.org/books/gt3m/}.} Our conformal manifold $\mathcal{M}$ is topologically a two-dimensional disk with three corner reflectors of order $(2,3,3)$.\footnote{If one does not include complex conjugation in the duality group, the resulting orbifold $\mathbf{CP}^1 / A_4$ is topologically a sphere $S^2$ with three elliptic points of order $(2,3,3)$. This space is a double cover of $\mathbf{CP}^1 / S_4$ via the identification $S_4 = A_4 \rtimes \mathbb{Z}_2$.} As shown in \cite{thurston1978geometry} its orbifold Euler characteristic is $\chi(\mathcal{M}) = 1/12$ and its orbifold fundamental group is $\pi_1(\mathcal{M}) = S_4$, i.e. it coincides with the duality symmetry group.

We now describe the enhanced flavor symmetries at each of the special points.

\noindent $\bullet\quad \tau=0$: the superpotential is 
\es{XYZW}{
W_{\tau=0} \propto X_1X_2X_3\,,
}
which describes the so called XYZ model. This theory has an enhanced $U(1)\times U(1)\rtimes S_3$ flavor symmetry, where $S_3$ permutes $X_i$ and $U(1)\times U(1)$ is generated by 
\es{XYZspecial}{
U(1)\times U(1):\quad \begin{pmatrix}
e^{{\rm i}\theta_1}&0&0\\
0&e^{{\rm i}(-\theta_1+\theta_2)}&0\\
0&0&e^{-{\rm i}\theta_2}\\
\end{pmatrix}\quad\text{s.t.}\quad \theta_1,\theta_2\in[0,2\pi)\,.
}
The $S_3$ in \eqref{special} is a subgroup of $U(1)\times U(1)\rtimes S_3$. The quadratic operator $\cO_{\bold2,0}$ in \eqref{bilinears} forms the lowest component of the current supermultiplets for each $U(1)$, and so has dimension
\es{xyzdim}{
\tau=0:\qquad\Delta_{\bold2,0}=d-2\,.
}

\noindent $\bullet\quad \tau=1$: to describe this point, it is more convenient to use the duality transformation
\begin{equation}
	\tau \to \frac{\tau + 2}{\tau - 1}~,
\end{equation}
which identifies this theory with the Wess-Zumino model at $\tau=\infty$, with superpotential 
\es{isingW}{
W_{\tau=\infty} \propto X_1^3+X_2^3+X_3^3\,.
}
This superpotential describes three decoupled critical Wess-Zumino models (cWZ$^3$). From \eqref{bilinears}, we see that the quadratic operators $\cO_{\bold2',0}$, $\cO_{\bold2'',0}$, and $\cO_{\bold2''',0}$ in this model are composites of a chiral field from two different cWZ's, and so their scaling dimensions are simply the sum of the two component chiral fields, i.e. $2\frac{d-1}{3}$. As we will see shortly, this implies that the operators in the dual $\tau=1$ theory then have scaling dimensions 
\es{ising}{
 \tau=1:\qquad\Delta_{\bold2,0}=\Delta_{\bold2'',0}=\Delta_{\bold2''',0}=2\frac{d-1}{3}\,.
 }

\noindent $\bullet\quad \tau=(1-\sqrt{3})\omega^2$: the superpotential has no special form and does not correspond to any well studied theory. We will refer to this theory by its extra $\mathbb{Z}_2\times \mathbb{Z}_2$ symmetry.

\subsubsection{Duality action on the operator spectrum}
\label{Morse}

Since the duality action is given by field redefinitions, we can determine explicitly how operators transform under duality. In this section we focus on the quadratic operators defined in \eqref{bilinears} and derive some interesting consequences on physical observables. Since $\mathcal{O}_{\mathbf{1},0}$ is invariant under complex conjugation and generic $U(3)$ transformations, it transforms into itself under all duality transformations, i.e. it is a self-dual operator. On the other hand, the four operators $\mathcal{O}_{\mathbf{2},0},\mathcal{O}_{\mathbf{2}',0},\mathcal{O}_{\mathbf{2}'',0}$ and $\mathcal{O}_{\mathbf{2}''',0}$ are permuted in the obvious way by $S_4$. This can be understood from the fact that $S_4$, seen as the group of outer automorphisms of $G$, acts by permuting the four two-dimensional representations. We also notice that along the boundary of the fundamental domain, the $\mathbb{Z}_2$ reflections that leave the three segments invariant relate operators in different pairs of doublet irreps, as indicated in Figure \ref{fund}.

The first consequence of this is the presence of monodromies as the coupling constant is adiabatically varied along non-trivial loops in the conformal manifold. Such loops are classified by $\pi_1(\mathcal{M})=S_4$ and the operators that mix under such motion are precisely those in the four two-dimensional representations. This is a global version of the Berry phase, which has recently been studied for infinitesimal loops in the conformal manifold in \cite{Baggio:2017aww}.

The other consequence is that the fixed points are critical points for the conformal dimensions of operators in the theory. We illustrate this phenomenon explicitly for the conformal dimension $\Delta_{\mathbf{1}}$ of the singlet operator $\mathcal{O}_{\mathbf{1},0}$. This operator is self-dual, so at a fixed point $\tau^* = d(\tau^*)$, where $d$ is a holomorphic duality transformation (that is, belonging to $A_4 \subset S_4$), we have
\begin{equation}
\label{Morse1}
	\frac{\partial d}{\partial \tau}(\tau^*) \partial_\tau \Delta_{\mathbf{1}}(\tau^*) = \partial_\tau \Delta_{\mathbf{1}}(\tau^*)~.
\end{equation} 
It is easy to check that for all the three inequivalent fixed points, there is a duality transformation such that $\frac{\partial d}{\partial \tau}(\tau^*) \neq 1$, which implies $\partial_\tau \Delta_{\mathbf{1}}(\tau^*) = 0$. We conclude that the fixed points are critical points for $\Delta_{\mathbf{1}}(\tau)$. The singlet conformal dimension is then a function on $\mathbf{CP}^1$ with $14= 4+6+4$ critical points, corresponding to the four XYZ points, the six $\mathbb{Z}_2 \times \mathbb{Z}_2$ points and the four cWZ${}^3$ points.\footnote{We assume that the only critical points are those predicted by the dualities and that they are non-degenerate.} Using Morse inequalities, we conclude that four of these points are minima, four are maxima and six are saddles.\footnote{More precisely, both minima and maxima must be present because the 0th and 2nd Betti numbers of the 2-sphere are non-vanishing ($b_0 = b_2 = 1$). Then the only way to correctly reproduce the Euler characteristic of the 2-sphere with these critical points is $\chi(\mathbf{CP}^1) = 4(-1)^0 + 6(-1)^1 + 4(-1)^2 = 2$, showing that the six $\mathbb{Z}_2 \times \mathbb{Z}_2$ points must be saddles.} Therefore the $\mathbb{Z}_2 \times \mathbb{Z}_2$ self-dual points are saddles for the singlet conformal dimension, while the XYZ/cWZ${}^3$ points are either minima/maxima or maxima/minima.\footnote{This analysis applies to all self-dual quantities in the CFT, including the appropriately contracted OPE coefficient that we compute and compare in the $4-\varepsilon$-expansion and in the numerical bootstrap, see equation \eqref{OPEeps} and Figure \ref{bootBoundOPE}.}
From the bootstrap results of \cite{Bobev:2015vsa,Chester:2015lej}, we conclude that the XYZ points are minima and the cWZ${}^3$ points are maxima for $\Delta_{\mathbf{1}}(\tau)$. Thus we have arrived at a qualitative picture of the behavior of the function $\Delta_{\mathbf{1}}(\tau)$ entirely based on non-perturbative arguments. This qualitative analysis is indeed confirmed by a fourth order $4-\varepsilon$-expansion computation shown in Figure \ref{epsSing} as well as a numerical conformal bootstrap analysis.

The discussion above can be repeated almost verbatim for the doublet operators. To illustrate this it is sufficient to study the operator $\mathcal{O}_{\mathbf{2},0}$ defined in \eqref{bilinears}. The results then extend directly to the other three doublet operators in \eqref{bilinears} by applying duality transformations. Focusing on the $A_4$ part of the duality group, it is possible to show that the operator $\mathcal{O}_{\mathbf{2},0}$ is left invariant only by the transformation $u_1$ defined in \eqref{dualMat}. This duality transformation acts on the coupling $\tau$ as $d_1(\tau)=\omega\tau$, see \eqref{dualPot}. As a consequence, $\Delta_{\mathbf{2}}$ obeys the equation (analogous to \eqref{Morse1} above)
\begin{equation}
\frac{\partial d_1}{\partial \tau}(\tau^*) \partial_\tau
\Delta_{\mathbf{2}}(\tau^*) = \partial_\tau
\Delta_{\mathbf{2}}(\tau^*)~.
\end{equation}
Here $\tau^*$ are the points on the complex $\tau$ plane left invariant by the action of $d_1$, namely $\tau=0$ and $\tau=\infty$, which are XYZ and cWZ${}^3$ points respectively. Since $\frac{\partial d_1}{\partial \tau}(\tau^*) \neq 1$ at these two fixed points, we find that the function $\Delta_{\mathbf{2}}(\tau)$ exhibits critical points at $\tau=0$ and $\tau=\infty$. Assuming that there are no other critical points on $\mathbf{CP}^1$, Morse inequalities imply that one of them is a minimum and the other is a maximum. Since we know that at $\tau=0$ the operator $\mathcal{O}_{\mathbf{2},0}$ is the lowest component of the $U(1)\times U(1)$ current multiplet, its conformal dimension $\Delta_{\mathbf{2}}(0)=1$ saturates the BPS bound, so this critical point is a minimum. It then immediately follows that $\tau=\infty$ is a maximum. This qualitative behavior of the conformal dimensions of the quadratic operators in the two-dimensional representations is indeed realized as we show explicitly using the $4-\varepsilon$-expansion and the numerical bootstrap, see Figure \ref{epsDub}.

\subsection{Supersymmetric localization results}
\label{subsec:localization}

It is possible to calculate certain quantities at the IR fixed point of the model \eqref{eq:superpotential} using the technique of supersymmetric localization (see \cite{Pestun:2016zxk} for a recent review and a list of references).  For this model, in $d>2$, we are aware of two quantities that can be computed exactly.\footnote{In $d=2$ it is possible to also calculate the Zamolodchikov metric on the conformal manifold (see for example Section 4.2 in \cite{Gomis:2012wy}) as well as correlation functions of other chiral and anti-chiral operators following the 4d approach of \cite{Gerchkovitz:2016gxx}.  However, these quantities can also be computed exactly using the description of the IR SCFT as a $\Z_3$ orbifold theory \cite{Lerche:1989cs,Verlinde:1991ci,Lin:2016gcl,Lerche:1989uy,Cecotti:1990wz}.}

The first quantity is the coefficient $C_T$ that appears in the two point function of the canonically normalized stress-energy tensor which in Euclidean $\mathbb{R}^3$ is given by:
 \es{cTDef}{
    \langle T_{\mu\nu}(\vec{x}) T_{\rho \sigma}(0) \rangle = \frac{C_T}{96} \left(P_{\mu\rho} P_{\nu \sigma} + P_{\nu \rho} P_{\mu \sigma} - P_{\mu\nu} P_{\rho\sigma} \right) \frac{1}{16 \pi^2 \vec{x}^2} \,, \qquad P_{\mu\nu} \equiv \delta_{\mu\nu} \partial^2 - \partial_\mu \partial_\nu \,.
 }
It was shown in \cite{Closset:2012ru,Nishioka:2013gza} that $C_T$ can be determined by differentiating the supersymmetric partition function of the three-dimensional $\mathcal{N}=2$ SCFT on a squashed $S^3$, which in turn was computed by localization in \cite{Hama:2011ea,Imamura:2011wg}. The result is given by the concise formula\footnote{We use the same normalization as in \cite{Bobev:2015jxa}. For a free chiral multiplet one has $C_T=6$, thus $C_T=18$ for three free chiral multiplets.}
\begin{equation}
C_T = \frac{48}{\pi^2}\frac{\partial^2F(b)}{\partial b^2}\bigg|_{b=1}\;.
\end{equation}
Here $F(b) = -\log Z(b)$ is the squashed sphere supersymmetric free energy.  The quantity $b$ is the squashing parameter controlling the deviation of the $S^3$ metric from the Einstein one.  The round sphere is obtained for $b=1$. For a theory of a single chiral multiplet of R-charge $\Delta$, one finds the following compact integral expression  \cite{Nishioka:2013gza}
\begin{equation}\label{F3dchiral}
F_{\rm chiral}(b) = - \int_{0}^{\infty} \frac{dx}{2x} \left( \frac{\sinh(2(1-\Delta)\hat{\omega} x)}{\sinh(bx)\sinh(x/b)} - \frac{2\hat{\omega}(1-\Delta)}{x}\right)\;,
\end{equation}
where $\hat{\omega} \equiv \frac{b+b^{-1}}{2}$. In the case of interest to us, $F(b) = 3 F_{\rm chiral}(b)$, where $F_{\rm chiral}(b)$ is evaluated with $\Delta=2/3$. This gives \cite{Witczak-Krempa:2015jca}
\es{cTloc}{
C_T=\frac{32}{27}\left(16-\frac{9\sqrt{3}}{\pi}\right)\approx 13.0821\;.
}
Note that the value of $C_T$ does not depend on the value of the marginal coupling $\tau$ parameterizing the conformal manifold.

The second quantity that can be computed exactly using supersymmetric localization is the coefficient $C_J$ that appears in the two-point function of canonically normalized conserved currents at the XYZ point.  The XYZ theory has two $U(1)$ flavor symmetries that act on the $X_i$ with charges $(1, -1, 0)$ and $(0, 1, -1)$ (see \eqref{XYZspecial}).  The two-point function of each of the two $U(1)$ canonically-normalized conserved currents takes the form\footnote{With this definition, a free massless chiral multiplet has $C_J  =1$ for the $U(1)$ flavor symmetry under which it has charge $1$.}
 \es{CurrentTwo}{
  \langle J_\mu(\vec{x}) J_{\nu}(0) \rangle = \frac{C_J}{128 \pi^2} P_{\mu\nu} \frac{1}{\vec{x}^2} \,.
 }
The coefficient $C_J$ can be computed from the second derivative of the free energy on a round $S^3$ in the presence of a real mass parameter $m$:
\es{cJ}{
  C_J = \frac{2}{\pi^2} \frac{d^2 F}{dm^2} \bigg|_{m=0}\,.
 }
For a chiral multiplet of R-charge $\Delta$ and charge $q$ under the $U(1)$ symmetry associated with the real mass $m$, we have
 \es{Fm}{
  F_{\rm chiral}(m) = -\ell (1 - \Delta + {\rm i} q m) \,, 
 }
where
\begin{equation}
\ell(z) \equiv - z \log (1 - e^{2 \pi {\rm i}z} ) + \frac{{\rm i}}{2} \left( \pi z^2 + \frac 1\pi \text{Li}_2(e^{2 \pi {\rm i} z}) \right) - \frac{{\rm i} \pi}{12}\;.
\end{equation}
For any of the two $U(1)$ currents mentioned above, we have $F = -\ell (1/3 + {\rm i} m) - \ell (1/3 - {\rm i} m) - \ell(1/3)$, which gives \cite{Chester:2015lej,Chester:2015qca}
\es{GotcJ}{
  C_J = \frac{16}{9} - \frac{4}{\sqrt{3} \pi} \approx 1.043 \,.
 }
%

\section{Results in $d = 4-\varepsilon$}
\label{sec:epsilon}

The $4 -\varepsilon$ expansion \cite{Wilson:1971dc} has been extremely successful in computing observables of strongly coupled theories in three dimensions. The idea is to compute the physical quantities of interest in dimension $d=4-\varepsilon$, express them as power series in $\varepsilon$, and then use an extrapolation method in order to evaluate them at $\varepsilon = 1$, which corresponds to the 3d theory. In this section we use the $4 - \varepsilon$ expansion to compute the scaling dimensions of the non-protected quadratic operators to order $\varepsilon^4$ and the structure constants of the chiral ring at order $\varepsilon^2$. We also provide the Zamolodchikov metric up to order $\varepsilon^2$.

\subsection{Generalities}

In a generic cubic model
\begin{align}\label{WK}
W =\frac{1}{6} h^{ijk}X_i X_j X_k\;,
\end{align}
the beta function for the physical coupling\footnote{We can think of \eqref{betafh} as written in a non-holomorphic scheme where the K\"ahler potential is canonical $K = Y_i \overline Y^i$ and not renormalized. This is related to the holomorphic scheme that we use in the rest of this section by the identifications $X_i = M_i{}^j  Y_j$ and $\overline{X}^i = \overline{M}^i{}_j \overline{Y}^j$, such that the superpotential $W = \frac16 h^{ijk} X_i X_j X_k$ is not renormalized ($M^i{}_j$ can be taken to be the inverse square root of the wave-function normalization matrix $Z^i{}_j$, which is positive and Hermitian).  The beta function in \eqref{betafh} is then the beta function of the physical coupling $ h^{ijk}_\text{phys} = h^{\ell mn} M^i{}_\ell M^j{}_m M^k{}_n$.  To see that the beta function takes the form~\eqref{betafh}, note that the anomalous dimension matrix of the $X_i$ is $\gamma^i{}_j = \mu \frac{\partial \log(M)^i{}_j }{\partial \mu}$.  Thus, the logarithmic running of $h^{ijk}_\text{phys}$ is given by the sum between the classical logarithmic running of $h^{ijk}$ (the first term in \eqref{betafh}) and the running of $M^i{}_j$ (the last three terms in \eqref{betafh}).} has the following general expression
\begin{equation}
\label{betafh}
\beta^{ijk} \equiv \mu \frac{\partial {h}^{ijk}}{\partial \mu} = -\frac{\varepsilon}{2} {h}^{ijk} + {\gamma^i}_m {h}^{mjk} +  {\gamma^j}_m {h}^{imk} + {\gamma^k}_m {h}^{ijm}~,
\end{equation}
where ${\gamma^i}_j$ is the matrix of anomalous dimensions for the chiral superfields $X_i$. At a fixed point $\beta^{ijk}(h_*) = 0$, and the discrete symmetry group $G$ implies that the matrix of anomalous dimensions is proportional to the identity ${\gamma^i}_j(h_*) = \gamma(h_*) {\delta^i}_j$. From \eqref{betafh} it immediately follows that
\begin{equation}
\label{eq:gamma1diag}
\gamma(h_*) = \frac{\varepsilon}{6}~,
\end{equation}
or equivalently the conformal dimensions of the $X_i$'s are
\begin{equation}\label{deltaX}
\Delta_{X_i}=\frac{d-2}{2}+\frac{\varepsilon}{6}=\frac{d-1}{3}\;.
\end{equation}
This result is of course compatible with the general expectation for the value of $\Delta_{X_i}$ at a superconformal  fixed point \eqref{Deltaqrel}.

It is worthwhile to provide another argument for the existence of a one-dimensional conformal manifold for our model. The existence of a fixed point imposes ten complex equations $\beta^{ijk}= 0$ for the ten complex couplings $h^{ijk}$. Nine of these couplings can be eliminated by a $U(3)$ field redefinition. In addition, since the anomalous dimension matrix ${\gamma^i}_j$ is Hermitian, it is clear from \eqref{betafh} that there are only 9 independent conditions to ensure that the $\beta$ functions vanish. Thus in general we expect a one complex parameter family of solutions of the fixed point equations  $\beta^{ijk}= 0$. This is a variation of the argument of Leigh-Strassler for the existence of fixed points in 4d $\mathcal{N}=1$ gauge theories \cite{Leigh:1995ep}. Notice that in 3d $\mathcal{N}=2$ theories the couplings $h^{ijk}$ are not marginal as was assumed in \cite{Leigh:1995ep}. Nevertheless, as appreciated in \cite{Strassler:1998iz}, the essence of the argument in  \cite{Leigh:1995ep} relies on linear relations between the beta functions, which indeed exist in our model.

\subsection{A line of fixed points}

The beta function for the generic cubic model with superpotential given in  \eqref{WK} is known to four loops \cite{Ferreira:1996ug,Ferreira:1996az}. The fixed point is determined by solving the algebraic relation in  \eqref{eq:gamma1diag}. It is convenient to parametrize the coupling constant space with the coordinates\footnote{The overall phase of $h_1$ and $h_2$ can be changed by an $R$-symmetry transformation, so it is a redundant coupling and does appear in the $\beta$ function.}
\begin{align}
r^2 & = 2|h_1|^2 + |h_2|^2~,\\
\tau & = \frac{h_2}{h_1}~.
\end{align}
The anomalous dimension is then given by
\begin{equation}\label{betaexpansionr}
\gamma(r,\tau,\bar{\tau}) =  f_1(\tau,\bar{\tau}) r^2 + f_2(\tau,\bar{\tau}) r^4 + f_3(\tau,\bar{\tau}) r^6 + f_4(\tau,\bar{\tau}) r^8 + O(r^{10})~,
\end{equation}
where
\begin{align}\label{betarvarex}
f_1(\tau,\bar{\tau}) & = \frac{1}{2^5\pi^2}~,\\
f_2(\tau,\bar{\tau}) & = -\frac{1}{2^9\pi^4}~,\\
f_3(\tau,\bar{\tau}) & = \frac{1}{2^{13} \pi^6}\left(\frac{5}{4} + 3\zeta(3) \frac{\left(\left(\tau ^3+2\right) \left(\bar{\tau }^3+2\right)+18 \tau  \bar{\tau }\right)}{ (2+\tau \bar{\tau})^3}\right)~,\\
f_4(\tau,\bar{\tau}) & = -\frac{1}{2^{17} \pi^8}\left[\frac{9}{4} + \left(15 \zeta(3)-\frac{9}{2}\zeta(4)\right)\frac{\left(\left(\tau ^3+2\right) \left(\bar{\tau }^3+2\right)+18 \tau  \bar{\tau }\right)}{ (2+\tau \bar{\tau})^3 }\right.\nonumber\\
&  \phantom{ = } \left.\qquad\qquad~+ 20\zeta(5)\frac{\left(\left(\tau  \bar{\tau }+2\right)^4-8 \left(1-\tau ^3\right) \left(1-\bar{\tau }^3\right)\right)}{(2+\tau\bar{\tau})^4}\right]~.
\end{align}
By equating $\gamma(r, \tau, \bar \tau) = \varepsilon / 6$, we can compute the fixed point couplings up to order $\varepsilon^4$:
\begin{equation}\label{rsqrd}
r^2 = a_1(\tau,\bar{\tau}) \varepsilon + a_2(\tau,\bar{\tau}) \varepsilon^2 + a_3(\tau,\bar{\tau}) \varepsilon^3 + a_4(\tau,\bar{\tau}) \varepsilon^4 + O(\varepsilon^{5})~,
\end{equation}
with
\begin{align}
a_1 & = \frac{1}{6f_1}~,&
a_2 & = -\frac{f_2}{6^2f_1^3}~,&
a_3 & = \frac{2 f_2^2 - f_1 f_3}{6^3 f_1^5}~,&
a_4 & = -\frac{5f_2^3 - 5 f_1 f_2 f_3 + f_1^2 f_4}{6^4 f_1^7}~.&
\end{align}
A simple computation shows that the functions $f_i$'s (and consequently $a_i$'s) are invariant under the duality transformations generated by 
\eqref{dualPot}, \eqref{d3def}. Since $(\tau, \bar \tau)$ are arbitrary parameters, we thus have a one complex-dimensional manifold of fixed points, i.e.~our conformal manifold.

\subsection{Conformal dimensions of quadratic operators}
\label{subsec:confquad}

It is possible to use the $4-\varepsilon$ expansion to compute the scaling dimensions of the quadratic operators in our model as a function of $\tau$. There are 21 real quadratic operators, of which the six given in \eqref{chiralBis} and their complex conjugates belong to chiral or anti-chiral multiplets and thus have protected scaling dimensions. The scaling dimensions of the remaining nine operators of zero R-charge given in equation \eqref{bilinears} are not protected by supersymmetry and depend on the marginal coupling $\tau$. 

These scaling dimensions can be computed directly from the matrix of anomalous dimensions for the fundamental fields $X_i$ \cite{Jack:1998iy}. Indeed, the beta functions for the couplings $(m^2)^i{}_j X_i \overline{X}^j$ can be computed as
 \es{betam2}{
  \beta_{(m^2)^i{}_j} = \mu \frac{d(m^2)^i{}_j}{d\mu}
   =  - 2(m^2)^i{}_j + \left[ \left[ (m^2)^l{}_p h^{pmn} + (m^2)^m{}_p h^{lpn}  + (m^2)^n{}_p h^{lmp}  \right] \frac{\partial \gamma^i{}_j}{ \partial h^{lmn}} + \text{c.c.} \right] \,.
 }
The scaling dimensions of the quadratic operators are the eigenvalues of the $9 \times 9$ matrix
 \es{Matrix99}{
  \Delta^i{}_j{}^l{}_k = (4 - \varepsilon) \delta^i_k \delta^l_j+ \frac{\partial \beta_{(m^2)^i{}_j}}{\partial (m^2)^k{}_l} \,,
 } 
where we think of the indices $(i, j)$ as the row indices and $(k, l)$ as the column indices. The operators in \eqref{bilinears} directly provide a basis of eigenvectors for the resulting matrix, from which the anomalous dimensions can be immediately extracted.

Plugging in the couplings $h^{ijk}$ as well as the anomalous dimension matrix $\gamma^i{}_j$ corresponding to the model \eqref{eq:superpotential}, and using the results of the previous subsection, we can then compute the conformal dimensions of all the unprotected quadratic operators up to order $\varepsilon^4$. The conformal dimension $\Delta_{\mathbf{1}}$ of the singlet operator $X_i \overline{X}^i$ reads
\begin{eqnarray}\label{Delta1}\centering
\Delta_{\mathbf{1}} &= & 2-\frac{1}{3}\varepsilon ^2+\frac{1}{3} \left[\frac{1}{6}+2 \zeta (3)\frac{ \left(\left(\tau ^3+2\right) \left(\bar{\tau }^3+2\right)+18 \tau  \bar{\tau }\right)}{\left(2+\tau  \bar{\tau }\right)^3}\right]\varepsilon^3- \nonumber\\
&&-\frac{1}{9}\left[\frac{7}{12}-\left(7\zeta(3)-\frac{9}{2}\zeta(4)\right)\frac{ \left(\left(\tau ^3+2\right) \left(\bar{\tau }^3+2\right)+18 \tau  \bar{\tau }\right)}{\left(2+\tau  \bar{\tau }\right)^3}+\right.\\
&&\left. \qquad\qquad\qquad\qquad+20 \zeta (5) \frac{\left(\left(\tau  \bar{\tau }+2\right)^4-8 \left(1-\tau ^3\right) \left(1-\bar{\tau }^3\right)\right)}{ \left(2+\tau \bar{\tau }\right)^4}\right]\varepsilon^4 + O(\varepsilon^5)~.\nonumber
\end{eqnarray}
It is pleasing to see that at each order in $\varepsilon$ the conformal dimension is invariant under the duality group and exhibits critical points at the three inequivalent self-dual points, as predicted from the considerations of the previous section.

For the quadratic operator $\mathcal{O}_{\mathbf{2},0}$ in \eqref{bilinears} one finds the conformal dimension
\begin{eqnarray}\label{Delta2}\centering
\Delta_{\mathbf{2}}&=& 2-\frac{2}{2+\tau  \bar{\tau }}\varepsilon+\frac{1}{3}\frac{\tau  \bar{\tau } \left(1-\tau  \bar{\tau }\right)}{ \left(2+\tau  \bar{\tau }\right)^2}\varepsilon^2\notag\\
&&+\frac{1}{9}\frac{\tau  \bar{\tau } }{ \left(2+\tau  \bar{\tau }\right)^3}\left[\frac{(10-\tau\bar{\tau})(1-\tau\bar{\tau})}{2}+6\zeta(3)\frac{3(1-\tau\bar{\tau})^2+(1-\tau^3)(1-\bar{\tau}^3)}{2+\tau\bar{\tau}}\right]\varepsilon^3\notag\\
&&+\frac{1}{27}\frac{\tau  \bar{\tau } }{ \left(2+\tau  \bar{\tau }\right)^4}\left[\frac{(100-26\tau\bar{\tau}+7\tau^2\bar{\tau}^2)(1-\tau\bar{\tau})}{4}\right.\\
&&\left.-3\zeta(3)\frac{2(1-\tau\bar{\tau})(2+\tau\bar{\tau})^2+(2+7\tau\bar{\tau})[3(1-\tau\bar{\tau})^2+(1-\tau^3)(1-\bar{\tau}^3)]}{2+\tau\bar{\tau}}\right.\notag\\
&& \left.+\frac{27}{2}\zeta(4)[3(1-\tau\bar{\tau})^2+(1-\tau^3)(1-\bar{\tau}^3)] -20\zeta(5)\frac{3\tau\bar{\tau}(2+\tau\bar{\tau})(1-\tau\bar{\tau})^2+8(1-\tau^3)(1-\bar{\tau}^3)}{2+\tau\bar{\tau}}\right]\varepsilon^4 \notag \\
&&+ O(\varepsilon^5) \notag.
\end{eqnarray}
The conformal dimensions of the other doublet operators \eqref{bilinears} can be obtained from $\Delta_{\mathbf{2}}$ above by the following substitutions 
\begin{equation}\label{Delta2123}
\begin{split}
\Delta_{\mathbf{2}} &\to \Delta_{\mathbf{2}'}\;, \qquad \qquad \tau \to \frac{\tau+2}{\tau-1}\;, \qquad \bar{\tau} \to \frac{\bar{\tau}+2}{\bar{\tau}-1}\;,\\
\Delta_{\mathbf{2}} &\to \Delta_{\mathbf{2}''}\;, \qquad \qquad \tau \to \frac{\omega\tau+2}{\omega\tau-1}\;, \qquad \bar{\tau} \to \frac{\bar{\omega}\bar{\tau}+2}{\bar{\omega}\bar{\tau}-1}\;,\\
\Delta_{\mathbf{2}} &\to \Delta_{\mathbf{2}'''}\;, \qquad \qquad \tau \to \frac{\omega^2\tau+2}{\omega^2\tau-1}\;, \qquad \bar{\tau} \to \frac{\bar{\omega}^2\bar{\tau}+2}{\bar{\omega}^2\bar{\tau}-1}\;.
\end{split}
\end{equation}
We emphasize that these results are obtained using the $4-\varepsilon$ expansion without using the duality properties of our model. The fact that the results for these conformal dimensions are compatible with the duality transformations constitutes a strong consistency check of our calculations.

When the results for the conformal dimensions in \eqref{Delta1}, \eqref{Delta2}, and \eqref{Delta2123} are restricted to order $\varepsilon^2$ we find agreement with the two loops results presented in \cite{Fei:2016sgs,Zerf:2016fti,Chester:2015lej,Chester:2015qca}.

\subsubsection*{Resummation}

\begin{figure}[t!]
\begin{center}
   \includegraphics[width=0.85\textwidth]{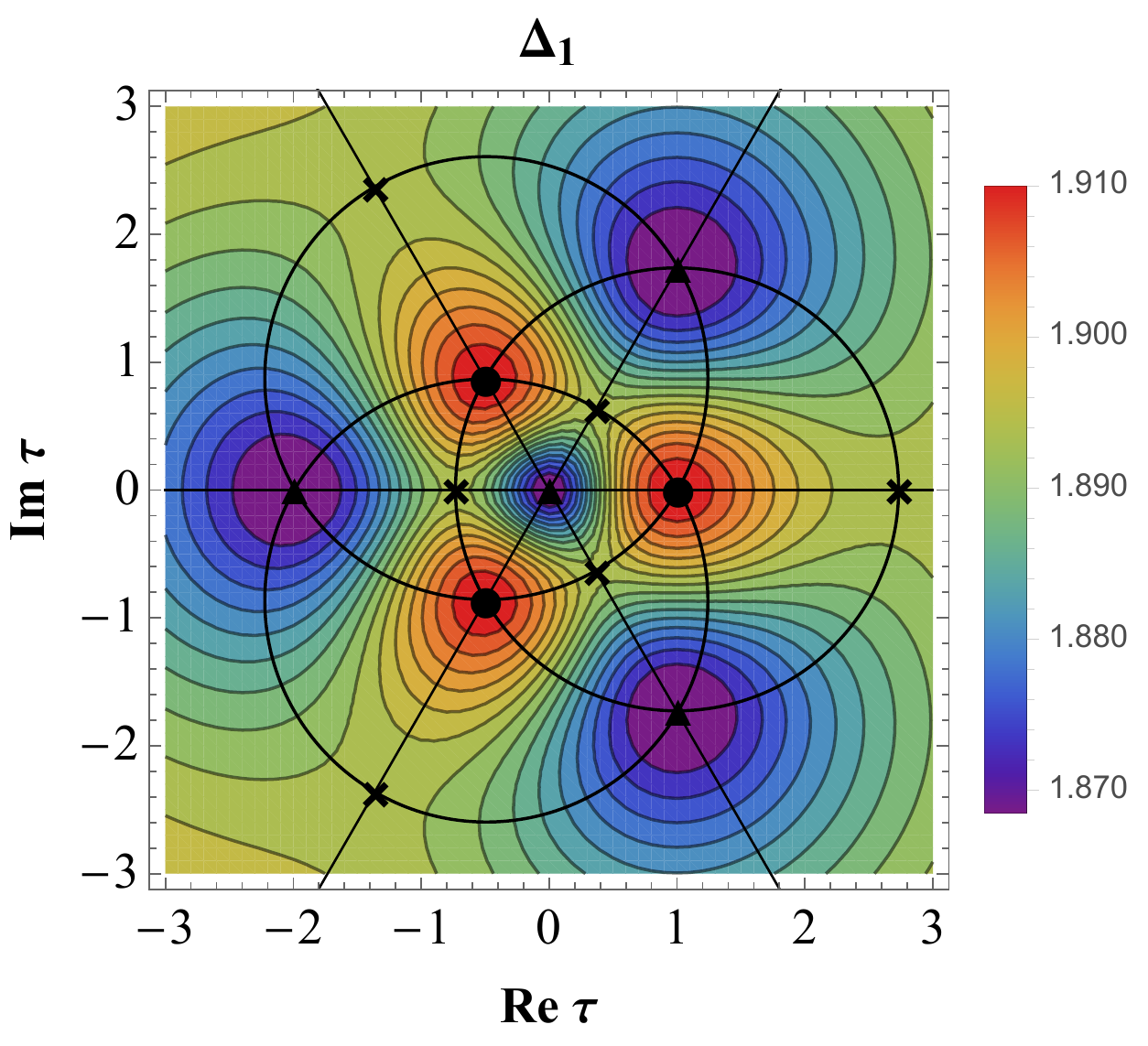}
 \caption{ The 3-loop Pad\'e[1,2] resummed $4-\varepsilon$-expansion values for the singlet. The cross, circle, and triangle denote values of $\tau$ that correspond to the $\mathbb{Z}_2\times\mathbb{Z}_2$, cWZ$^3$, and XYZ models, respectively.}
\label{epsSing}
\end{center}
\end{figure}

\begin{figure}[t!]
\begin{center}
   \includegraphics[width=0.85\textwidth]{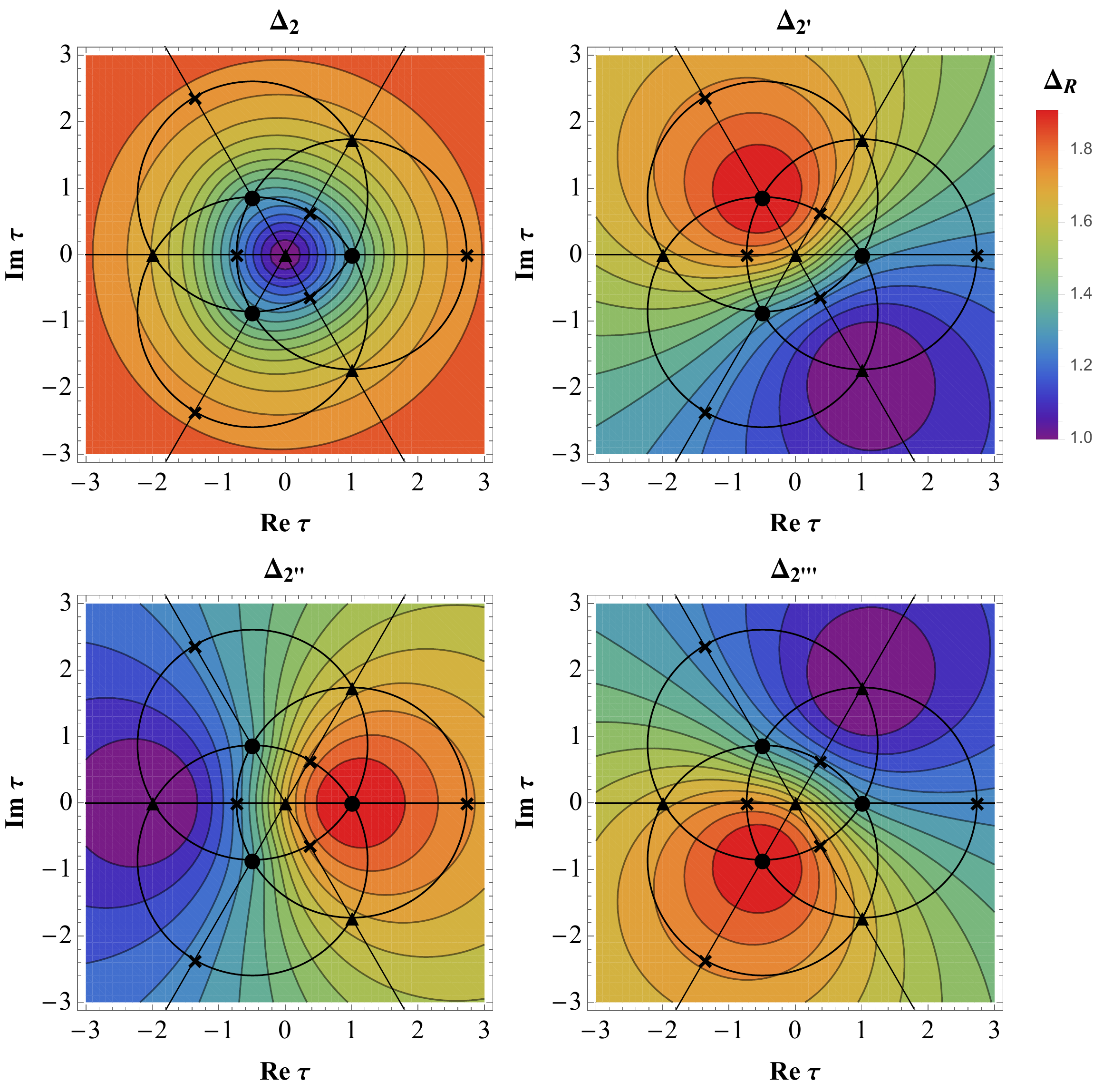}
 \caption{ The 3-loop Pad\'e[1,2] resummed $4-\varepsilon$-expansion values for the doublets. The cross, circle, and triangle denote values of $\tau$ that correspond to the $\mathbb{Z}_2\times\mathbb{Z}_2$, cWZ$^3$, and XYZ models, respectively.}
\label{epsDub}
\end{center}
\end{figure}

In order to find meaningful results when $\varepsilon  = 1$, we need to employ a resummation method. For the scaling dimensions, which are known to 4-loops, we have used the Pad\'e approximation method, which has been shown to be successful in related examples. We find that the results that match the numerical bootstrap the best are given by Pad\'e[1,2], which only uses the 3-loop result. We plot these doublet and singlet scaling dimension in Figures \ref{epsSing} and \ref{epsDub}, which demonstrate how these operators transform under the dualities. For the OPE coefficients computed in the next section we only find 2-loop results and thus do not use any resummation.

\subsection{The chiral ring and the Zamolodchikov metric}
\label{sec:chiralring}

In this subsection we discuss the structure of the chiral ring of the theory \cite{Lerche:1989uy}. Chiral operators are obtained by taking combinations of the form $X_{i_1} X_{i_2} \cdots X_{i_n}$. However, most of these combinations are not superconformal primaries due to the equations of motion that schematically read 
\begin{equation}
\label{eq:descendantrelation}
\overline{D}^2 \overline{X}^i = \frac{\partial W}{\partial X_i}~.
\end{equation} 
Therefore these operators do not belong to the chiral ring. As a consequence, in order to find the spectrum of chiral operators we can set to zero the descendant combinations $\frac{\partial W}{\partial X_i} \sim 0$.

At a generic point $\tau$, it is easy to show that the chiral ring consists of finitely many operators. Indeed, there are three independent conditions involving chiral quadratic operators:
\begin{align}
\label{eq:descendants}
\mathcal{W}_1 \equiv X_2X_3 + \frac{\tau}{2} X_1^2 & \sim 0~, & \mathcal{W}_2 \equiv X_1 X_3 + \frac{\tau}{2} X_2^2 & \sim 0~, & \mathcal{W}_3 \equiv X_1 X_2 + \frac{\tau}{2} X_3^2 & \sim 0~.
\end{align}
At the cubic level, multiplying these relations with the three chiral fields $X_i$ gives nine relations that remove all but one of the ten possible chiral combinations. All the quartic or higher operators are removed as well. We can count the chiral operators more systematically by using the generating function
\begin{equation}
P(t) = \mathrm{Tr}\,t^{J_R}~,
\end{equation}
where $J_R$ is the generator of the $R$-symmetry current and the trace is taken over the space of chiral primaries. Each of the three chiral operators contributes a factor $\frac{1}{1-t^{2/3}}$, while the chiral ring relations contribute a factor $(1-t^{4/3})^3$. All in all, we have
\begin{equation}
P(t) = \left(\frac{1-t^{4/3}}{1-t^{2/3}}\right)^3 = 1 + 3\, t^{2/3} + 3\, t^{4/3} + t^2~,
\end{equation}
so in addition to the three chiral fields $X_i$, we find three quadratic chiral primary operators $\mathcal{Q}_I$ and one cubic chiral primary operator $\mathcal{O}$. The quadratic and cubic chiral operators are given by\footnote{These combinations are orthogonal to the chiral ring relations at tree level. We have checked explicitly that they remain chiral primaries to order $\varepsilon$, but their explicit expressions in terms of the fundamental fields receive corrections at higher orders in $\varepsilon$. The corrections to $\mathcal{Q}_I$ and $\mathcal{O}$ can be chosen to be proportional to $\mathcal{W}_I$ and $W$ respectively, thus such corrections will not contribute to the quantities that we compute below.}
\begin{align}
\label{eq:chirfirst}
\mathcal{Q}_1 & \equiv X_1^2 - \bar{\tau}\, X_2 X_3 + O(\varepsilon^2)~,\\
\mathcal{Q}_2 & \equiv X_2^2 - \bar{\tau}\, X_1 X_3 + O(\varepsilon^2)~,\\
\mathcal{Q}_3 & \equiv X_3^2 - \bar{\tau}\, X_1 X_2 + O(\varepsilon^2)~,\\
\label{eq:chirlast}
\mathcal{O} & \equiv X_1^3 + X_2^3 + X_3^3 - 3\bar{\tau} X_1 X_2 X_3 + O(\varepsilon^2)~.
\end{align}
In this basis, it is easy to compute the chiral ring structure constants, defined by
 \es{OPEs}{
X_i(x)X_j(0) & = C_{ij}^K \mathcal{Q}_K(0) + \ldots~,\\
X_i(x)\mathcal{Q}_J(0) & = C_{iJ} \mathcal{O}(0) + \ldots~,
 }
where the ellipses denote terms that go to zero as $x \to 0$. The non-vanishing structure constants are given by
 \es{OPEcoeff}{
C_{11}^1 = C_{22}^2 = C_{33}^3 & = \frac{2}{2+\tau \bar{\tau}}~,\\
C_{12}^3 = C_{13}^2 = C_{23}^1 & = -\frac{\tau}{2+\tau \bar{\tau}}~,\\
C_{11} = C_{22} = C_{33} & = \frac{1}{3}~.
 }
The three-point functions are then obtained by computing the two-point functions of the operators in \eqref{eq:chirfirst}-\eqref{eq:chirlast}. The discrete symmetries imply that
 \es{2ptfunchiral}{
\left<X_i(x)\, \overline{X}^j(0)\right> & = G_1(\tau,\bar{\tau})\,\frac{{\delta_i}^j}{|x|^{2(d-1)/3}}~,\\
\left<\mathcal{Q}_I(x)\, \overline{\mathcal{Q}}^J(0)\right> & = G_2(\tau,\bar{\tau})\,\frac{{\delta_I}^J}{{|x|^{4(d-1)/3}}}~,\\
\left<\mathcal{O}(x)\,\overline{\mathcal{O}}(0)\right> & = G_3(\tau,\bar{\tau})\,\frac{1}{{|x|^{2(d-1)}}}~.
 }
The functions $G_i$ transform very simply under the duality group generated by \eqref{dualMat}, \eqref{dualPot}. The only generator acting non-trivially on them is $d_2(\tau) = \frac{\tau+2\omega^2}{\omega\tau-1}$, which gives
\begin{align}
d_2: \quad G_1 & \rightarrow G_1~,\nonumber\\
\label{eq:Gitransf}
G_2 & \rightarrow \frac{3}{(\omega^2\tau-1)(\omega \bar{\tau} - 1)}\,G_2~,\\\nonumber
G_3 & \rightarrow \frac{3}{(\omega^2\tau-1)(\omega \bar{\tau} - 1)}\,G_3~.
\end{align}

It is possible to calculate the two-point functions in \eqref{2ptfunchiral} using perturbation theory in $\varepsilon$. The details of this calculation to order $\varepsilon^2$ are outlined in Appendix \ref{sec:G1comp}. The calculation of $G_1(\tau,\bar{\tau})$ is quite standard and we find that the result is independent of $\tau$ to this order, while the computation of $G_2$ and $G_3$ to order $\varepsilon^2$ is more complicated and involves the same Feynman diagrams that were computed in \cite{Baggio:2014ioa} (see Section 5.1 in \cite{Baggio:2014ioa}). In order to discuss normalization and scheme independent quantities it is useful to present the final result as the following ratios:
 \es{GRatio}{
\frac{G_2(\tau,\bar{\tau})}{G_1(\tau,\bar{\tau})^2} & = (2+\tau \bar{\tau}) + \frac{2\,\zeta(3)}{3} \, \left(\frac{(1-\tau \bar{\tau})^3+(1-\tau^3)(1-\bar{\tau}^3)}{(2+\tau \bar{\tau})^2}\right)\, \varepsilon^2 + O(\varepsilon^3)\;,\\
\frac{G_3(\tau,\bar{\tau})}{G_1(\tau,\bar{\tau})G_2(\tau,\bar{\tau})}  & = 9 + 12\,\zeta(3)\, \left(\frac{(1-\tau \bar{\tau})^3+(1-\tau^3)(1-\bar{\tau}^3)}{(2+\tau \bar{\tau})^3}\right)\, \varepsilon^2 + O(\varepsilon^3)\;.
 }
%

\subsubsection{The four point functions in the chiral channel}

In Section~\ref{sec:bootstrap} we use the numerical conformal bootstrap to study the constraints imposed on our model by using crossing symmetry and unitarity of the four-point function of 2 chiral and 2 anti-chiral operators, $X_i$ and $\overline{X}^j$. In anticipation of this analysis, it is useful to study this four-point function in perturbation theory. To compare to the bootstrap results, it is useful to define properly normalized operators $\hat{X}$ and $\hat{\mathcal{Q}}$, i.e. operators with unit two-point functions. To this end we define
\begin{equation}
    \hat X_i \equiv \frac{X_i}{\sqrt{G_1}}~, \qquad \hat{\mathcal{Q}}_I \equiv \frac{\mathcal{Q}_I}{\sqrt{G_2}}~.
\end{equation}
We then consider the four point functions
\begin{align}
    & \left<\hat{X}_i\, \hat{X}_j\, \hat{\overline{X}}^k\, \hat{\overline{X}}^\ell \right>~,
    & \left<\hat{X}_i\, \hat{\mathcal{Q}}_J\, \hat{\overline{X}}^k\, \hat{\overline{\mathcal{Q}}}^L \right>~,
\end{align}
and expand them in the chiral channel. The dominant contribution in the conformal block expansion of these four-point functions is controlled by the following contractions of properly normalized OPE coefficients (we denote complex conjugation by $^*$)
\begin{align}
    & (C_{ij}^K \delta_{KM} C_{kl}^{*M})\, \frac{G_2}{G_1^2}~,\label{G2G12}\\
    & (C_{iJ} C_{kL}^{*})\, \frac{G_3}{G_1\,G_2}\label{G3G13}~.
\end{align}
The quantities appearing above are well defined, independent of normalization choices and can therefore be meaningfully compared to the bootstrap results. Using \eqref{OPEs}--\eqref{2ptfunchiral} we find for \eqref{G2G12} (sum over repeated indices implied):
\begin{align}\label{resOPE2}
    (C_{ij}^K C_{ij}^{*K})\, \frac{G_2}{G_1^2} =\frac{6}{(2+\tau\bar{\tau})}\frac{G_2}{G_1^2} \;.
\end{align}
Similarly for \eqref{G3G13} we have
\begin{align}\label{resG3OPE2}
    (C_{iJ} C_{kL}^{*})\, \frac{G_3}{G_1\,G_2} = \frac{1}{9}\frac{G_3}{G_1\,G_2} \delta_{iJ}\delta_{kL}\;.
\end{align}
Combining \eqref{GRatio} and \eqref{resOPE2} and multiplying by $2^{\frac{4(d-1)}{3}}$ to match our conventions in the bootstrap section, we get
\es{OPEeps}{
|\lambda_{\bar{\bold3},2\frac{d-1}{3},0}|^2=2^{\frac{4(3-\varepsilon)}{3}} \left[1+ \frac{2\,\zeta(3)}{3} \, \left(\frac{(1-\tau \bar{\tau})^3+(1-\tau^3)(1-\bar{\tau}^3)}{(2+\tau \bar{\tau})^3}\right)\, \varepsilon^2 + O(\varepsilon^3)\right]\,.
}

\subsubsection{The Zamolodchikov metric}

The operator $\mathcal{O}$ in \eqref{eq:chirfirst} has quantum numbers $\Delta = d-1$, $q = 2$ and it is the lowest component of a protected supermultiplet that contains the marginal operator associated to the exactly marginal coupling $\tau$. The two-point function of the exactly marginal operator determines the so-called Zamolodchikov metric on the conformal manifold \cite{Zamolodchikov:1986gt}. Superconformal Ward identities in turn relate this metric to the two-point function of the operator $\mathcal{O}$ itself. It may seem that we have already defined this two-point function in \eqref{2ptfunchiral} and computed it to order $\varepsilon^2$ in \eqref{GRatio}. However there is an important subtlety. To arrive at an object that transforms as a metric on the conformal manifold we need to work with an operator $\mathcal{O}_\tau$, proportional to $\mathcal{O}$, which is normalized such that an infinitesimal transformation of the superpotential along the conformal manifold yields
\begin{equation}
\delta W = \mathcal{O}_\tau \delta \tau~.
\end{equation}
To determine this normalization consider varying the coupling constants $h_i$ subject to the condition 
\begin{equation}
	\delta \beta(h_1,h_2) = 0~.
\end{equation}
The variation of the superpotential then reads
\begin{equation}
\delta W = r\left(\frac{1}{3(2+\tau \bar{\tau})^{3/2}} \left(X_1^3 + X_2^3 + X_3^3 - 3 \bar{\tau} X_1 X_2 X_3\right) \delta \tau +O(\varepsilon^2)\right)~,
\end{equation}
where the order $\varepsilon^2$ correction is proportional to the superpotential $W$ and so does not affect the Zamolodchikov metric to order $\varepsilon^2$. We observe that this operator is proportional to the chiral combination $\mathcal{O}$ in \eqref{eq:chirlast}, with an important prefactor that ensures that its two-point function transforms as a metric.
As a consequence, the Zamolodchikov metric $G(\tau,\bar{\tau})$ is given by
\begin{equation}
G(\tau,\bar{\tau}) \equiv |x|^{2(d-1)} \langle \mathcal{O}_\tau(x) \overline{\mathcal{O}}_{\bar{\tau}}(0)\rangle = \frac{r^2}{9(2+\tau \bar{\tau})^{3}} G_3(\tau,\bar{\tau})~.
\end{equation}
Using \eqref{eq:Gitransf}, one can show that $G(\tau,\bar{\tau})$ indeed transforms as a metric under a duality $\tau' = d(\tau)$:
\begin{equation}
G(\tau,\bar{\tau}) \rightarrow \left(\frac{\partial d}{\partial \tau}\right)^{-1}\left(\frac{\partial \bar{d}}{\partial \bar{\tau}}\right)^{-1} G(\tau,\bar{\tau})~.
\end{equation}
Using equations \eqref{GRatio}, we get the explicit expression
\begin{equation}
\label{eq:Zmetric}
	G(\tau,\bar{\tau}) = \frac{r_{*}^2}{(2+\tau \bar{\tau})^2}G_1^3 \left(1 + 2\,\zeta(3)\, \left(\frac{(1-\tau \bar{\tau})^3+(1-\tau^3)(1-\bar{\tau}^3)}{(2+\tau \bar{\tau})^3}\right)\, \varepsilon^2 + O(\varepsilon^3)\right)~,
\end{equation}
where $r_{*}^2$ is the value of $r^2$ at the fixed point given in \eqref{rsqrd} and $G_1$ is defined in \eqref{eq:G1eps} and is independent of $\tau$ to this order in perturbation theory.\footnote{While $r^2$ and $G_1$ are separately scheme dependent, the combination $r^2 G_1^3$ is in fact scheme independent. One way to show this is to notice that each propagator is accompanied by a factor of $\mu^{-\varepsilon/3}$ as in equation \eqref{eq:scalar2pt}, while the dimensionless coupling constant $r^2$ is accompanied by a factor of $\mu^{\varepsilon}$. Therefore the right hand side of \eqref{eq:Zmetric} is independent of $\mu$. A more detailed discussion of scheme dependence in the $\varepsilon$ expansion can be found in Section 9.3 of \cite{Kleinert:2001ax}.} It is pleasing to see that the leading order result is just the Fubini-Study metric on $\mathbf{CP}^1$, whose volume shrinks to zero as $\varepsilon \to 0$ as expected. Furthermore, we notice that the $\varepsilon^2$ correction does not exhibit singularities in the $\tau$ plane.

\section{Conformal bootstrap}
\label{sec:bootstrap}

We now show how to constrain $\mathcal{N}=2$ theories along the conformal manifold parameterized by the marginal coupling $\tau$ using the conformal bootstrap technique. We will focus on the 4-point function of two chiral operators $X_i$ and two anti-chiral operators $\overline{X}^i$. First we will compute the crossing equations, and then we will use them to bound the scaling dimensions and OPE coefficients of some scalar operators in $d=3$, and compare to the $4-\varepsilon$-expansion results.  See also Appendix~\ref{2d} for a summary of results on the scaling dimensions and OPE coefficients in our model in 2d.

\subsection{Crossing equations}
\label{cross}

In this section we will compute the crossing equations for various values of $\tau$.  As explained in Section~\ref{model}, for generic values of $\tau$, the theory \eqref{eq:superpotential} has an order 54 discrete flavor symmetry group $G=(\mathbb{Z}_3\times \mathbb{Z}_3)\rtimes S_3$.  In this case, we will derive 15 crossing equations. We then specialize to the cases $\Im\, \tau = 0$ (the boundary of the fundamental domain), $\tau=0$ (the XYZ point), and $\tau=(1-\sqrt{3})$ (the $\Z_2 \times \Z_2$ point), in which the symmetry is enhanced by $\mathbb{Z}_2$, $S_3$, and $\mathbb{Z}_2\times\mathbb{Z}_2$, respectively.  We will find 12 crossing equations for $\Im\, \tau = 0$, and 9 crossing equations for $\tau=0$ and $\tau=(1-\sqrt{3})$. We write all these crossing equations for arbitrary spacetime dimension $d$. 

\subsubsection{Symmetry group $G$ (generic point)}
\label{Gcross}

For general $\tau$, let us begin by describing the representations of the operators that appear in the 4-point function. We can decompose the representation of operators that appear in the OPEs $X_i\times \overline{X}^j$ and $X_i\times X_j$ as
\es{Gproducts}{
    &\bold3\otimes\bar{\bold3}=\bold1\oplus\bold2\oplus\bold2'\oplus\bold2''\oplus\bold2'''\,,\\
    &\bold3\otimes{\bold3}=\bar{\bold3}_s\oplus\bar{\bold3}_s\oplus\bold3'_a\,,
}
where $s/a$ denotes the symmetric/antisymmetric product. Operators in $X_i\times X_j$ that appear in the symmetric/antisymmetric product are restricted by Bose symmetry to have even/odd spins. Note that two copies of $\bar{\bold 3}$ appear in the symmetric product of $\bold3\otimes{\bold3}$, so we will denote operators in each copy separately. By taking the OPEs $X_i\times X_j$  and $X_i\times \overline{X}^j$ twice, we can now express the 4-point function in the $s$ and $t$ channels as
\es{4pt}{
\langle X_i\overline X^jX_k\overline X^l\rangle=&\frac{1}{|\vec x_{12}|^{2\frac{d-1}{3}}|\vec x_{34}|^{2\frac{d-1}{3}}}\sum_{\Delta,\ell}\mathcal{G}_{\Delta,\ell}(u,v)\sum_{R\in\bold3\otimes\bar{\bold3}}T_R{}_i{}^j{}_k{}^l|\lambda_{R,\Delta,\ell}|^2\,,\\
\langle X_i X_j\overline X^k\overline X^l\rangle=&\frac{1}{|\vec x_{12}|^{2\frac{d-1}{3}}|\vec x_{34}|^{2\frac{d-1}{3}}}\sum_{\Delta,\ell}{G}_{\Delta,\ell}(u,v)\left[T_{\bold3'}{}_i{}_j{}^k{}^l|\lambda_{\bold3',\Delta,\ell}|^2+\sum_{\alpha,\beta=1,2}T_{\bar{\bold3}_{\alpha\beta}}{}_i{}_j{}^k{}^l\lambda_{\bar{\bold3}_\alpha,\Delta,\ell}\bar\lambda_{\bar{\bold3}_\beta,\Delta,\ell}\right]\,,\\
}
where the complex OPE coefficients $\lambda_{R,\Delta,\ell}$ are labeled by $G$ irrep $R$, scaling dimension $\Delta$, and spin $\ell$, and the conformal blocks $G_{\Delta,\ell}(u,v)$\footnote{We use the normalization of the conformal blocks in \cite{Kos:2013tga}.  Specifically, in the $r$ and $\eta$ coordinates introduced in \cite{Hogervorst:2013sma}, we have $G_{\Delta, \ell} = r^\Delta P_\ell(\eta) + \ldots$, as $r \to 0$ with $\eta$ kept fixed.} are functions of $\ell$, $\Delta$, and conformal cross ratios
\es{uv}{
u\equiv\frac{|\vec x_{12}|^2|\vec x_{34}|^2}{|\vec x_{13}|^2|\vec x_{24}|^2}\,,\qquad v\equiv\frac{|\vec x_{14}|^2|\vec x_{23}|^2}{|\vec x_{13}|^2|\vec x_{24}|^2}\,,
}
while the superconformal blocks $\mathcal{G}_{\Delta,\ell}(u,v)$, originally derived in \cite{Bobev:2015jxa}, are defined in our conventions \cite{Chester:2015qca} as
 \es{SuperConf}{
\cG_{\Delta,\ell} &= G_{\Delta ,\ell}  +\frac{2(\ell+d-2)(\Delta +\ell) }{(2\ell+d-2)(\Delta +\ell+1)} G_{\Delta +1,\ell+1} + \frac{2\ell (\Delta - \ell+2-d)}{ (2\ell+d-2) (\Delta-\ell-d+3 )}G_{\Delta +1,\ell-1} \\
&+\frac{\Delta(\Delta+3-d) (\Delta - \ell +2-d) (\Delta +\ell) }{\left( \Delta+2-\frac d 2 \right) \left(\Delta+1-\frac d 2 \right)(\Delta-\ell+3-d ) (\Delta +\ell+1)} G_{\Delta +2,\ell} \,. 
 }
The tensor structures $T_R{}_i{}^j{}_k{}^l$ are computed in terms of the eigenvectors of the projectors \eqref{projectorEig} as
\es{T}{
T_R{}_i{}^j{}_k{}^l=P^{RR}_{\bold1}{}_{rs}v_{R,r}{}_i{}^jv_{R,s}{}_k{}^l\,,\qquad T_R{}_i{}_j{}^k{}^l=P^{R\bar R}_{\bold1}{}_{rs}v_{R,r}{}_i{}_jv_{\bar R,s}{}^k{}^l\,,
}
where $P^{RR}_{\bold1}{}_{rs}$ and $P^{R\bar R}_{\bold1}{}_{rs}$ are projectors from $R\otimes R$ and $R\otimes \bar R$ to the singlet, respectively, which can be constructed analogously to \eqref{projectors}, or simply by inspection of the bilinears \eqref{bilinears}---for more details, see Appendix~\ref{sec:discretegroup}.  Note that there are two OPE coefficients $\lambda_{\bar{\bold3}_{\alpha},\Delta,\ell}$, with $\alpha=1,2$, because $\bar{\bold3}$ appears twice in $\bold3\otimes\bold3$, so there are four possible tensor structures $T_{\bar{\bold3}_{\alpha\beta}}{}_i{}^j{}_k{}^l$ for that irrep, with $\alpha = 1, 2$ that get multiplied by quadratic combinations of two OPE coefficients $\lambda_{\bar{\bold3}_\alpha,\Delta,\ell}$ and their conjugates. 

A very important ingredient in our analysis is the fact that we can relate the marginal coupling $\tau$ to the CFT data.  This is achieved as follows.  Note that the chiral ring relations following from \eqref{eq:superpotential} imply that the linear combinations of OPEs
 \es{chiralRel}{
  &\frac{\tau}{2} X_1 \times X_1 + X_2 \times X_3 \,, \qquad
  \frac{\tau}{2} X_2 \times X_2 + X_3 \times X_1 \,, \qquad
  \frac{\tau}{2} X_3 \times X_3 + X_1 \times X_2 
 }
do not contain scalar chiral operators of dimension $\Delta = 2 (d-1)/3$.  A similar statement holds for the complex conjugate of \eqref{chiralRel}, which should not contain any anti-chiral operators with this scaling dimension.  This information implies that in the decomposition of the linear combinations of four-point functions
 \es{FourPointLinear}{
  &\frac{\tau}{2} \langle X_1 X_1 \overline X^k \overline X^l \rangle + \langle X_2 X_3 \overline X^k \overline X^l \rangle \,, \\
  &\frac{\tau}{2} \langle X_2 X_2 \overline X^k \overline X^l \rangle + \langle X_3 X_1 \overline X^k \overline X^l \rangle\,, \\
  &\frac{\tau}{2} \langle X_3 X_3 \overline X^k \overline X^l \rangle + \langle X_1 X_2 \overline X^k \overline X^l \rangle\,,
 }
into conformal blocks there should be no contribution from a conformal primary of scaling dimension $2(d-1)/3$.  Using \eqref{SuperConf} as well as the definitions of the tensor structures $T_R{}_i{}^j{}_k{}^l$, one finds the following relation between the OPE coefficients of scalar operators in the $\bar{\bold 3}$ representation:
\es{BootstrapTau}{
-\frac{1}{2}\tau\lambda_{\bar{\bold3}_{1},2\frac{d-1}{3},0}=\lambda_{\bar{\bold3}_{2},2\frac{d-1}{3},0}\,.
}
This relation encodes the way the CFT data depends on the marginal coupling $\tau$.

We now equate the two different channels of the 4-point function \eqref{4pt} as
\es{channels}{
&\langle X_i\overline X^jX_k\overline X^l\rangle=\langle X_k\overline X^jX_i\overline X^l\rangle\,,\\
&\langle X_i X_j\overline X^k\overline X^l\rangle=\langle \overline X^k X_jX_i\overline X^l\rangle\,,
}
which yields the crossing equations
\es{crossing}{
0=&\sum_{\substack{\text{all }\ell\\R\in\{{\bold1},\bold2,\bold2',\bold2'',\bold2'''\}}}|\lambda_{R,\Delta,\ell}|^2\vec V_{R,\Delta,\ell}+\sum_{\text{odd }\ell}|\lambda_{\bold3',\Delta,\ell}|^2\vec V_{\bold3',\Delta,\ell}+\sum_{\text{even }\ell}\vec\lambda_{\bar{\bold3},\Delta,\ell}\vec{ \bold V}_{\bar{\bold3},\Delta,\ell}\vec{\lambda}^T_{\bar{\bold3},\Delta,\ell}\,,\\
&+\left(\frac{|\tau|^2+2}{4}\right)^{-1}|\lambda_{\bar{\bold3}_1,2\frac{d-1}{3},0}|^2\vec V_{\bar{\bold3}_1,2\frac{d-1}{3},0}(\tau)\,,\\
\vec\lambda_{\bar{\bold3},\Delta,\ell}=&\begin{pmatrix}\Re\lambda_{\bar{\bold3}_1,\Delta,\ell}&\Im\lambda_{\bar{\bold3}_1,\Delta,\ell}&\Re\lambda_{\bar{\bold3}_2,\Delta,\ell}&\Im\lambda_{\bar{\bold3}_2,\Delta,\ell}\end{pmatrix}\,,
}
where $\vec V_{R,\Delta,\ell}$ and $\vec{ \bold V}_{\bar{\bold3},\Delta,\ell}$ are 15-dimensional vectors of scalars and $4\times4$ matrices,\footnote{We require $4\times4$ matrices so that the matrix is real and symmetric, which is required for the numerics.} respectively, given in Appendix \ref{bootApp}. We have separated out the contribution of the $\lambda_{\bold3_\alpha,2\frac{d-1}{3},0}$ term, for which we used the chiral ring relation \eqref{BootstrapTau} to write the scalar constraint $\vec V_{\bar{\bold3}_1,2\frac{d-1}{3},0}(\tau)$. To ensure that $|\lambda_{\bar{\bold3}_1,2\frac{d-1}{3},0}|^2$ is a duality-invariant quantity, we have included the factor
\es{tauFac}{
\frac{|\tau|^2+2}{4}=T_{\bar{\bold3}_{11}}{}_i{}^j{}_j{}^i+\frac{|\tau|^2}{4}T_{\bar{\bold3}_{22}}{}_i{}^j{}_j{}^i-\frac{\tau}{2}T_{\bar{\bold3}_{21}}{}_i{}^j{}_j{}^i-\frac{\bar\tau}{2}T_{\bar{\bold3}_{12}}{}_i{}^j{}_j{}^i\,,
}
where the tensor structures are defined in \eqref{T}.

The operator spectrum is further constrained due to the $\mathcal{N}=2$ supersymmetry \cite{Bobev:2015vsa,Bobev:2015jxa}. Generalizing the results for the XYZ model \cite{Chester:2015qca}, we find that the following operators may appear:
\es{opList}{
\bold1\,,\bold2\,,\bold2'\,,\bold2''\,,\bold2'':\qquad &\Delta\geq \ell+d-2\quad\text{for all $\ell$}\,,\\
\bar{\bold3}:\qquad &\Delta\geq \abs{2\frac{d-1}{3}-(d-1)}+d-1+\ell\quad\text{for even $\ell$}\,,\\
&\Delta=2\frac{d-1}{3}+\ell\quad\text{for even $\ell$}\,,\\
&\Delta=d-2\frac{d-1}{3}\quad\text{for $\ell=0$}\,,\\
{\bold3'}:\qquad &\Delta\geq \abs{2\frac{d-1}{3}-(d-1)}+d-1+\ell\quad\text{for odd $\ell$}\,,\\
&\Delta=2\frac{d-1}{3}+\ell\quad\text{for odd $\ell$}\,.\\
}

\subsubsection{Symmetry group $G\rtimes \mathbb{Z}_2$ (the boundary of the fundamental domain)}
\label{GZ2cross}
We will now specialize to the boundary of the conformal manifold, which has an enhanced $\mathbb{Z}_2$ symmetry. For simplicity, let us focus on the duality frame where $\mathbb{Z}_2$ acts as conjugation. We can now combine $X_i$ and $\overline X^j$ into a single operator $\tilde X_I=\{X_i,\overline X^j\}$ where $I=1,\dots,6$, where $\tilde X_I$ transforms in the real representation $\bold6$ of $G\rtimes \mathbb{Z}_2$. We can then decompose the representation of operators that appear in the OPE $\tilde X_I\times \tilde X_J$ as
\es{GZ2products}{
\bold6\otimes\bold6=\bold1_s^E+\bold1_a^O+\bold2_a^O+\bold2_s^E+{\bold2'_a}^O+{\bold2'_s}^E+\bold4_s+\bold4_a+\bold6_s+\bold6_s+\bold6'_a\,,
}
where $s/a$ denotes the symmetric/antisymmetric product. Operators that appear in the symmetric/antisymmetric product are restricted by Bose symmetry to even/odd spins. The notation $R^{E,O}$ denotes two different representations, where $E,O$ denotes that operators in this representation only appear with even/odd spins. Comparing \eqref{GZ2products} to \eqref{Gproducts}, we find that $\bold2''$ and $\bold2'''$ have combined into $\bold4$, the conjugate representations $\bold3$ and $\bar{\bold3}$ have combined into $\bold6$, and similarly $\bold3'$ and $\bar{\bold3'}$ have combined into $\bold6'$. As in $G$, two copies of ${\bold 6}$ appear in the symmetric product of $\bold6\otimes{\bold6}$, so we will denote operators in each copy separately. By taking the OPE $\tilde X_I\times \tilde X_J$  twice, we can now express the 4-point function in the $s$ channel as
\es{4ptZ2}{
\langle \tilde X_I  \tilde X_J \tilde X_K \tilde X_L\rangle=&\frac{1}{|\vec x_{12}|^{2\frac{d-1}{3}}|\vec x_{34}|^{2\frac{d-1}{3}}}\sum_{\Delta,\ell}\mathcal{G}_{\Delta,\ell}(u,v)\left[\sum_{\substack{R\in\bold6\otimes{\bold6}\\ R\neq\bold6}}T_{R,}{}_{IJKL}\lambda_{R,\Delta,\ell}^2+\sum_{\alpha,\beta=1,2}T_{{\bold6}_{\alpha\beta}}{}_{,IJKL}\lambda_{{\bold6}_\alpha,\Delta,\ell}\lambda_{{\bold6}_\beta,\Delta,\ell}\right]\,,\\
}
where the OPE coefficients $\lambda_{R,\Delta,\ell}$ are now real, and the tensor structures $T_{R,IJKL}$ are constructed as in \eqref{T} except using the projectors \eqref{projectorsZ2} for $G\rtimes \mathbb{Z}_2$. As in $G$, there are two OPE coefficients $\lambda_{{\bold6}_{\alpha},\Delta,\ell}$, with $\alpha=1,2$, because ${\bold6}$ appears twice in $\bold6\otimes\bold6$, so there are four possible tensor structures $T_{{\bold6}_{\alpha\beta}}{}_{IJKL}$ for that irrep. For the case $\ell=0$ and $\Delta = 2 (d-1)/3$, we can again use the chiral ring relation to relate these OPE coefficients as
\es{BootstrapTau2}{
-\frac{1}{2}\tau\lambda_{{\bold6}_{1},2\frac{d-1}{3},0}=\lambda_{{\bold6}_{2},2\frac{d-1}{3},0}\,.
}

We now equate the two different channels of the 4-point function \eqref{4ptZ2} as
\es{channelsZ2}{
&\langle \tilde X_I  \tilde X_J  \tilde X_K  \tilde X_L\rangle=\langle \tilde X_K  \tilde X_J  \tilde X_I  \tilde X_L\rangle\,,
}
which yields the crossing equations
\es{crossingZ2}{
0=&\sum_{\substack{\text{even }\ell\\R\in\{{\bold1}^E,\bold2^E,\bold2'^E,\bold4\}}}\lambda_{R,\Delta,\ell}^2\vec V_{R,\Delta,\ell}+\sum_{\substack{\text{odd }\ell\\R\in\{{\bold1}^O,\bold2^O,\bold2'^O,\bold4,\bold6'\}}}\lambda_{R,\Delta,\ell}^2\vec V_{R,\Delta,\ell}\\
&+\left(\frac{|\tau|^2+2}{4}\right)^{-1}\lambda_{\bold6_1,2\frac{d-1}{3},0}^2\vec V_{\bold6_1,2\frac{d-1}{3},0}(\tau)+\sum_{\text{even }\ell}\vec\lambda_{{\bold6},\Delta,\ell}\vec{ \bold V}_{{\bold6},\Delta,\ell}\vec{\lambda}^T_{{\bold6},\Delta,\ell}\,,\\\vec\lambda_{{\bold6},\Delta,\ell}=&\begin{pmatrix}\lambda_{{\bold6}_1,\Delta,\ell}&\lambda_{{\bold6}_2,\Delta,\ell}\end{pmatrix}\,,
}
where $\vec V_{R,\Delta,\ell}$ and $\vec{ \bold V}_{\bar{\bold3},\Delta,\ell}$ are 12-dimensional vectors of scalars and $2\times2$ matrices, respectively, which are given explicitly in Appendix \ref{bootApp}\@. We have separated out the contribution of the $\lambda_{\bold6_\alpha,2\frac{d-1}{3},0}$ term, for which we used the chiral ring relation \eqref{BootstrapTau2} to write the scalar constraint $\vec{  V}_{{\bold6}_1,\Delta,0}(\tau)$ and included the $\tau$-dependent factor defined in \eqref{tauFac}.

Just as with the $G$ crossing equations, the operator spectrum is further constrained due to the $\mathcal{N}=2$ supersymmetry so that the following operators may appear:
\es{opListZ2}{
\bold1^{E,O}\,,\bold2^{E,O}\,,\bold2'^{E,O}\,,\bold4:\qquad &\Delta\geq \ell+d-2\quad\text{for all $\ell$}\,,\\
{\bold6}:\qquad &\Delta\geq \abs{2\frac{d-1}{3}-(d-1)}+d-1+\ell\quad\text{for even $\ell$}\,,\\
&\Delta=2\frac{d-1}{3}+\ell\quad\text{for even $\ell$}\,,\\
&\Delta=d-2\frac{d-1}{3}\quad\text{for $\ell=0$}\,,\\
{\bold6'}:\qquad &\Delta\geq \abs{2\frac{d-1}{3}-(d-1)}+d-1+\ell\quad\text{for odd $\ell$}\,,\\
&\Delta=2\frac{d-1}{3}+\ell\quad\text{for odd $\ell$}\,.\\
}

\subsubsection{Symmetry group $G\rtimes (\mathbb{Z}_2\times\mathbb{Z}_2)$ (the $\Z_2 \times \Z_2$ point)}
\label{GZ2Z2cross}
Let us now further specialize to the point on the boundary of the conformal manifold that has an enhanced $\mathbb{Z}_2\times \mathbb{Z}_2$ symmetry. For simplicity, we will choose the point $\tau=1-\sqrt{3}$, so that one of the $\mathbb{Z}_2$'s acts as conjugation and the chiral operator $\tilde X_I$ transforms in the real representation $\bold6^1$ of $G\rtimes \mathbb{Z}_2$. We can then decompose the representation of operators that appear in the OPE $\tilde X_I\times \tilde X_J$ as
\es{GproductsZ2Z2}{
\bold6^1\otimes\bold6^1=\bold1_s^E+\bold1_a^O+\bold4_s^E+\bold4_a^O+{\bold4'_s}^E+{\bold4'_a}^O+\bold6_{s}^1+\bold6_{s}^2+{\bold6'_{a}}^1\,,
}
where $s/a$ denotes the symmetric/antisymmetric product. Operators that appear in the symmetric/antisymmetric product are restricted by Bose symmetry to even/odd spins. As with $G\rtimes \mathbb{Z}_2$, the notation $R^{E,O}$ denotes two different representations, where $E,O$ denotes that operators in this representation only appear with even/odd spins. Comparing \eqref{GproductsZ2Z2} to \eqref{GZ2products}, we find that $\bold2$ and $\bold2'$ have combined into $\bold4'$, and now the two 6-dimensional irreps that appear in the symmetric product belong to different irreps $\bold6^1$ and $\bold6^2$. By taking the OPE $\tilde X_I\times \tilde X_J$  twice, we can now express the 4-point function in the $s$ channel as
\es{4ptZ2Z2}{
\langle \tilde X_I  \tilde X_J \tilde X_K \tilde X_L\rangle=&\frac{1}{|\vec x_{12}|^{2\frac{d-1}{3}}|\vec x_{34}|^{2\frac{d-1}{3}}}\sum_{\Delta,\ell}\mathcal{G}_{\Delta,\ell}(u,v)\sum_{\substack{R\in\bold6^1\otimes{\bold6^1}}}T_{R,}{}_{IJKL}\lambda_{R,\Delta,\ell}^2\,,\\
}
where the OPE coefficients $\lambda_{R,\Delta,\ell}$ are again real, and the tensor structures $T_{R,IJKL}$ are constructed as in \eqref{T} except using the projectors for $G\rtimes( \mathbb{Z}_2\times \mathbb{Z}_2)$. 

We now equate the two different channels of the 4-point function \eqref{4ptZ2Z2} as
\es{channelsZ2Z2}{
&\langle \tilde X_I  \tilde X_J  \tilde X_K  \tilde X_L\rangle=\langle \tilde X_K  \tilde X_J  \tilde X_I  \tilde X_L\rangle\,,
}
which yields the crossing equations
\es{crossingZ2Z2}{
0=&\left[1+\frac{\sqrt{3}}{3}\right]\lambda_{\bold6^2,2\frac{d-1}{3},0}^2\vec V_{\bold6^2,2\frac{d-1}{3},0}+\sum_{\substack{\text{even }\ell\\R\in\{{\bold1}^E,\bold4^E,\bold4'^E,\bold6^1,\bold6^2\}}}\lambda_{R,\Delta,\ell}^2\vec V_{R,\Delta,\ell}+\sum_{\substack{\text{odd }\ell\\R\in\{{\bold1}^O,\bold4^O,\bold4'^O,\bold6'\}}}\lambda_{\bold6',\Delta,\ell}^2\vec V_{\bold6',\Delta,\ell}\,,
}
where $\vec V_{R,\Delta,\ell}$ are 9-dimensional vectors of scalars, which are given explicitly in Appendix \ref{bootApp}, and we have included the factor next to $\lambda_{\bold6^2,2\frac{d-1}{3},0}^2$ so that it equals $\lambda_{\bar{\bold3}_1,2\frac{d-1}{3},0}^2$ in \eqref{crossing} when $\tau=1-\sqrt{3}$ or any other duality related value.

Just as in the previous cases, the operator spectrum is further constrained due to the $\mathcal{N}=2$ supersymmetry so that the following operators may appear:
\es{opListZ2Z2}{
\bold1\,,\bold4\,,\bold4':\qquad &\Delta\geq \ell+d-2\quad\text{for all $\ell$}\,,\\
{\bold6^1}:\qquad &\Delta\geq \abs{2\frac{d-1}{3}-(d-1)}+d-1+\ell\quad\text{for even $\ell$}\,,\\
&\Delta=2\frac{d-1}{3}+\ell\quad\text{for even $\ell\geq2$}\,,\\
&\Delta=d-2\frac{d-1}{3}\quad\text{for $\ell=0$}\,,\\
{\bold6^2}:\qquad &\Delta\geq \abs{2\frac{d-1}{3}-(d-1)}+d-1+\ell\quad\text{for even $\ell$}\,,\\
&\Delta=2\frac{d-1}{3}+\ell\quad\text{for even $\ell$}\,,\\
{\bold6'}^1:\qquad &\Delta\geq \abs{2\frac{d-1}{3}-(d-1)}+d-1+\ell\quad\text{for odd $\ell$}\,,\\
&\Delta=2\frac{d-1}{3}+\ell\quad\text{for odd $\ell$}\,.\\
}

\subsubsection{Symmetry group $G\rtimes S_3$ (the XYZ point)}
\label{GS3cross}
Let us now discuss the point on the boundary of the conformal manifold that has an enhanced $S_3$ symmetry. Note that for the XYZ model, this $S_3$ is just a subgroup of the full flavor symmetry $U(1)\times U(1)\rtimes S_3$, but including the full group would require a numerically unfeasible number of crossing equations, so here we just use an $S_3$ subgroup. For $\mathcal{N}=2$ crossing equations that use just the $U(1)\times U(1)$ subgroup see \cite{Chester:2015lej}. For simplicity, we will choose the point $\tau=0$, so that $\mathbb{Z}_2\subset S_3$ acts as conjugation and the chiral operator $\tilde X_I$ transforms in the real representation $\bold6^1$ of $G\rtimes S_3$. We can then decompose the representation of operators that appear in the OPE $\tilde X_I\times \tilde X_J$ as
\es{GproductS3}{
\bold6^1\otimes\bold6^1=\bold1^E_s+\bold1^O_a+\bold2^E_s+\bold2^O_a+\bold6^E_s+\bold6^O_a+\bold6_{s}^1+\bold6_{s}^2+{\bold6'}_{a}^1
}
where $s/a$ denotes the symmetric/antisymmetric product. Operators that appear in the symmetric/antisymmetric product are restricted by Bose symmetry to even/odd spins. As with $G\rtimes \mathbb{Z}_2$, the notation $R^{E,O}$ denotes two different representations, where $E,O$ denotes that operators in this representation only appear with even/odd spins. Comparing \eqref{GproductS3} to \eqref{GZ2products}, we find that $\bold2'$ and $\bold4$ have combined into $\bold6$, and now the two 6-dimensional irreps that appear in the symmetric product belong to different irreps $\bold6^1$ and $\bold6^2$. By taking the OPE $\tilde X_I\times \tilde X_J$  twice, we can now express the 4-point function in the $s$ channel as
\es{4ptS3}{
\langle \tilde X_I  \tilde X_J \tilde X_K \tilde X_L\rangle=&\frac{1}{|\vec x_{12}|^{2\frac{d-1}{3}}|\vec x_{34}|^{2\frac{d-1}{3}}}\sum_{\Delta,\ell}\mathcal{G}_{\Delta,\ell}(u,v)\sum_{\substack{R\in\bold6^1\otimes{\bold6^1}}}T_{R,}{}_{IJKL}\lambda_{R,\Delta,\ell}^2\,,\\
}
where the OPE coefficients $\lambda_{R,\Delta,\ell}$ are again real, and the tensor structures $T_{R,IJKL}$ are constructed as in \eqref{T} except using the projectors for $G\rtimes S_3$. 

We now equate the two different channels of the 4-point function \eqref{4ptS3} as
\es{channelsS3}{
&\langle \tilde X_I  \tilde X_J  \tilde X_K  \tilde X_L\rangle=\langle \tilde X_K  \tilde X_J  \tilde X_I  \tilde X_L\rangle\,,
}
which yields the crossing equations
\es{crossingS3}{
0=&2\lambda_{\bold6^2,2\frac{d-1}{3},0}^2\vec V_{\bold6^2,2\frac{d-1}{3},0}+\sum_{\substack{\text{even }\ell\\R\in\{{\bold1}^E,\bold2^E,\bold6^E,\bold6^1,\bold6^2\}}}\lambda_{R,\Delta,\ell}^2\vec V_{R,\Delta,\ell}+\sum_{\substack{\text{odd }\ell\\R\in\{{\bold1}^O,\bold2^O,\bold6^O,\bold6'\}}}\lambda_{\bold6',\Delta,\ell}^2\vec V_{\bold6',\Delta,\ell}\,,
}
where $\vec V_{R,\Delta,\ell}$ are 9-dimensional vectors of scalars, which are given explicitly in Appendix \ref{bootApp}, and we have included the factor of 2 next to $\lambda_{\bold6^2,2\frac{d-1}{3},0}^2$ so that it equals $\lambda_{\bar{\bold3}_1,2\frac{d-1}{3},0}^2$ in \eqref{crossing} when $\tau=0$ or any other duality related value.

Just as in the previous cases, the operator spectrum is further constrained due to the $\mathcal{N}=2$ supersymmetry so that the following operators may appear:
\es{opListS3}{
\bold1\,,\bold2\,,\bold6:\qquad &\Delta\geq \ell+d-2\quad\text{for all $\ell$}\,,\\
{\bold6^1}:\qquad &\Delta\geq \abs{2\frac{d-1}{3}-(d-1)}+d-1+\ell\quad\text{for even $\ell$}\,,\\
&\Delta=2\frac{d-1}{3}+\ell\quad\text{for even $\ell\geq2$}\,,\\
&\Delta=d-2\frac{d-1}{3}\quad\text{for $\ell=0$}\,,\\
{\bold6^2}:\qquad &\Delta\geq \abs{2\frac{d-1}{3}-(d-1)}+d-1+\ell\quad\text{for even $\ell$}\,,\\
&\Delta=2\frac{d-1}{3}+\ell\quad\text{for even $\ell$}\,,\\
{\bold6'}^1:\qquad &\Delta\geq \abs{2\frac{d-1}{3}-(d-1)}+d-1+\ell\quad\text{for odd $\ell$}\,,\\
&\Delta=2\frac{d-1}{3}+\ell\quad\text{for odd $\ell$}\,.\\
}

\subsection{Numerical bootstrap setup}
\label{numericSetup}

We now describe how to compute bounds on scaling dimensions and OPE coefficients with the crossing equations defined above. Recall that for the case of general $\tau$, this parameter appears explicitly in the crossing equations.

We can find upper or lower bounds on a given OPE coefficient of an operator $\cO^*$ that belongs to an isolated representation of the superconformal algebra\footnote{For a representation that is not isolated, we can only find upper bound this way.} of the four point functions by considering linear functionals $\vec{\alpha}$ satisfying the following conditions:
\es{lowerBound}{
&\vec{\alpha}(\vec{V}_{\cO^*})=s \,\qquad\text{$s=1$ for upper bounds, $s=-1$ for lower bounds} \,,\\
&\vec{\alpha}(\vec{V}_{R, \Delta,\ell}(\tau))\geq0, \quad \text{for all chiral $\cO\notin\{\cO_{\bold1,0,0},\cO^*\}$ with fixed $\Delta$} \,,\\
&\vec{\alpha}(\vec{V}_{R,\Delta, \ell})\geq0, \quad \text{for all non-chiral $\cO$ with $\Delta \geq\ell+d-2$} \,.\\
}
If such a functional $\vec{\alpha}$ exists, then this $\alpha$ applied to \eqref{crossing} along with the positivity of all $|\lambda_{\cO}|^2$ except, possibly, for that of $|\lambda_{\cO^*}|^2$ implies that
 \es{UpperOPE}{
  &\text{if $s=1$, then}\quad |\lambda_{\cO^*}|^2 \leq - \alpha (\vec{V}_{\bold1,0,0})\,, \\
  &\text{if $s=-1$, then}\quad |\lambda_{\cO^*}|^2 \geq  \alpha (\vec{V}_{\bold1,0,0})\,. \\
   }
To obtain the most stringent  bounds on $|\lambda_{\cO^*}|^2$, one should then minimize the RHS of \eqref{UpperOPE} under the constraints \eqref{lowerBound}.

To find upper bounds on the scaling dimensions of non-chiral operators $\cO^*_{R^*,\Delta^*,\ell^*}$, we consider linear functionals $\vec{\alpha}$ satisfying the following conditions:
\es{upperBound}{
&\vec{\alpha}(\vec{V}_{\bold1,0,0})=1 \,, \\
&\vec{\alpha}(\vec{V}_{R,\Delta, \ell}(\tau))\geq0, \quad \text{for all chiral $\cO$ with fixed $\Delta$} \,,\\
&\vec{\alpha}(\vec{V}_{R, \Delta,\ell})\geq0, \quad \text{for all non-chiral $\cO\neq\cO^*$ with $\Delta \geq\ell+d-2$} \,,\\
&\vec{\alpha}(\vec{V}_{R^*, \Delta,\ell^*})\geq0, \quad \text{for all non-chiral $\cO$ with $\Delta \geq\Delta_{R^*, \ell^*}^*$} \,.\\
}
The existence of any such $\vec{\alpha}$ would contradict \eqref{crossing}, and thereby would allow us to find an upper bound on the lowest-dimension $\Delta^*_{R^*, \ell^*}$ of the spin-$\ell^*$ superconformal primary in irrep $R^*$.

The numerical implementation of the above problems requires two truncations: one in the number of derivatives used to construct $\vec{\alpha}$ and one in the range of spins $\ell$ that we consider, whose contributions to the conformal blocks are exponentially suppressed for large spin $\ell$. The truncated constraint problem can then be rephrased as a semidefinite programing problem using the method developed in \cite{Rattazzi:2010yc}. We will implement this semi-definite programming using SDPB \cite{Simmons-Duffin:2015qma}, for which we use the parameters specified in the first column of Table 1 in the \texttt{SDPB} manual \cite{Simmons-Duffin:2015qma}, and consider spins up to 25 and derivative parameter $\Lambda=19$ for the $G$ and $G\rtimes \mathbb{Z}_2$ cases, and spins up to 35 and derivative parameter $\Lambda=27$ for the $G\rtimes \mathbb{Z}_2\times\mathbb{Z}_2$ and $G\rtimes {S}_3$ cases.

\subsection{Numerical results}
\label{num}
We now give numerical results computed using the crossing equations derived above, and compare them to the $4-\varepsilon$ expansion. For cWZ$^3$, since this model consists of three non-interacting copies of cWZ, we can compute some of its CFT data analytically and some using the numerical bootstrap study previously performed in \cite{Bobev:2015jxa}.

\subsubsection{cWZ$^3$}
\label{numcWZ}

We will first show how some CFT data can be computed analytically for this model. For convenience, we work in the duality frame $\tau\to\infty$ with superpotential \eqref{isingW}, where each chiral field $X_i$ belongs to a different decoupled cWZ. By inspection of \eqref{bilinears} and \eqref{chiralBis}, we see that the bilinear operators $\cO_{\bold2',0}$, $\cO_{\bold2'',0}$, $\cO_{\bold2''',0}$, and $\cO_{\bar{\bold3}_2,0}$\footnote{$\cO_{\bar{\bold3}_1,0}$ is a descendant due to the chiral ring relation.} are formed of chiral and anti-chiral operators from different non-interacting copies of cWZ, so their scaling dimensions and OPE coefficients can be computed exactly. In \eqref{isingW} we gave their scaling dimensions, which are just twice the value of a single chiral field. By similar reasoning, we can compute their OPE coefficients in terms of 2-point functions of a single chiral field. In particular, we can write the 4-point function \eqref{4pt} in each channel as
\es{3ising4pnt}{
\langle X_i\overline X^j X_k\overline X^l\rangle&=\frac{1}{|\vec x_{12}|^{2\frac{d-1}{3}}|\vec x_{34}|^{2\frac{d-1}{3}}}\left[\delta_{ik}\delta^{jl}\langle XX\overline X\overline X\rangle+\langle X_i\overline X^j\rangle \langle X_k\overline X^l\rangle+\langle X_i\overline X^l\rangle \langle X_k\overline X^j\rangle\right]\\
&=\frac{1}{|\vec x_{12}|^{2\frac{d-1}{3}}|\vec x_{34}|^{2\frac{d-1}{3}}}\left[\delta_{ik}\delta^{jl}\langle XX\overline X\overline X\rangle+\delta_i^j\delta_k^l+\left(\frac uv\right)^{\frac{d-1}{3}}\delta_i^l\delta_k^j\right]\,,\\
\langle X_iX_j\overline X^k\overline X^l\rangle&=\frac{1}{|\vec x_{12}|^{2\frac{d-1}{3}}|\vec x_{34}|^{2\frac{d-1}{3}}}\left[\delta_{ij}\delta^{kl}\langle XX\overline X\overline X\rangle+\langle X_i\overline X^k\rangle \langle X_j\overline X^l\rangle+\langle X_i\overline X^l\rangle \langle X_j\overline X^k\rangle\right]\\
&=\frac{1}{|\vec x_{12}|^{2\frac{d-1}{3}}|\vec x_{34}|^{2\frac{d-1}{3}}}\left[\delta_{ij}\delta^{kl}\langle XX\overline X\overline X\rangle+u^{\frac{d-1}{3}}\delta_i^k\delta_j^l+\left(\frac uv\right)^{\frac{d-1}{3}}\delta_i^l\delta_j^k\right]\,,
}
were $\langle XX\overline X\overline X\rangle$ is the unknown 4-point function of each cWZ with itself, and the second and third terms factorize into the different non-interacting cWZ 2-point functions. We can now compare \eqref{3ising4pnt} to \eqref{4pt} and expand $u$, $v$, and the scalar conformal blocks as 
\es{rtheta}{
v&=\left(\frac{1+r^2-2r\eta}{1+r^2+2r\eta}\right)^2\,,\qquad u=\left(\frac{4r}{1+r^2+2r\eta}\right)^2\,,\qquad G_{\Delta,0}=r^\Delta\left[1+O(r^2)\right]\,,
}
to extract some OPE coefficients that do not depend on $\langle XX\overline X\overline X\rangle$:
\es{isingOPEs}{
&|\lambda_{\bar{\bold3}_2,2\frac{d-1}{3},0}|^2=2^{\frac{4(d-1)}{3}}\,,\qquad |\lambda_{{\bold2},2\frac{d-1}{3},0}|^2=|\lambda_{{\bold2'},2\frac{d-1}{3},0}|^2=|\lambda_{{\bold2'''},2\frac{d-1}{3},0}|^2=\frac132^{\frac{4(d-1)}{3}}\,.
}

For the bilinears $\cO_{\bold2,0}$ and $\cO_{\bold1,0}$ in \eqref{bilinears} that are composed of chiral and anti-chiral operators from the same cWZ, we can use the numerical results that were computed for that model in \cite{Bobev:2015jxa}. In particular, we will use the scaling dimensions $\Delta_{\bold1,0}$ and $\Delta_{\bold2,0}$, which are in fact the same because both operators are just linear combinations of a singlet bilinear for a single cWZ. The scaling dimension of this operator was found in \cite{Bobev:2015jxa} to be
\es{BobevResults}{
\Delta_{\bold1,0}=\Delta_{\bold2,0}=1.9098(20)\,.
}

\subsubsection{Symmetry group $G\rtimes(\mathbb{Z}_2\times\mathbb{Z}_2)$ (the $\Z_2 \times \Z_2$ point)}
\label{numZ2Z2}
Next, we describe numerical bounds for the point on the conformal manifold in $d=3$ with $G\rtimes (\mathbb{Z}_2\times\mathbb{Z}_2)$ symmetry, using the crossing equations derived in Section~\ref{GZ2Z2cross}. There are three unprotected scalar scaling dimensions: $\Delta_{\bold1,0}$, $\Delta_{\bold4,0}$, and $\Delta_{\bold4',0}$. On the right of Figure \ref{XYZscal} we show the numerical bounds for these quantities, which form a rectangle. We conjecture that the $\mathbb{Z}_2\times\mathbb{Z}_2$ model lives at the nontrivial corner of this rectangle, so that $(\Delta_{\bold1,0}\,,\Delta_{\bold4,0},\Delta_{\bold4',0})\approx(1.898\,,1.259\,,1.727)$, where in terms of $G$ irreps $\Delta_{\bold4,0}=\Delta_{\bold2,0}=\Delta_{\bold2',0}$ and $\Delta_{\bold4',0}=\Delta_{\bold2'',0}=\Delta_{\bold2''',0}$.

Independently of this conjecture, we can also use the  $G\rtimes (\mathbb{Z}_2\times\mathbb{Z}_2)$ crossing equations to compute upper and lower bounds on the chiral bilinear OPE coefficient squared $\lambda^2_{\bold6_1,\frac{4}{3},0}$. We find
\es{Z2Z2OPEbound}{
6.339\leq\lambda^2_{\bold6_1,\frac{4}{3},0}\leq6.997\,,
}
where in terms of $G$ irreps $\lambda^2_{\bold6_1,\frac{4}{3},0}=|\lambda_{\bold3_1,\frac{4}{3},0}|^2$.

\subsubsection{Symmetry group $G\rtimes S_3$ (the XYZ point)}
\label{numXYZ}
\begin{figure}[t!]
\begin{center}
   \includegraphics[width=0.47\textwidth]{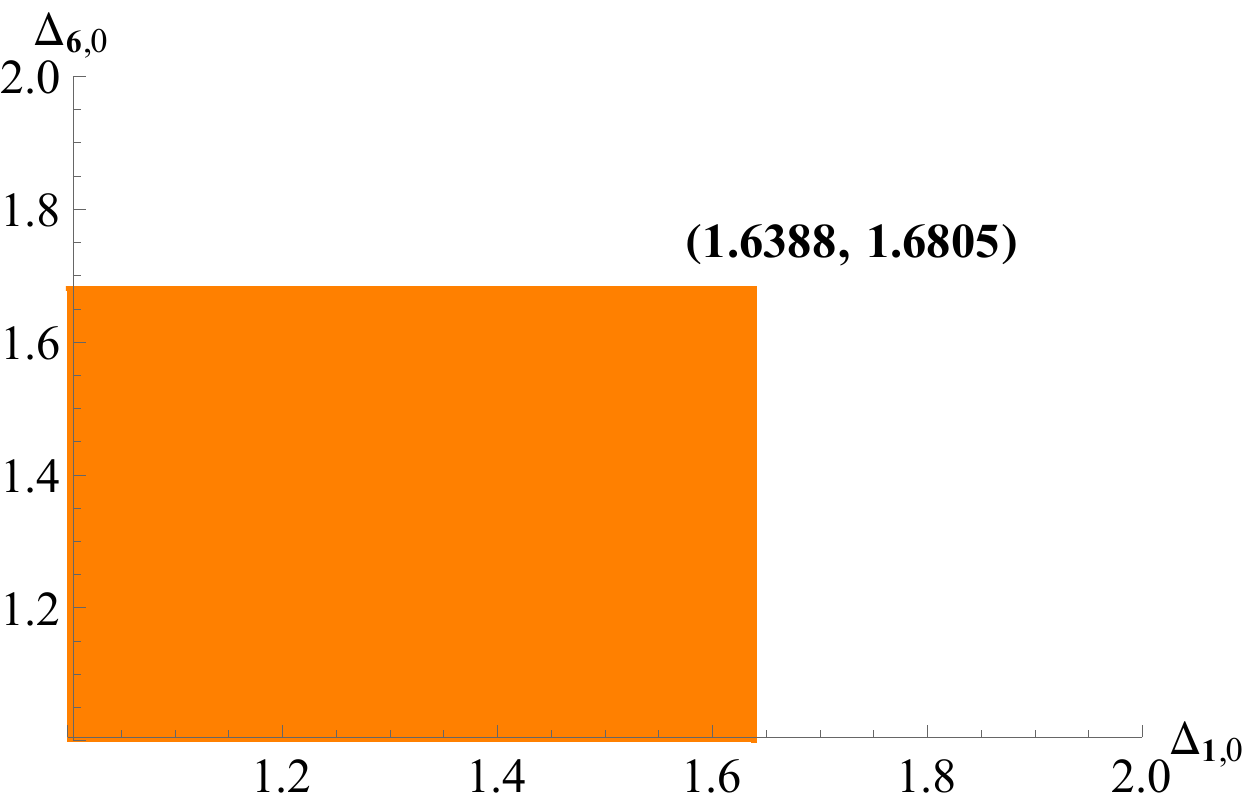}
   \includegraphics[width=0.47\textwidth]{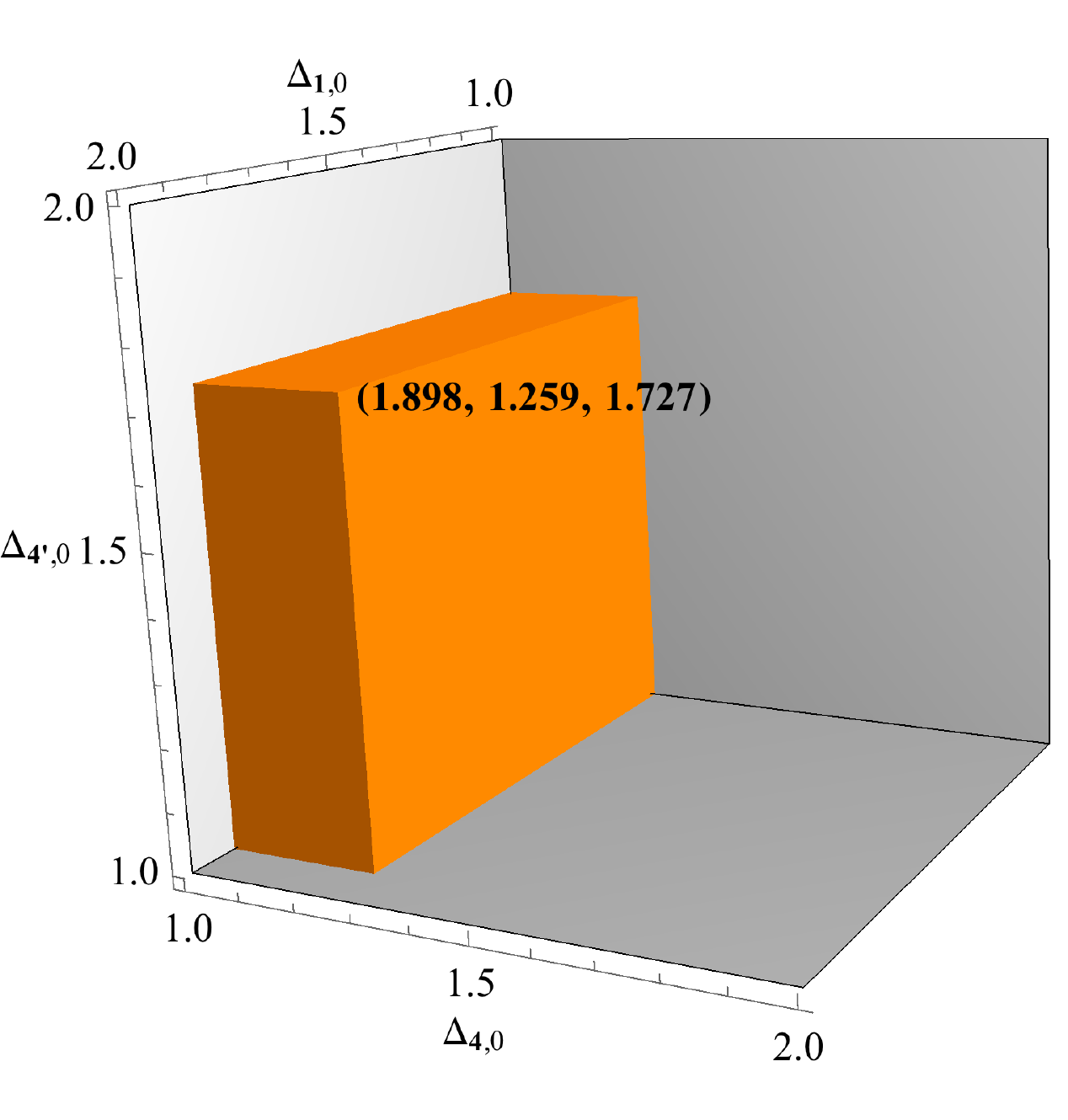}
 \caption{{\bf Left:} Bounds on the unprotected scaling dimensions $\Delta_{\bold6,0}$ and $\Delta_{\bold1,0}$ for the XYZ model in $d=3$, computed using $G\rtimes S_3$ flavor symmetry crossing equations. {\bf Right:} Bounds on the unprotected scaling dimensions  $\Delta_{\bold1,0}$, $\Delta_{\bold4,0}$, and $\Delta_{\bold4',0}$ for the $\mathbb{Z}_2\times\mathbb{Z}_2$ model in $d=3$, computed using $G\rtimes (\mathbb{Z}_2\times\mathbb{Z}_2)$ flavor symmetry crossing equations. In terms of the two-dimensional irreps of $G$, we have $\Delta_{\bold6,0}=\Delta_{\bold2',0}=\Delta_{\bold2'',0}=\Delta_{\bold2'',0}$, $\Delta_{\bold4,0}=\Delta_{\bold2,0}=\Delta_{\bold2',0}$, and $\Delta_{\bold4',0}=\Delta_{\bold2'',0}=\Delta_{\bold2''',0}$. In both plots the orange denotes the allowed region, and we conjecture that the theory lives at the corner. These bounds were computed with $\Lambda=27$.}
\label{XYZscal}
\end{center}
\end{figure}

Next, we describe numerical bounds for the XYZ model in $d=3$ that were computed using the $G\rtimes S_3$ crossing equations derived in Section~\ref{GS3cross}. Recall that the full flavor symmetry of the XYZ model is $U(1)\times U(1)\rtimes S_3$, so we are just using a fraction of the symmetry. From the $G\rtimes S_3$ perspective, the only effect of the $U(1)\times U(1)$ symmetry is to fix $\Delta_{\bold2,0}=d-2$, because this operator is the superconformal primary of the $U(1)\times U(1)$ conserved current multiplets. There are just two unprotected scalar scaling dimensions then: $\Delta_{\bold1,0}$ and $\Delta_{\bold6,0}$. On the left of Figure \ref{XYZscal} we show the numerical bounds for these quantities, which form a rectangle. We conjecture that the XYZ model lives at the nontrivial corner of this rectangle, so that $(\Delta_{\bold1,0}\,,\Delta_{\bold6,0})\approx(1.6388,1.6805)$, where in terms of $G$ irreps $\Delta_{\bold6,0}=\Delta_{\bold2',0}=\Delta_{\bold2'',0}=\Delta_{\bold2'',0}$.

We can compare these results to those of \cite{Chester:2015lej}, which studied 3d $\mathcal{N}=2$ theories with $U(1)\times O(N)$ flavor symmetry. For the case $N=2$, this describes the XYZ model, although it still only uses a fraction of the symmetry, as it neglects the $S_3$ permutation symmetry. That study found an upper bound $\Delta_{\bold1,0}\leq 1.70$, which is weaker than our bounds.

Independently of whether the XYZ model saturates the bounds in Figure~\ref{XYZscal}, we can also use the $G\rtimes S_3$ crossing equations to compute upper and lower bounds on the chiral bilinear OPE coefficient squared $\lambda^2_{\bold6_1,\frac{3}{4},0}$. We find
\es{XYZOPEbound}{
6.743\leq\lambda^2_{\bold6_1,\frac{4}{3},0}\leq8.533\,,
}
where in terms of $G$ irreps $\lambda^2_{\bold6_1,\frac{4}{3},0}=|\lambda_{\bold3_1,\frac{4}{3},0}|^2$.

\subsubsection{Symmetry group $G\rtimes \mathbb{Z}_2$ (the boundary of the fundamental domain)}
\label{numBound}

We now describe the numerical bounds for points on the boundary of the fundamental domain in $d=3$, which has $G\rtimes\mathbb{Z}_2$ flavor symmetry, using the crossing equations derived in Section \ref{GZ2cross}. For convenience, we choose the duality frame where $\Im\, \tau = 0$, so our plots will be functions of real $\tau$. In order to view all three bounding curves of the conformal manifold in a single plot, we will use the range $1-\sqrt{3}\leq\tau\leq1+\sqrt{3}$, which as shown on the LHS of Figure \ref{fund} involves the fundamental domain $\mathbb{F}$ defined on the RHS of Figure \ref{fund}, as well as two adjacent domains for $1-\sqrt{3}\leq\tau\leq0$ and $1\leq\tau\leq1+\sqrt{3}$. When we map these fundamental domains to $\mathbb{F}$, some of the doublets are permuted by the duality group $S_4$, as we show in Table \ref{perm}.

\begin{table}[!h]
\begin{center}
\begin{tabular}{c||c|c|c|c}
$\tau\in\mathbb{F}$ & $\cO_{\bold2,0}$& $\cO_{\bold2',0}$ & $\cO_{\bold2'',0}$ & $\cO_{\bold2''',0}$  \\
 \hline 
$1-\sqrt{3}\leq\tau\leq0$   &  $\cO_{\bold2,0}$  &$\cO_{\bold2''',0}$   &  $\cO_{\bold2'',0}$  & $\cO_{\bold2',0}$ \\
\hline
$1\leq\tau\leq1+\sqrt{3}$ & $\cO_{\bold2'',0}$ &   $\cO_{\bold2',0}$  &  $\cO_{\bold2,0}$  &   $\cO_{\bold2''',0}$\\
 \hline
\end{tabular}
\caption{Relation of doublets in $1-\sqrt{3}\leq\tau\leq0$ and $1\leq\tau\leq1+\sqrt{3}$ to doublets in the fundamental domain $\mathbb{F}$ used in this paper.\label{perm}}
\end{center}
\end{table}

\begin{figure}[t!]
\begin{center}
   \includegraphics[width=0.85\textwidth]{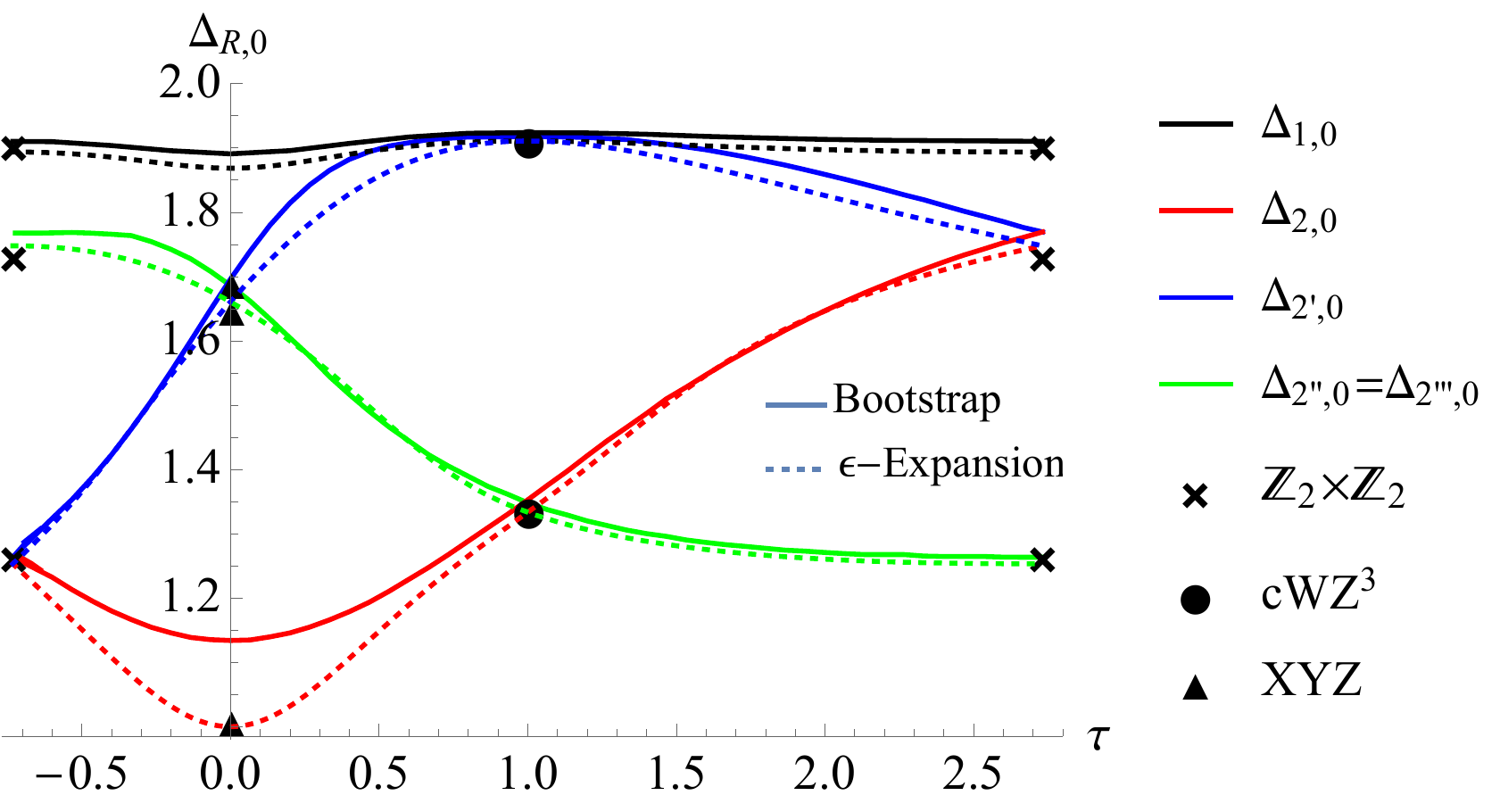}
 \caption{Upper bounds on the unprotected scaling dimensions of the scalar singlet and doublets for real $1-\sqrt{3}\leq\tau\leq1+\sqrt{3}$, computed using the $G\rtimes \mathbb{Z}_2$ flavor symmetry crossing equations. The cross, circle, and triangle denote the results from the previous sections for the enhanced symmetry points $\tau=1\pm \sqrt{3}, 1,0$ for the $\mathbb{Z}_2\times\mathbb{Z}_2$, cWZ$^3$, and XYZ models respectively. (For the XYZ model, the top and bottom triangles correspond to the doublets while the middle one corresponds to the singlet. See also Table~\ref{finalResults} in the Discussion section.) The dotted lines show the 3-loop resummed $4-\varepsilon$-expansion results. These bounds were computed with $\Lambda=19$.}
\label{bootBound}
\end{center}
\end{figure}

In Figure \ref{bootBound} we show upper bounds on scaling dimensions of the singlet and doublets as a function of real $\tau$. The different colors correspond to the singlet and different doublets, where $\Delta_{\bold{2}'',0}=\Delta_{\bold{2}''',0}$ due to the enhanced $\mathbb{Z}_2$ symmetry. The cross, circle, and triangle denote the results from the previous sections for the enhanced symmetry points $\tau=1\pm \sqrt{3}, 1,0$ for the $\mathbb{Z}_2\times\mathbb{Z}_2$, cWZ$^3$, and XYZ models respectively. Note that the results $\Delta_{\bold{2},0}=1$ at $\tau=0$ and $\Delta_{\bold{2},0}=\Delta_{\bold{2}'',0}=\Delta_{\bold{2}'',0}=\frac43$ at $\tau=1$ are analytical, while the rest are numerical upper bounds. The dotted lines show the 3-loop  Pad\'e[1,2] resummation of the $4-\varepsilon$-expansion results in \eqref{Delta1}, \eqref{Delta2}, and \eqref{Delta2123}. These $4-\varepsilon$-expansion results for the doublets are very close the bootstrap upper bounds for most of the plot, so we conjecture that at infinite numerical precision the CFT saturates the upper bounds. The singlet bounds appear to not be as well converged, as they differ significantly from the perturbative and exact results throughout the manifold. The bootstrap results appear to be less converged near the $\tau=0$ XYZ point. For instance, the bootstrap upper bound gives $\Delta_{\bold{2},0}\leq1.14$, which is weaker than the analytic value $\Delta_{\bold{2},0}=1$.

\begin{figure}[t!]
\begin{center}
   \includegraphics[width=0.47\textwidth]{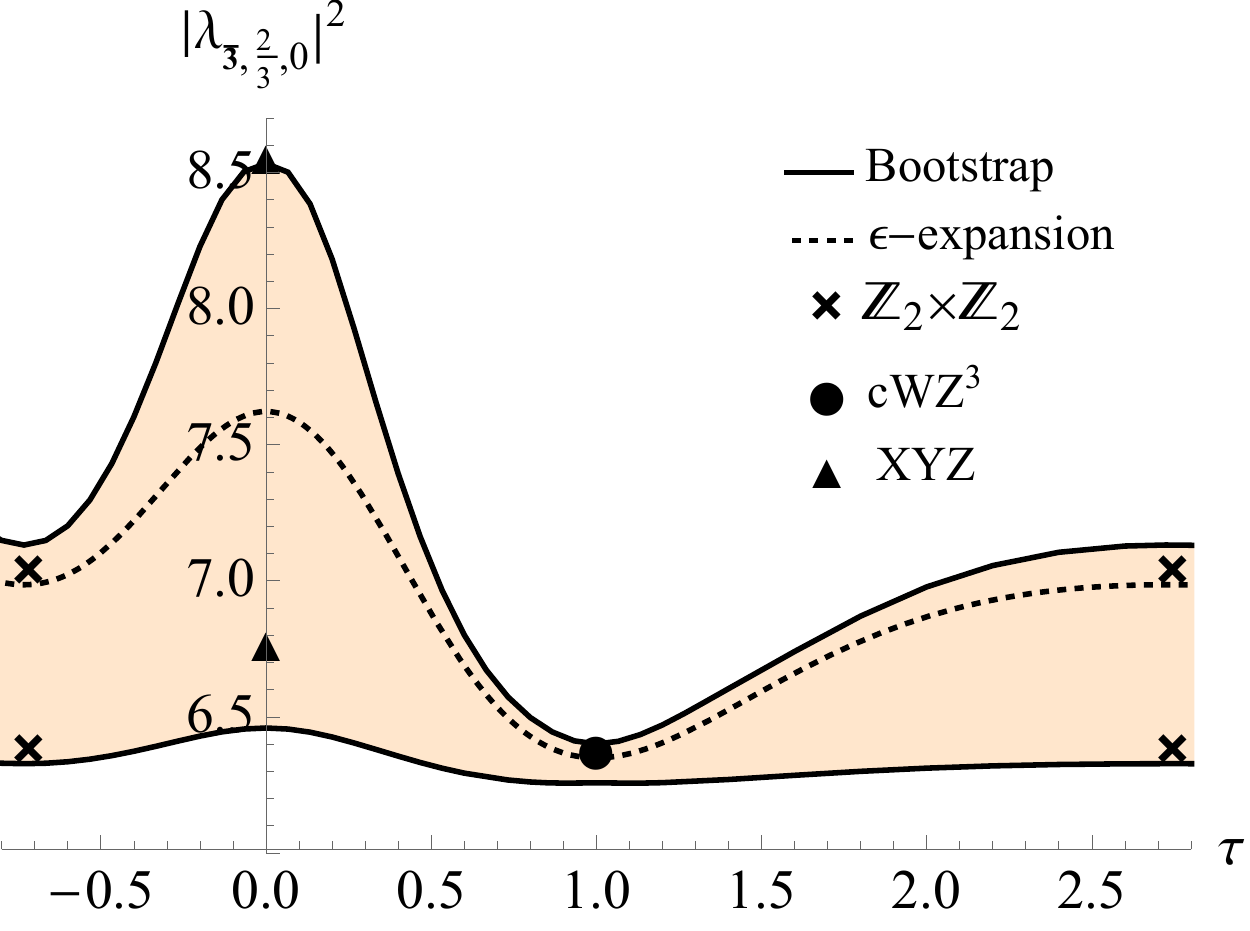}
      \includegraphics[width=0.47\textwidth]{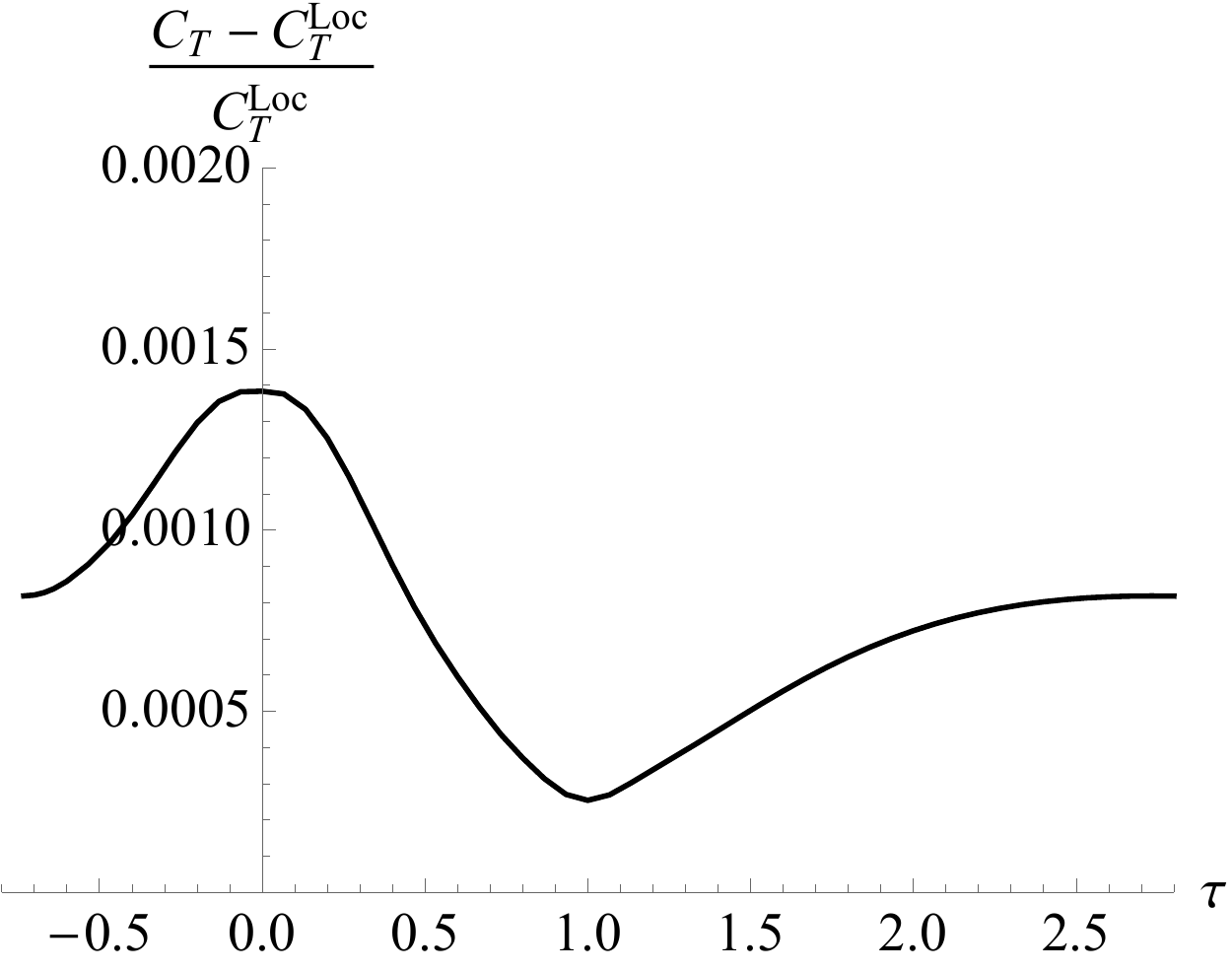}
 \caption{{\bf Left:} Upper and lower bounds on the chiral bilinear OPE coefficient squared $\lambda^2_{\bold6_1,\frac{4}{3},0}$ for real $1-\sqrt{3}\leq\tau\leq1+\sqrt{3}$. The cross, circle, and triangle denote the results from the previous sections for the enhanced symmetry points $\tau=1\pm \sqrt{3}, 1,0$ for the $\mathbb{Z}_2\times\mathbb{Z}_2$, cWZ$^3$, and XYZ models respectively. The dotted lines show the 2-loop $4-\varepsilon$-expansion results. {\bf Right:} Upper bounds on $C_T$  for real $1-\sqrt{3}\leq\tau\leq1+\sqrt{3}$ compared to the $\tau$-independent localization value in \eqref{cTloc}. Both plots were computed using the $G\rtimes \mathbb{Z}_2$ flavor symmetry crossing equations with $\Lambda=19$.}
\label{bootBoundOPE}
\end{center}
\end{figure}

On the left of Figure \ref{bootBoundOPE} we show upper and lower bounds on the chiral bilinear OPE coefficient squared $|\lambda_{\bar{\bold3}_1,\frac{3}{4},0}|^2$ as a function of real $\tau$. Again, the cross, circle, and triangle denote the results from the previous sections for the enhanced symmetry points $\tau=1\pm \sqrt{3}, 1,0$ for the $\mathbb{Z}_2\times\mathbb{Z}_2$, cWZ$^3$, and XYZ models respectively. Note that only the result $|\lambda_{\bar{\bold3}_1,\frac{3}{4},0}|^2=2^{8/3}$ at $\tau=1$ is analytical, while the rest are numerical upper and lower bounds. The dotted lines show the 2-loop $4-\varepsilon$-expansion result in \eqref{OPEeps}. As with the scaling dimension plots, the $4-\varepsilon$-expansion results are close to the bootstrap results everywhere except near the $\tau=0$ XYZ point. 

As a further check on the accuracy of the bootstrap bounds, on the right of Figure \ref{bootBoundOPE} we compare the upper bounds on $C_T$ as a function of real $\tau$ versus the exact $\tau$-independent value computed using supersymmetric localization in \eqref{cTloc}, where $C_T$ is computed in terms of CFT data in our conventions as
\es{cTdata}{
C_T=\frac{128}{3}|\lambda_{\bold1,2,2}|^2\,.
}
For all $\tau$ the upper bound is close to saturating the exact value, but the match is more precise away from the $\tau=0$ XYZ point.

\subsubsection{Symmetry group $G$ (generic point)}
\label{numBulk}

\begin{figure}[t!]
\begin{center}
   \includegraphics[width=0.85\textwidth]{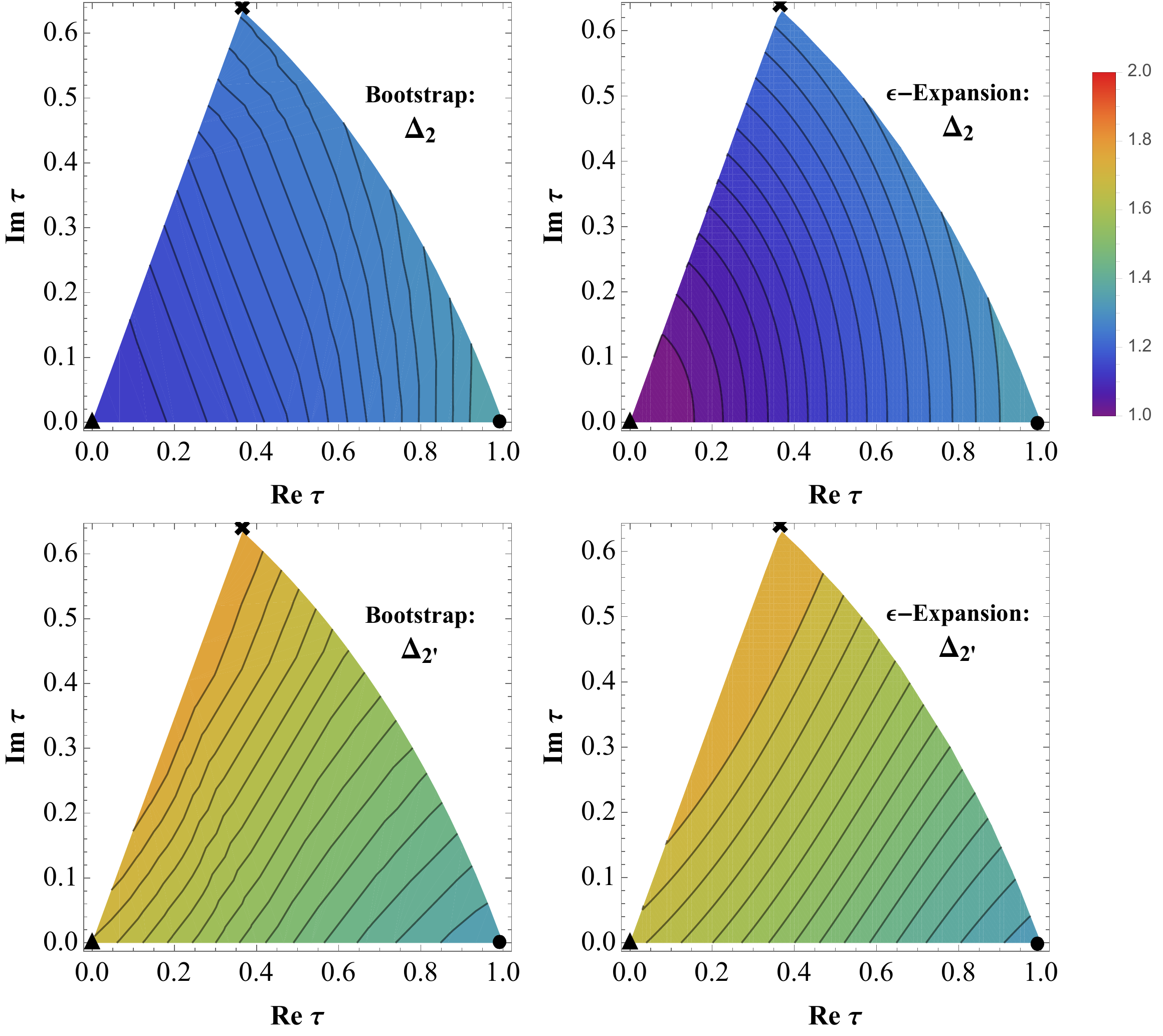}
 \caption{{\bf Left:} Upper bounds on the doublet scaling dimensions $\Delta_\bold2$ and $\Delta_{\bold2'}$ for all $\tau$ in the fundamental domain $\mathbb{F}$ defined in Figure \ref{fund}, computed using the $G$ flavor symmetry crossing equations with $\Lambda=19$. {\bf Right:} Resummed 3-loop $4-\varepsilon$-expansion values for these same quantities. In all plots the cross, circle, and triangle denote the enhanced symmetry points $\tau=1\pm \sqrt{3}, 1,0$ for the $\mathbb{Z}_2\times\mathbb{Z}_2$, cWZ$^3$, and XYZ models respectively.}
\label{dub1}
\end{center}
\end{figure}

\begin{figure}[t!]
\begin{center}
   \includegraphics[width=0.85\textwidth]{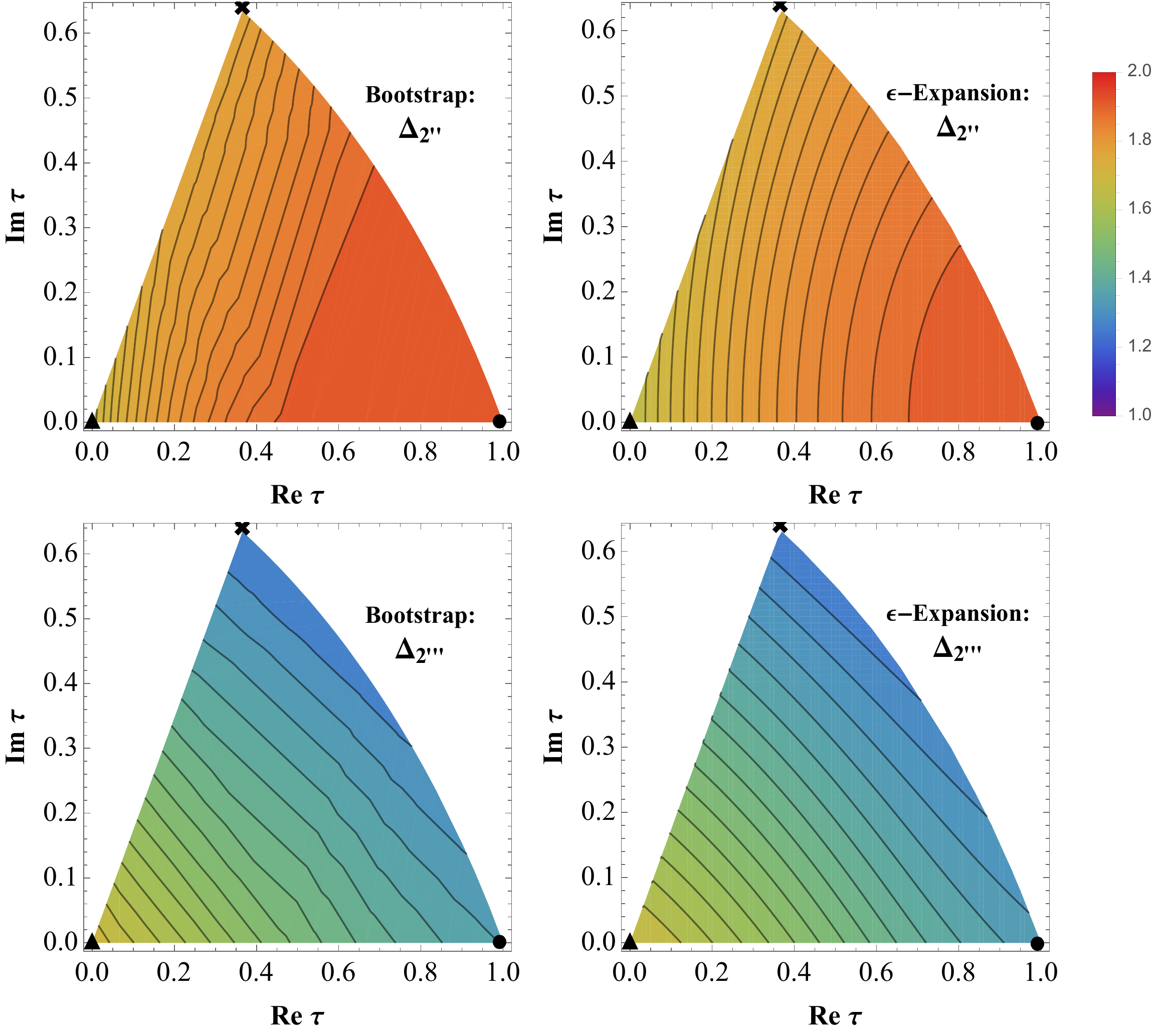}
 \caption{{\bf Left:} Upper bounds on the doublet scaling dimensions $\Delta_{\bold2''}$ and $\Delta_{\bold2'''}$ for all $\tau$ in the fundamental domain $\mathbb{F}$ defined in Figure \ref{fund}, computed using the $G$ flavor symmetry crossing equations with $\Lambda=19$. {\bf Right:} Resummed 3-loop $4-\varepsilon$ expansion values for these same quantities. In all plots the cross, circle, and triangle denote the enhanced symmetry points $\tau=1\pm \sqrt{3}, 1,0$ for the $\mathbb{Z}_2\times\mathbb{Z}_2$, cWZ$^3$, and XYZ models respectively.}
\label{dub2}
\end{center}
\end{figure}

\begin{figure}[t!]
\begin{center}
   \includegraphics[width=0.85\textwidth]{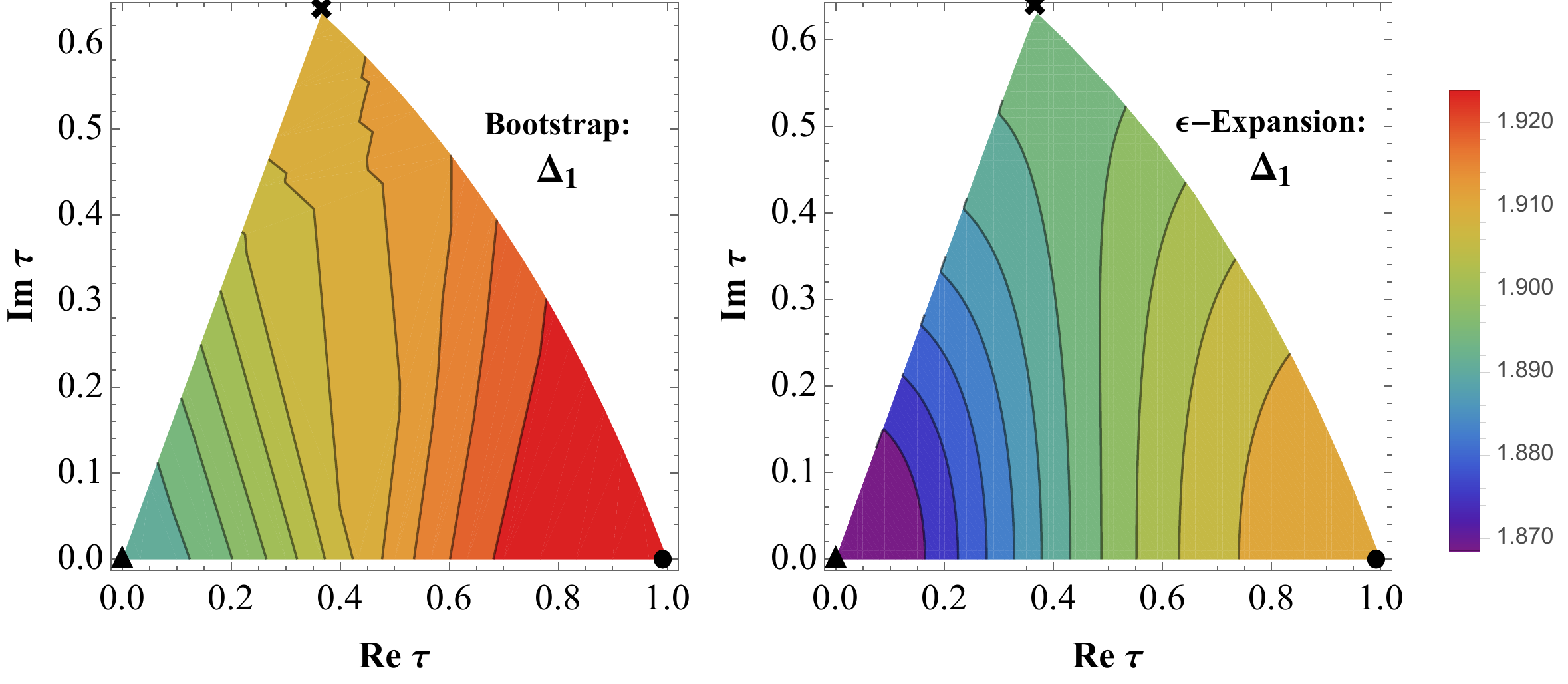}
 \caption{{\bf Left:} Upper bounds on the singlet scaling dimension $\Delta_{\bold1}$ for all $\tau$ in the fundamental domain $\mathbb{F}$ defined in Figure \ref{fund}, computed using the $G$ flavor symmetry crossing equations with $\Lambda=19$. {\bf Right:} Resummed 3-loop $4-\varepsilon$ expansion values for this same quantity. In all plots the cross, circle, and triangle denote the enhanced symmetry points $\tau=1\pm \sqrt{3}, 1,0$ for the $\mathbb{Z}_2\times\mathbb{Z}_2$, cWZ$^3$, and XYZ models respectively.}
\label{dub3}
\end{center}
\end{figure}

\begin{figure}[t!]
\begin{center}
   \includegraphics[width=0.85\textwidth]{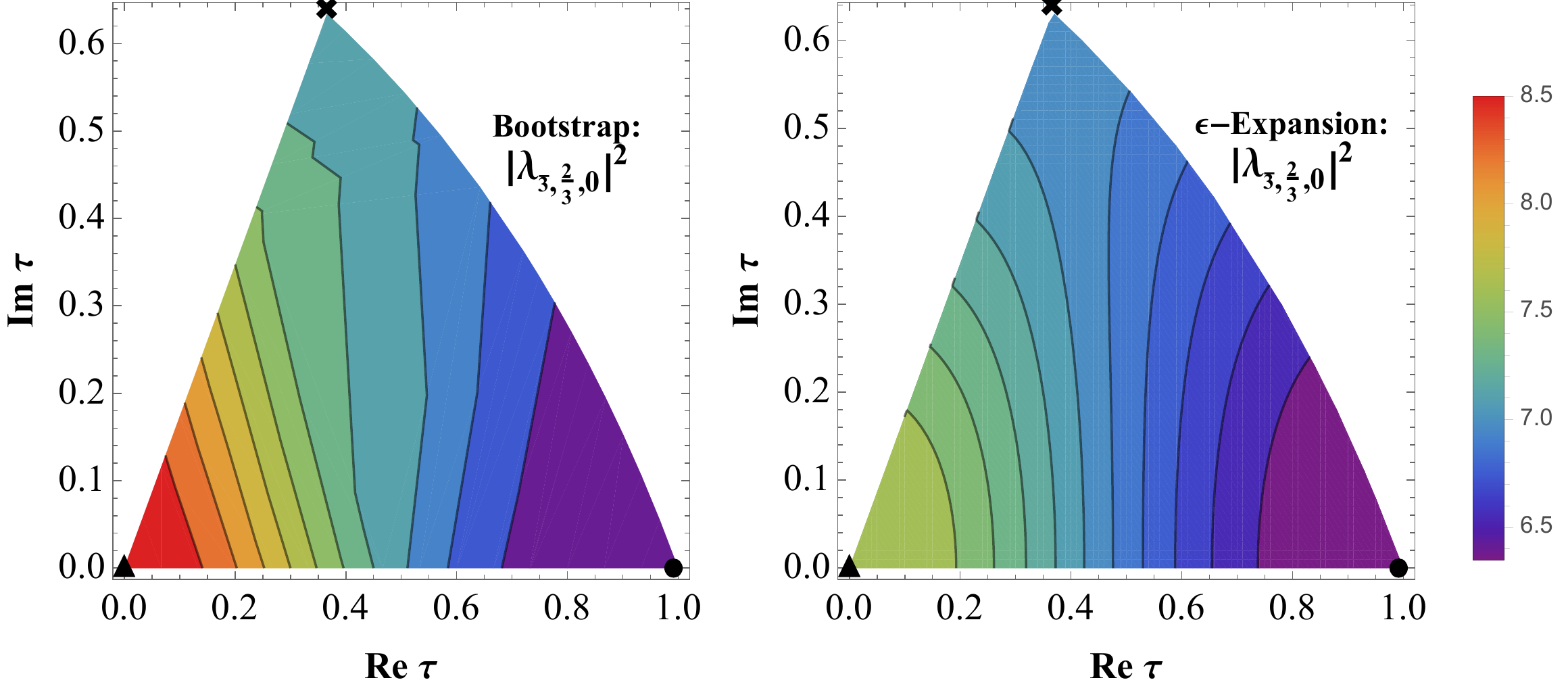}
 \caption{{\bf Left:} Upper bounds on the chiral bilinear OPE coefficient squared $\lambda^2_{\bold6_1,\frac{4}{3},0}$ for all $\tau$ in the fundamental domain $\mathbb{F}$ defined in Figure \ref{fund}, computed using the $G$ flavor symmetry crossing equations with $\Lambda=19$. {\bf Right:} 2-loop $4-\varepsilon$-expansion values for this quantity. In all plots the cross, circle, and triangle denote the enhanced symmetry points $\tau=1\pm \sqrt{3}, 1,0$ for the $\mathbb{Z}_2\times\mathbb{Z}_2$, cWZ$^3$, and XYZ models respectively.}
\label{dub4}
\end{center}
\end{figure}

We now describe the numerical bounds for general points on the conformal manifold in $d=3$, which has $G$ flavor symmetry, using the crossing equations derived in Section \ref{GZ2cross}. Without loss of generality we restrict $\tau$ to the fundamental domain $\mathbb{F}$ defined in Figure \ref{fund}. In Figures \ref{dub1}, \ref{dub2}, and \ref{dub3} we show upper bounds on the doublet and singlet scaling dimensions along with the 3-loop Pad\'e[1,2] resummation of the $4-\varepsilon$-expansion results in \eqref{Delta1}, \eqref{Delta2}, and \eqref{Delta2123}. As was the case with the boundary of the fundamental domain, the $4-\varepsilon$-expansion results are very close to the numerical upper bounds for most of the doublets plots except near the XYZ model at $\tau=0$, while the agreement for the singlet plots is somewhat less precise. In Figure \ref{dub4} we show upper bounds for chiral bilinear OPE coefficient squared $|\lambda_{\bar{\bold3}_1,\frac{3}{4},0}|^2$ along with the 2-loop $4-\varepsilon$-expansion results in \eqref{OPEeps}, and again find similar values for each away from the $\tau=0$ point.

\section{Discussion}
\label{sec:discussion}

In this work, we have uncovered the structure of a relatively simple example of a 3d $\mathcal{N}=2$ conformal manifold using duality, perturbative tools, as well as the numerical implementation of the conformal bootstrap. In particular, we  find that the $3$-loop $4-\varepsilon$-expansion results for the scaling dimensions of scalar bilinears as a function of the complex conformal manifold parameter $\tau$ match the upper bounds from the bootstrap to high precision everywhere on the manifold away from the XYZ point. For a quantitative comparison, in Table~\ref{finalResults} we summarize these results at the three points on the manifold with enhanced symmetry (XYZ, cWZ$^3$, and $\mathbb{Z}_2\times\mathbb{Z}_2$) for the operators whose scaling dimensions are not already fixed by symmetry.  See also Figures~\ref{dub1}, \ref{dub2}, and~\ref{dub3}. We have also computed the OPE coefficient of the bilinear chiral operator to 2-loops in the $4-\varepsilon$-expansion.  Comparing it to our bootstrap results,  we again find a good match away from the XYZ point, just as in the case of the scaling dimensions---see Figure~\ref{bootBoundOPE}. 

In the future, it would be interesting to see if this match becomes more precise as we push the $4-\varepsilon$ expansion and bootstrap to higher precision.  In particular, it would be interesting to know if there is a fundamental reason why the match is worse near the XYZ point, perhaps having to do with the existence of the continuous global symmetry at that point.

\begin{table}[!h]
\begin{center}
\begin{tabular}{c||c|c|c|c|c}
 & $\Delta_{\bold1}$& $\Delta_{\bold2}$ & $\Delta_{\bold2'}$ & $\Delta_{\bold2''}$ & $\Delta_{\bold2'''}$ \\
 \hline 
 \hline
XYZ from Bootstrap   &  $1.639$ &  $1\ {}^*$  & $1.681$ &$1.681$ & $1.681$ \\
\hline
XYZ from $4-\varepsilon$-expansion  &  $1.869$ &  $1\ {}^*$   & $1.661$ &$1.661$ & $1.661$\\
 \hline
 \hline
 cWZ$^3$ from Bootstrap   &  $1.910$ &  $4/3\ {}^*$   & $1.910$ &$4/3\ {}^*$ & $4/3\ {}^*$\\
\hline
 cWZ$^3$ from $4-\varepsilon$-expansion &  $1.911$ &  $4/3\ {}^*$   & $1.911$ &$4/3\ {}^*$ & $4/3\ {}^*$\\
 \hline
 \hline
 $\mathbb{Z}_2\times \mathbb{Z}_2$ from Bootstrap   &  $1.898$ &  $1.259$   & $1.259$ &$1.727$ & $1.727$ \\
\hline
 $\mathbb{Z}_2\times \mathbb{Z}_2$ from $4-\varepsilon$-expansion  &  $1.894$ &  $1.253$   & $1.253$ &$1.748$ & $1.748$\\
\end{tabular}
\caption{Summary of results for the doublet and singlet scaling dimensions that are not fixed by symmetry in 3d for the XYZ, cWZ$^3$, and $\mathbb{Z}_2\times\mathbb{Z}_2$ theories at $\tau=0,1,(1-\sqrt{3})\omega^2$, respectively, from the numerical bootstrap and the resummed 3-loop $4-\varepsilon$-expansion.  The results marked with a ${}^*$ are exact.\label{finalResults}}
\end{center}
\end{table}

In Section~\ref{sec:chiralring}, we derived the Zamolodchikov metric up to 2-loops in the $4-\varepsilon$-expansion.  This quantity cannot be compared to the bootstrap analysis we performed here, because no operators in the same multiplet as the marginal operator used to define the Zamolodchikov metric appear in any OPE channel of the four-point function we study.   To circumvent this problem, one would have to perform a bootstrap analysis of more correlators.   For instance, if one were to analyze a system of four-point functions of the chiral/anti-chiral operators of dimension $2/3$ (namely $X_i$ and $\overline{X}^i$) that we study here as well as of the chiral/anti-chiral operators of dimension $4/3$, then the superconformal primary of the multiplet containing the marginal operator would appear in the OPE of the dimension $2/3$ and $4/3$ chiral operators. In order to extract the Zamolodchikov metric from these correlation functions, it would be useful to generalize the so called $tt^*$ equations to 3d theories with four supercharges, which would allows us to relate the OPE coefficient of the marginal operator to the Zamolodchikov curvature invariants.  Such a relation is currently understood in 2d  \cite{Cecotti:1991me} and 4d theories with eight supercharges \cite{Papadodimas:2009eu,Baggio:2014sna,Baggio:2014ioa}.

In the future it would be interesting to generalize our conformal manifold study to other setups, for example to 3d $\mathcal{N}=2$ theories with $N >3$ chiral superfields and a general cubic superpotential.  A simple calculation suggests that such a theory has a conformal manifold of complex dimension $N(N-1)(N-2)/6$.  It would be fascinating if the methods used in this paper could be applied to this more general class of theories.

When our model is taken at face value in four space-time dimensions the couplings $h_{1,2}$ are marginally irrelevant and thus the conformal manifold trivializes to a weakly coupled point in field theory space. However it should be noted that the superpotential in \eqref{superintro} looks superficially similar to the one on the $\mathcal{N}=1$ conformal manifold of 4d $\mathcal{N}=4$ SYM \cite{Leigh:1995ep}. (See also \cite{Aharony:2002hx} for a useful summary.)  Perhaps this similarity combined with our results can be used as leverage towards understanding this 4d conformal manifold in more detail. We should also emphasize that we managed to perform the numerical conformal bootstrap as a function of the marginal coupling $\tau$. One could hope that a similar analysis can be performed along the conformal manifold parameterizing the ${\cal N} = 1$-preserving exactly marginal deformations of 4d $\mathcal{N}=4$ SYM, thus extending the results in \cite{Beem:2013qxa,Beem:2016wfs}.

Lastly, let us mention that, as discussed in  \cite{Dimofte:2011ju} (see also \cite{Cecotti:2011iy}), certain 3d $\mathcal{N}=2$ QFTs can be realized as M5-branes wrapping hyperbolic 3-manifolds with a partial topological twist. Therefore there is a natural map between hyperbolic manifolds, and Chern-Simons theory on them, and many $\mathcal{N}=2$ QFTs\@. It is known that the XYZ model can be realized in this context.  However, it is established that the metric on hyperbolic 3-manifolds does not admit smooth deformations, which is a property known as Mostow rigidity \cite{Mostow} (see also \cite{thurston1978geometry}). This may naively suggest a tension with the existence of a conformal manifold stemming from the XYZ SCFT\@. However there is no theorem that forbids other, i.e.~non-metric, deformations of the twisted M5-brane theory to be compatible with $\mathcal{N}=2$ supersymmetry. It would be most interesting to identify a deformation that realizes the complex marginal parameter $\tau$ in our model and to understand the meaning of this marginal deformation from the point of view of the Chern-Simons theory on the hyperbolic manifold.

\section*{Acknowledgements }

\noindent We would like to thank Chris Beem, Francesco Benini, Alice Bernamonti, Matt Buican, Adam Bzowski, Jan de Boer, Clay C\'ordova, Lorenzo Di Pietro, Abhijit Gadde, Davide Gaiotto, Federico Galli, Fri\dh rik Gautason, Matthijs Hogervorst, Igor Klebanov, Dalimil Maz\'a\v c, Kyriakos Papadodimas, Gabriele Tartaglino-Mazzucchelli, and Yifan Wang for useful discussions. NB is grateful to the students at the 2016 Solvay Doctoral School in Amsterdam who made him think about 3d $\mathcal{N}=2$ conformal manifolds and to the organizers of the school for the inspiring atmosphere. EL would like to thank the Perimeter Institute for Theoretical Physics, where part of the project was carried out.
Research at Perimeter Institute is supported by the Government of Canada through Industry Canada and by the Province of Ontario through the Ministry of Research \& Innovation.  The work of MB is supported by the European Union's Horizon 2020 research and innovation programme under the
Marie Sk\l odowska-Curie grant agreement no.~665501 with the Research Foundation Flanders (FWO). MB is an FWO [PEGASUS]$^2$ Marie
Sk\l odowska-Curie Fellow. The work of NB is supported in part by the starting grant BOF/STG/14/032 from KU Leuven, by an Odysseus grant G0F9516N from the FWO, and by the KU Leuven C1 grant ZKD1118 C16/16/005. In addition MB, NB and EL acknowledges support by the Belgian Federal Science Policy Office through the Inter-University Attraction Pole P7/37, and by the COST Action MP1210 The String Theory Universe. EL is additionally supported by the European Research Council grant no. ERC-2013-CoG 616732 HoloQosmos, as well as the FWO  Odysseus  grants G.001.12 and G.0.E52.14N\@. SMC and SSP are supported in part by the Simons Foundation Grant No.~488651.  

\begin{appendices}

\appendix
\section{Details of flavor groups}
\label{sec:discretegroup}
In this appendix we collect some useful facts about the discrete flavor symmetry group $G=(\mathbb{Z}_3\times\mathbb{Z}_3)\rtimes S_3$ introduced in Section \ref{symmetries}, as well as the flavor groups $G\rtimes \mathbb{Z}_2$, $G\rtimes (\mathbb{Z}_2\times\mathbb{Z}_2)$, and $G\rtimes S_3$ that describe the points on the manifold with enhanced symmetry.

Let us begin with the group $G$. The conjugacy classes can be written in terms of the generators \eqref{G} as 
\es{conj}{
    C_1&=\{I\}\,,\\
    C_{9}&=\{g_1,\,g_1g_2,\,g_2g_1,\,g_1g_2g_3,\,g_1g_2g_3^2,\,g_1g_3^2g_1g_3g_1,\,g_2g_3g_1,\,(g_3g_1)^3,\,g_2g_1g_3(g_3g_1)^4\}\,,\\
    C'_1&=\{(g_3g_1)^4\}\,,\\
    \bar{C}'_1&=\{(g_3g_1)^2\}\,,\\
    C_6&=\{g_2,\,g_2^2,\,g_2(g_3g_1)^2,\,g_2^2(g_3g_1)^2,\,g_2(g_3g_1)^4,\,g_2^2(g_3g_1)^4\}\,,\\
    C'_6&=\{g^2_2g_3,\,g_2g_3^2,\,g^2_2g_3(g_3g_1)^2,\,g_2g_1g_3g_1,\,g_2^2g_3(g_3g_1)^4,\,g_2g_1(g_3g_1)^3\}\,,\\
    \bar{C}'_6&=\{g_2g_3,\,g^2_2g_3^2,\,g_2g_3(g_3g_1)^2,\,g_1g_2g_3g_1,\,g_2g_3(g_3g_1)^4,\,g_1g_2(g_3g_1)^3\}\,,\\
    C''_6&=\{g_3,\,g_3^2,\,g_3(g_3g_1)^2,\,g_1g_3g_1,\,g_3(g_3g_1)^4,\,g_1(g_3g_1)^3\}\,,\\
    C'_{9}&=\{g_1g_3,\,g_2g_1g_3^2,\,g_1^2g_3g_1,\,g_2g_1g_3(g_3g_1)^2,\,g_1(g_3g_1)^4,\,g_1g_2(g_3g_1)^4,\,g_2g_1(g_3g_1)^4,\,g_1g_2g_3(g_3g_1)^4,\,g_2^2(g_3g_1)^3\}\,,\\
    \bar{C}'_{9}&=\{g_1g_3^2,\,g_2g_1g_3,\,g_1(g_3g_1)^2,\,g_1g_2(g_3g_1)^2,\,g_2g_1(g_3g_1)^2,\,g_1g_2g_3(g_3g_1)^2,\,g_2^2g_3g_1,\,g_1g_3(g_3g_1)^4,\,g_2(g_3g_1)^3\}\,.
}
From these we determine the character table \ref{charTable}.
\begin{table}[htp]
    \begin{center}
        \begin{tabular}{c|c|c|c|c|c|c|c|c|c|c}
            $\chi_R(g_a)$  & $C_1$ & $C_9$ & $C'_1$ & $\bar{C}'_1$ & $C_6$  & $C'_6$ & $\bar{C}'_6$ & $C''_6$  & $C'_9$ & $\bar{C}'_9$ \\
            \hline\hline
            $\bold{1}$ & $1$ & $1$ & $1$ & $1$ & $1$ & $1$ & $1$ & $1$ & $1$& $1$  \\
            \hline
            $\bold{1}'$ & $1$ & $-1$ & $1$ & $1$ & $1$ & $1$ & $1$ & $1$ & $-1$& $-1$  \\
            \hline
            $\bold{2}$ & $2$ & $0$ & $2$ & $2$ & $-1$ & $-1$ & $-1$ & $2$ & $0$& $0$  \\
            \hline
            $\bold{2}'$ & $2$ & $0$ & $2$ & $2$ & $2$ & $-1$ & $-1$ & $-1$ & $0$& $0$  \\
            \hline
            $\bold{2}''$ & $2$ & $0$ & $2$ & $2$ & $-1$ & $-1$ & $2$ & $-1$ & $0$& $0$  \\
            \hline
            $\bold{2}'''$ & $2$ & $0$ & $2$ & $2$ & $-1$ & $2$ & $-1$ & $-1$ & $0$& $0$  \\
            \hline
            $\bold{3}$ & $3$ & $1$ & $3\omega^2$ & $3\omega$ & $0$ & $0$ & $0$ & $0$ & $\omega^2$& $\omega$  \\
            \hline
            $\bar{\bold{3}}$ & $3$ & $1$ & $3\omega$ & $3\omega^2$ & $0$ & $0$ & $0$ & $0$ & $\omega$& $\omega^2$  \\
            \hline
            $\bold{3}'$ & $3$ & $-1$ & $3\omega$ & $3\omega^2$ & $0$ & $0$ & $0$ & $0$ & $-\omega$& $-\omega^2$  \\
            \hline
            $\bar{\bold{3}'}$ & $3$ & $-1$ & $3\omega^2$ & $3\omega$ & $0$ & $0$ & $0$ & $0$ & $-\omega^2$& $-\omega$  \\
            \hline
        \end{tabular}
    \end{center}
    \caption{Character table for $G=(\mathbb{Z}_3\times \mathbb{Z}_3)\rtimes S_3$. The subscripts of the conjugacy classes indicate their size, and $\omega=e^{2\pi {\rm i}/3}$.}
    \label{charTable}
\end{table}

In the tensor products ${\bf 3} \otimes {\bf 3}$, $\bar {\bf 3} \otimes \bar {\bf 3}$, and ${\bf 3} \otimes \bar {\bf 3}$, the projector operators onto irrep $R$ are given by
\es{projectors}{
    &P_{R}{}_i{}^j{}_k{}^l=\frac{|R|}{|G|}\sum_{a,b}\chi_R(g_a\otimes g_b)g_{a}{}_i{}^j\otimes g_{b}{}_k{}^l\,,\\
    &P_{R}{}^i{}_j{}^k{}_l=\frac{|R|}{|G|}\sum_{a,b}\chi_R(\bar g_a\otimes \bar g_b)\bar g_{a}{}^i{}_j\otimes \bar g_{b}{}^k{}_l\,,\\
    &P_{R}{}_i{}^j{}^k{}_l=\frac{|R|}{|G|}\sum_{a,b}\chi_R(g_a\otimes \bar{g}_b)g_{a}{}_i{}^j\otimes \bar{g}_{b}{}^k{}_l\,,\\
}
where $a=1,\dots,|G|$, to compute the eigenvectors with unit eigenvalues
\es{projectorEig}{
    &P_{R}{}_i{}^j{}_k{}^l v_R{}_r{}_j{}_l=v_{R,r}{}_i{}_k\,,\qquad P_{R}{}^i{}_j{}^k{}_l v_R{}_r{}^j{}^l=v_{R,r}{}^i{}^k\,,\qquad P_{R}{}_i{}^j{}^k{}_l v_R{}_r{}_j{}^l=v_{R,r}{}_i{}^k\,,
}
where $r=1,\dots,|R|$.  For the irreps other than the ${\bf 3}'$, these eigenvectors can be identified with the operators in \eqref{bilinears} and \eqref{chiralBis}.

Let us now discuss the order 108 group $G\rtimes\mathbb{Z}_2$. For simplicity, let us focus on the duality frame where $\mathbb{Z}_2$ acts as conjugation. We can now combine $X_i$ and $\overline X^j$ into a single operator $\tilde X_I=\{X_i,\overline X^j\}$ where $I=1,\dots,6$, where $\tilde X_I$ transforms in the real representation $\bold6$ of $G\rtimes \mathbb{Z}_2$. In this representation, the elements $h\in G\rtimes\mathbb{Z}_2$ can be written as $4\times4$ matrices as
\es{Z2group}{
h\in G\rtimes\mathbb{Z}_2:\qquad\begin{pmatrix}
g&0\\
0&g\\
\end{pmatrix}\,,\qquad
\begin{pmatrix}
g&0\\
0&\bar g\\
\end{pmatrix}\,,
}
where $g\in G$. The character table for $G\rtimes\mathbb{Z}_2$ is given in Table \ref{charTableZ2}.
\begin{table}[htp]
\begin{center}
  \begin{tabular}{c|c|c|c|c|c|c|c|c|c|c|c}
  $\chi_R(g_a)$  & $C_1$ & $C_9$ & $C'_9$ & ${C}_6$ & $C'_6$  & $C_2$ & ${C}'''_9$ & $C_{18}$  & $C'_{18}$ & ${C}_{12}$ &$C''_{18}$ \\
 \hline\hline
 $\bold{1}^E$ & $1$ & $1$ & $1$ & $1$ & $1$ & $1$ & $1$ & $1$ & $1$& $1$ &$1$ \\
 \hline
 $\bold{1}^O$ & $1$ & $-1$ & $-1$ & $1$ & $1$ & $1$ & $1$ & $-1$ & $-1$& $1$ &$1$  \\
 \hline
 $\bold{1}'$ & $1$ & $-1$ & $1$ & $1$ & $1$ & $1$ & $-1$ & $-1$ & $1$& $1$ & $-1$  \\
 \hline
 $\bold{1}''$ & $1$ & $1$ & $-1$ & $1$ & $1$ & $1$ & $-1$ & $1$ & $-1$& $1$ & $-1$  \\
 \hline
 $\bold{2}^O$ & $2$ & $0$ & $-2$ & $2$ & $-1$ & $2$ & $0$ & $0$ & $1$& $-1$ & $0$  \\
 \hline
 $\bold{2}^E$ & $2$ & $0$ & $2$ & $2$ & $-1$ & $2$ & $0$ & $0$ & $-1$& $-1$ & $0$  \\
 \hline
 $\bold{2}'^O$ & $2$ & $-2$ & $0$ & $-1$ & $2$ & $2$ & $0$ & $1$ & $0$& $-1$ & $0$  \\
 \hline
 $\bold{2}'^E$ & $2$ & $2$ & $0$ & $-1$ & $2$ & $2$ & $0$ & $-1$ & $0$& $-1$ & $0$  \\
 \hline
  $\bold{4}$ & $4$ &$0$ & $0$ & $-2$ & $-2$ & $4$ & $0$ & $0$ & $0$ & $1$& $0$  \\
 \hline
   $\bold{6}$ & $6$ &$0$ & $0$ & $0$ & $0$ & $-3$ & $2$ & $0$ & $0$ & $0$& $-1$  \\
 \hline
    $\bold{6}'$ & $6$ &$0$ & $0$ & $0$ & $0$ & $-3$ & $-2$ & $0$ & $0$ & $0$& $1$  \\
 \hline
  \end{tabular}
\end{center}
\caption{Character table for $G\rtimes \mathbb{Z}_2$. The subscripts of the conjugacy classes indicate their size.}
\label{charTableZ2}
\end{table}
We can compute projectors onto a given irrep $R$ as
\es{projectorsZ2}{
    &P_{R}{}_{,IJKL}=\frac{|R|}{|G\rtimes\mathbb{Z}_2|}\sum_{a,b}\chi_R(h_a\otimes h_b)h_{a}{}_{IJ}\otimes h_{b}{}_{KL}\,,\\
}
where here $a=1,\dots,|G\rtimes \mathbb{Z}_2|$, and then compute the eigenvectors with unit eigenvalues as
\es{projectorEigZ2}{
    &P_{R}{}_{,IJKL} v_R{}_r{}_{JL}=v_{R,r}{}_{IK}\,,
}
where $r=1,\dots,|R|$.

The order 216 group $G\rtimes(\mathbb{Z}_2\times\mathbb{Z}_2)$ and the order 324 group $G\rtimes S_3$ can be described using a very similar formalism. For simplicity, we will choose the duality frames for each group with $\tau=1-\sqrt{3}$ and $\tau=0$, respectively. As with $G\rtimes \mathbb{Z}_2$, the chiral primary $X_I$ transform in a 6-dimensional irrep $\bold6^1$, where the superscript refers to the fact that several $\bold6$ irreps appear for these groups. In this representation, the elements of $ G\rtimes\mathbb{Z}_2$ and $ G\rtimes S_3$ can be written as $4\times4$ matrices as in \eqref{Z2group}, except where $g\in G\cup\{u_1u_2u_1^{-1}\}$ and $g\in G\cup\{u_2\}$, respectively, where $u_2$ and $u_1$ are defined in \eqref{dualMat}. The character tables for $ G\rtimes S_3$ and $ G\rtimes(\mathbb{Z}_2 \times \mathbb{Z}_2)$ are given in Tables \ref{charTableS3} and \ref{charTableZ2Z2}. The projectors can then be constructed as in \eqref{projectorsZ2} and \eqref{projectorEigZ2}.

\begin{table}[htp]
\begin{center}
  \begin{tabular}{c|c|c|c|c|c|c|c|c|c|c|c|c|c|c|c|c|c}
  $\chi_R(g_a)$  & $C_1$ & $C_{9}$ & $C_{27}$ & ${C}_6$ & $C_{18}$  & $C'_6$ & ${C}_{2}$ & $C'_{27}$  & $C'_{18}$ & ${C}''_{18}$ &$C_{54}$ &$C''_{6}$ &$C_{36}$ &$C'''_{6}$ &$C'_{54}$ &$C'''_{18}$ &$C''''_{18}$ \\
 \hline\hline
 $\bold{1}^E$ & $1$ & $1$ & $1$ & $1$ & $1$ & $1$ & $1$ & $1$ & $1$& $1$ &$1$ & $1$ &$1$ & $1$ & $1$ & $1$ & $1$ \\
 \hline
 $\bold{1}'$ & $1$ & $-1$ & $-1$ & $1$ & $1$ & $1$ & $1$ & $1$ & $-1$ & $-1$ & $-1$ & $1$ &$1$ & $1$ &$1$ & $-1$ &$-1$  \\
  \hline
 $\bold{1}''$ & $1$ & $-1$ & $1$ & $1$ & $1$ & $1$ & $1$ & $-1$ & $-1$ & $-1$ & $1$ & $1$ &$1$ & $1$ &$-1$ & $-1$ &$-1$  \\
  \hline
 $\bold{1}^O$ & $1$ & $1$ & $-1$ & $1$ & $1$ & $1$ & $1$ & $-1$ & $1$ & $1$ & $-1$ & $1$ &$1$ & $1$ &$-1$ & $1$ &$1$  \\
  \hline
 $\bold{2}^E$ & $2$ & $0$ & $-2$ & $2$ & $-1$ & $2$ & $2$ & $0$ & $0$& $0$ & $1$ & $2$ & $-1$ & $2$ & $0$ & $0$ & $0$ \\
 \hline
  $\bold{2}^O$ & $2$ & $0$ & $2$ & $2$ & $-1$ & $2$ & $2$ & $0$ & $0$& $0$ & $-1$ & $2$ & $-1$ & $2$ & $0$ & $0$ & $0$ \\
 \hline
  $\bold{2}''$ & $2$ & $-2$ & $0$ & $-1$ & $2$ & $2$ & $2$ & $0$ & $1$& $-2$ & $0$ & $-1$ & $-1$ & $-1$ & $0$ & $1$ & $1$ \\
 \hline
  $\bold{2}'''$ & $2$ & $2$ & $0$ & $-1$ & $2$ & $2$ & $2$ & $0$ & $-1$& $2$ & $0$ & $-1$ & $-1$ & $-1$ & $0$ & $-1$ & $-1$ \\
 \hline
  $\bold{4}$ & $4$ & $0$ & $0$ & $-2$ & $-2$ & $4$ & $4$ & $0$ & $0$& $0$ & $0$ & $-2$ & $1$ & $-2$ & $0$ & $0$ & $0$ \\
 \hline
  $\bold{6}^E$ & $6$ & $0$ & $0$ & $0$ & $0$ & $-3$ & $6$ & $-2$ & $0$& $0$ & $0$ & $0$ & $0$ & $0$ & $1$ & $0$ & $0$ \\
 \hline
  $\bold{6}^O$ & $6$ & $0$ & $0$ & $0$ & $0$ & $-3$ & $6$ & $2$ & $0$& $0$ & $0$ & $0$ & $0$ & $0$ & $-1$ & $0$ & $0$ \\
 \hline
 $\bold{6}'^1$ & $6$ & $-2$ & $0$ & $A$ & $0$ & $0$ & $-3$ & $0$ & $D$& $1$ & $0$ & $C$ & $0$ & $B$ & $0$ & $F$ & $E$ \\
 \hline
  $\bold{6}'^2$ & $6$ & $-2$ & $0$ & $C$ & $0$ & $0$ & $-3$ & $0$ & $F$& $1$ & $0$ & $B$ & $0$ & $A$ & $0$ & $E$ & $D$ \\
 \hline
   $\bold{6}'^3$ & $6$ & $-2$ & $0$ & $B$ & $0$ & $0$ & $-3$ & $0$ & $E$& $1$ & $0$ & $A$ & $0$ & $C$ & $0$ & $D$ & $F$ \\
 \hline
  $\bold{6}^1$ & $6$ & $2$ & $0$ & $A$ & $0$ & $0$ & $-3$ & $0$ & $-D$& $-1$ & $0$ & $C$ & $0$ & $B$ & $0$ & $-F$ & $-E$ \\
 \hline
  $\bold{6}^2$ & $6$ & $2$ & $0$ & $C$ & $0$ & $0$ & $-3$ & $0$ & $-F$& $-1$ & $0$ & $B$ & $0$ & $A$ & $0$ & $-E$ & $-D$ \\
 \hline
   $\bold{6}^3$ & $6$ & $2$ & $0$ & $B$ & $0$ & $0$ & $-3$ & $0$ & $-E$& $-1$ & $0$ & $A$ & $0$ & $C$ & $0$ & $-D$ & $-F$ \\
 \hline
  \end{tabular}
\end{center}
\caption{Character table for $G\rtimes S_3$. The subscripts of the conjugacy classes indicate their size, and $A=-\omega^{
\frac23}-2\omega^{
\frac43}-2\omega^{
\frac53}-\omega^{
\frac73}$, $B=-\omega^{
\frac23}+\omega^{
\frac43}+\omega^{
\frac53}-\omega^{
\frac73}$, $C=2\omega^{
\frac23}+\omega^{
\frac43}+\omega^{
\frac53}+2\omega^{
\frac73}$, $D=-\omega^{
\frac23}-\omega^{
\frac73}$, $E=\omega^{
\frac23}+\omega^{
\frac43}+\omega^{
\frac53}+\omega^{
\frac73}$, $F=-\omega^{
\frac43}-\omega^{
\frac53}$, where $\omega=e^{2\pi i/3}$.}
\label{charTableS3}
\end{table}

\begin{table}[htp]
\begin{center}
  \begin{tabular}{c|c|c|c|c|c|c|c|c|c|c|c|c|c}
  $\chi_R(g_a)$  & $C_1$ & $C_{18}$ & $C'_{18}$ & ${C}_9$ & $C_{12}$  & $C_2$ & ${C}''_{18}$ & $C_{36}$  & $C'_{36}$ & ${C}'''_{18}$ &$C'_{12}$ &$C''''_{18}$&$C'''''_{18}$ \\
 \hline\hline
 $\bold{1}^E$ & $1$ & $1$ & $1$ & $1$ & $1$ & $1$ & $1$ & $1$ & $1$& $1$ &$1$ & $1$ &$1$ \\
 \hline
 $\bold{1}^O$ & $1$ & $-1$ & $-1$ & $1$ & $1$ & $1$ & $1$ & $-1$ & $-1$& $1$ &$1$ & $1$ &$1$  \\
 \hline
 $\bold{1}'$ & $1$ & $-1$ & $1$ & $1$ & $1$ & $1$ & $-1$ & $-1$ & $1$& $1$ & $1$& $-1$ & $-1$  \\
 \hline
 $\bold{1}''$ & $1$ & $1$ & $-1$ & $1$ & $1$ & $1$ & $-1$ & $1$ & $-1$& $1$ & $1$ & $-1$ & $-1$  \\
 \hline
 $\bold{2}$ & $2$ & $0$ & $0$ & $-2$ & $2$ & $2$ & $0$ & $0$ & $0$& $-2$ & $2$ & $0$ & $0$  \\
 \hline
 $\bold{4}^O$ & $4$ & $-2$ & $0$ & $0$ & $1$ & $4$ & $0$ & $1$ & $0$ & $0$& $-2$ & $0$& $0$  \\
 \hline
   $\bold{4}^E$ & $4$ & $2$ & $0$ & $0$ & $1$ & $4$ & $0$ & $-1$ & $0$ & $0$& $-2$ & $0$& $0$  \\
 \hline
  $\bold{4}'^O$ & $4$ & $0$ & $-2$ & $0$ & $-2$ & $4$ & $0$ & $0$ & $1$ & $0$& $1$ & $0$& $0$  \\
 \hline
  $\bold{4}'^E$ & $4$ & $0$ & $2$ & $0$ & $-2$ & $4$ & $0$ & $0$ & $-1$ & $0$& $1$ & $0$& $0$  \\
 \hline
   $\bold{6}_1$ & $6$ &$0$ & $0$ & $2$ & $0$ & $-3$ & $0$ & $0$ & $0$ & $-1$& $0$ & $\sqrt{3}$& $-\sqrt{3}$ \\
    \hline
    $\bold{6}_2$ & $6$ &$0$ & $0$ & $2$ & $0$ & $-3$ & $0$ & $0$ & $0$ & $-1$& $0$ & $-\sqrt{3}$& $\sqrt{3}$ \\
 \hline
   $\bold{6}'_1$ & $6$ &$0$ & $0$ & $-2$ & $0$ & $-3$ & $-2$ & $0$ & $0$ & $1$& $0$ & $1$& $1$  \\
 \hline
    $\bold{6}'_2$ & $6$ &$0$ & $0$ & $-2$ & $0$ & $-3$ & $2$ & $0$ & $0$ & $1$& $0$ & $-1$& $-1$ \\
    \hline
  \end{tabular}
\end{center}
\caption{Character table for $G\rtimes \left(\mathbb{Z}_2\times\mathbb{Z}_2\right)$. The subscripts of the conjugacy classes indicate their size.}
\label{charTableZ2Z2}
\end{table}

\section{Perturbative calculations}
\label{sec:G1comp}

In this appendix we present details for the calculation of the chiral two-point functions defined in Subsection \ref{sec:chiralring}. We perform the computation in $d=4-\varepsilon$ in the minimal subtraction scheme. The Feynman rules can be easily derived from the Lagrangian
\begin{equation}\label{Toines}
\mathcal{L}=-\partial^\mu \overline{X}^i \partial_\mu X_i-\frac{1}{2}\bar{\chi}_i \gamma^\mu \partial_\mu \chi_i - W^i \overline{W}_i -\frac{1}{2}\overline{\chi}_i (P_L W_{ij}+P_R \overline{W}_{ij})\chi_j,
\end{equation}
where the $\chi_i$'s are the (four component) Majorana spinors in the supermultiplet of the fundamental chiral superfield $X_i$,\footnote{For notational simplicity, we denote both the chiral superfield and its bottom component as $X_i$.} the left/right chiral projectors are given by $P_{L/R} \equiv \frac{1}{2}(1\pm\gamma_5)$, and
\begin{align}
W^{i} &\equiv \frac{\partial W}{\partial X_i}~, &  W^{ij} & \equiv \frac{\partial^2 W}{\partial X_i \partial X_j}~.
\end{align} 
We then have the usual cubic Yukawa couplings proportional to $h^{ijk}$ and a quartic scalar vertex proportional to $h^{ijp}\bar{h}_{pk\ell}$.
\subsection*{Computation of $G_1({\tau,\bar{\tau}})$}
We begin by discussing the two point functions of the fundamental chiral fields of the model $X_i$. The computation is standard and can be found for example in \cite{Abbott:1980jk}, so we only sketch it here. The Feynman diagrams are given by\footnote{We used JaxoDraw \cite{Binosi:2003yf,Binosi:2008ig} to draw the Feynman diagrams in this paper. Solid lines denote scalars and dashed lines denote fermions.}
\begin{equation}
\vcenter{\hbox{\includegraphics[width=80px]{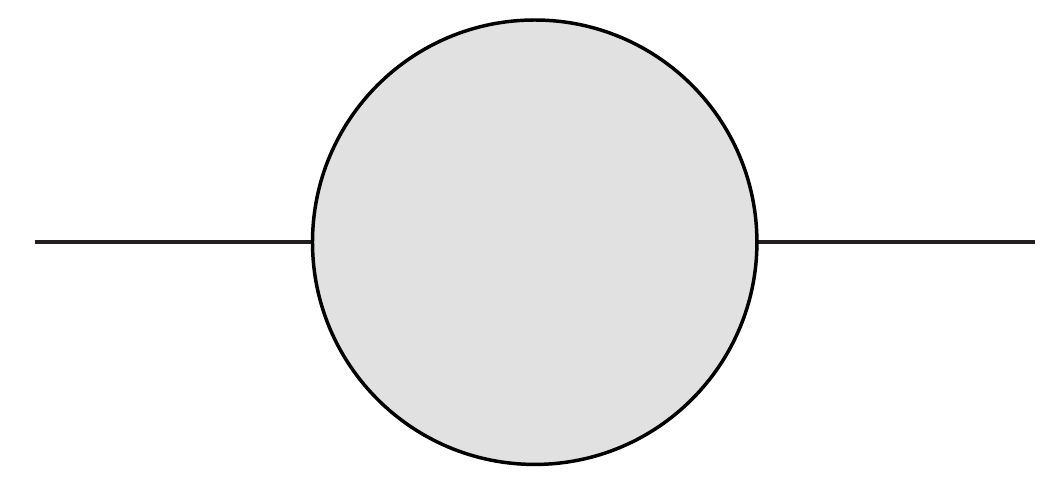}}} = \vcenter{\hbox{\includegraphics[width=80px]{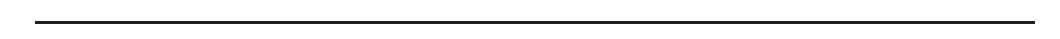}}}+\vcenter{\hbox{\includegraphics[width=80px]{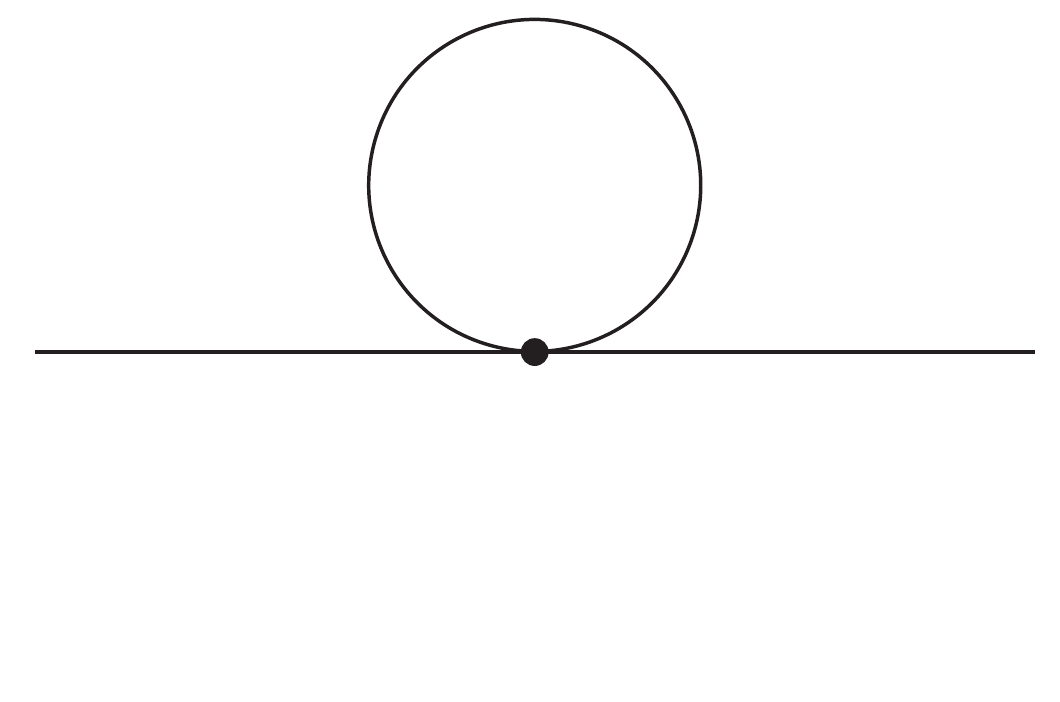}}}+\vcenter{\hbox{\includegraphics[width=80px]{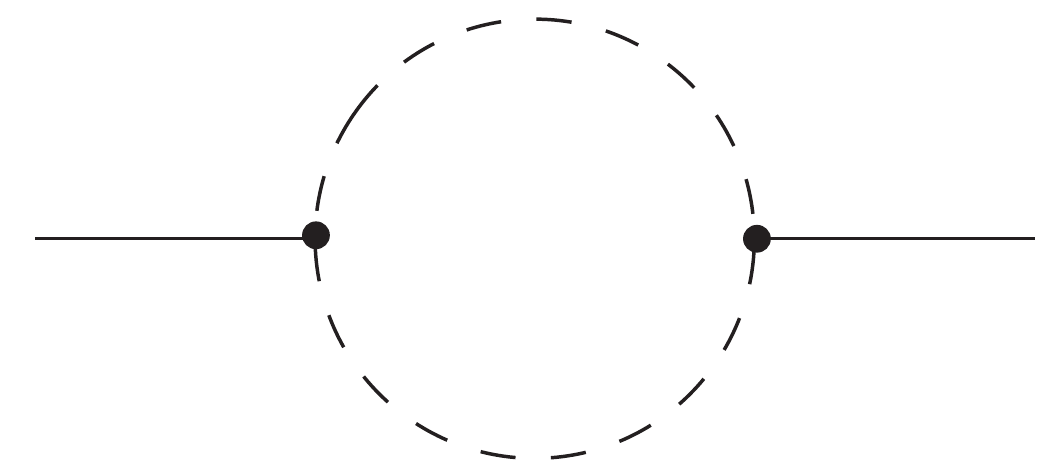}}}+\ldots~,
\end{equation}
where we are only showing diagrams to one loop for simplicity. After computing all the diagrams to two loops, using the appropriate counterterms to remove the divergences, and replacing the coupling constant with its value at the fixed point, we obtain
\begin{equation}
\label{eq:scalar2pt}
\langle X_i(x) \overline{X}^j(0) \rangle =  G_1(\tau,\bar{\tau})\,\frac{{\delta_i}^j}{|x|^{2-2\varepsilon/3}} \mu^{-\varepsilon/3}~,
\end{equation}
where $G_1(\tau,\bar{\tau})$ is given by
\begin{equation}
\label{eq:G1eps}
G_1(\tau,\bar{\tau}) =  \frac{1}{4\pi^{2-\varepsilon/3}\Gamma(1+\varepsilon/3)} \left(1 - \frac{1}{3}\varepsilon + \frac{1}{6^3}(5 \pi^2 - 6)\varepsilon^2+\ldots\right)~.
\end{equation}
We notice that \eqref{eq:scalar2pt} exhibits the correct behavior for a scalar field of dimension $\Delta = 1 -\varepsilon/3$. The presence of an explicit factor of $\mu$ in \eqref{eq:scalar2pt} reflects the scheme dependence of $G_1$. However, as explained in Section~\ref{sec:chiralring}, $G_1$ only appears in scheme-independent combinations in our final results.

\subsection*{Computation of $G_2(\tau,\bar{\tau})$}
The expansion in Feynman diagrams for the quadratic chiral operators has the following form:
\begin{align}
\label{eq:feynquadratic}
\langle X_i X_j(x) \overline{X}^k \overline{X}^\ell(0)\rangle & = \vcenter{\hbox{\includegraphics[width=80px]{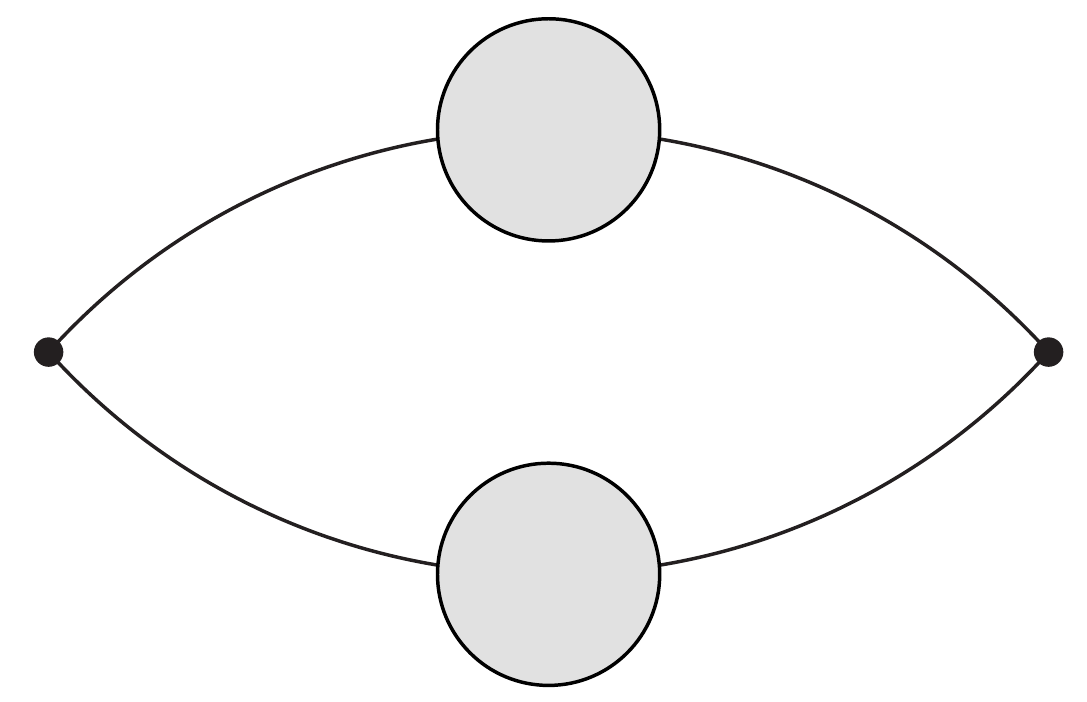}}}+\vcenter{\hbox{\includegraphics[width=80px]{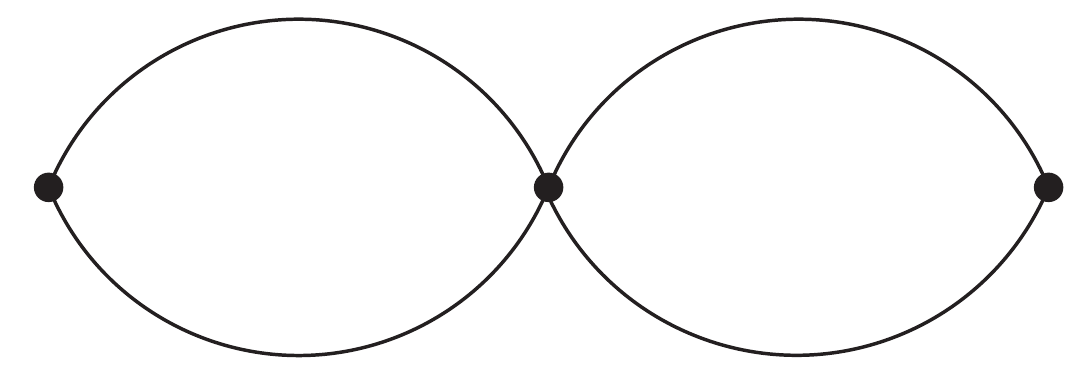}}}\\
& +\vcenter{\hbox{\includegraphics[width=80px]{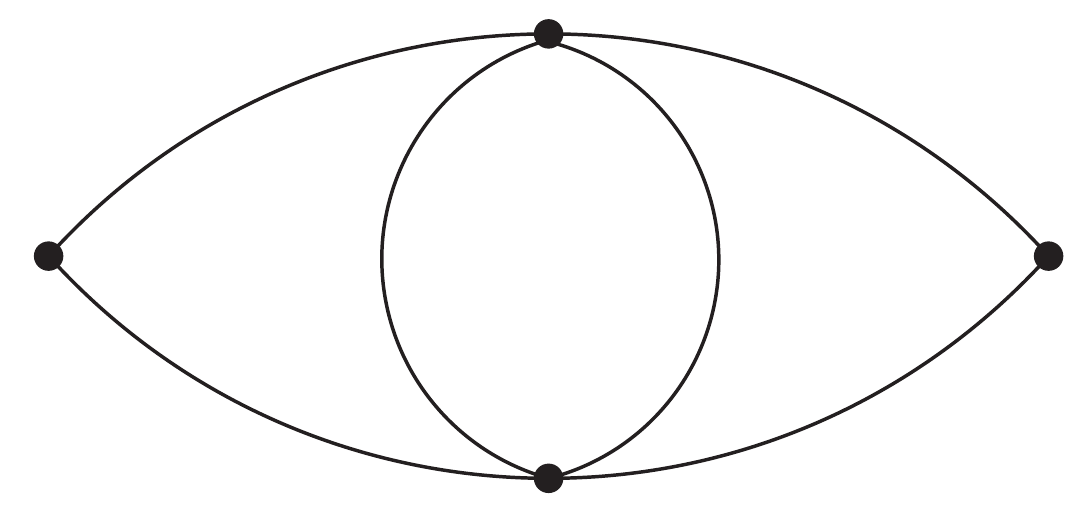}}}+\vcenter{\hbox{\includegraphics[width=80px]{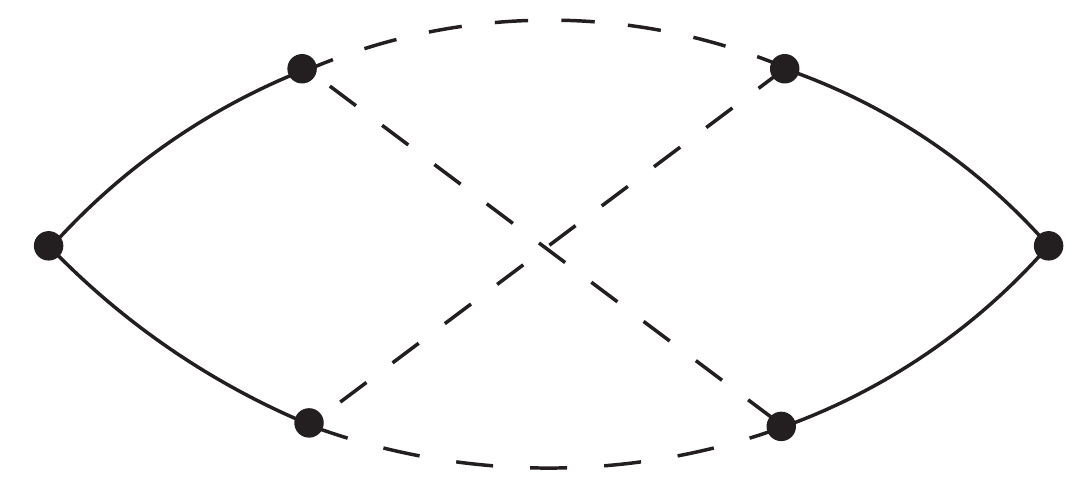}}}+\vcenter{\hbox{\includegraphics[width=120px]{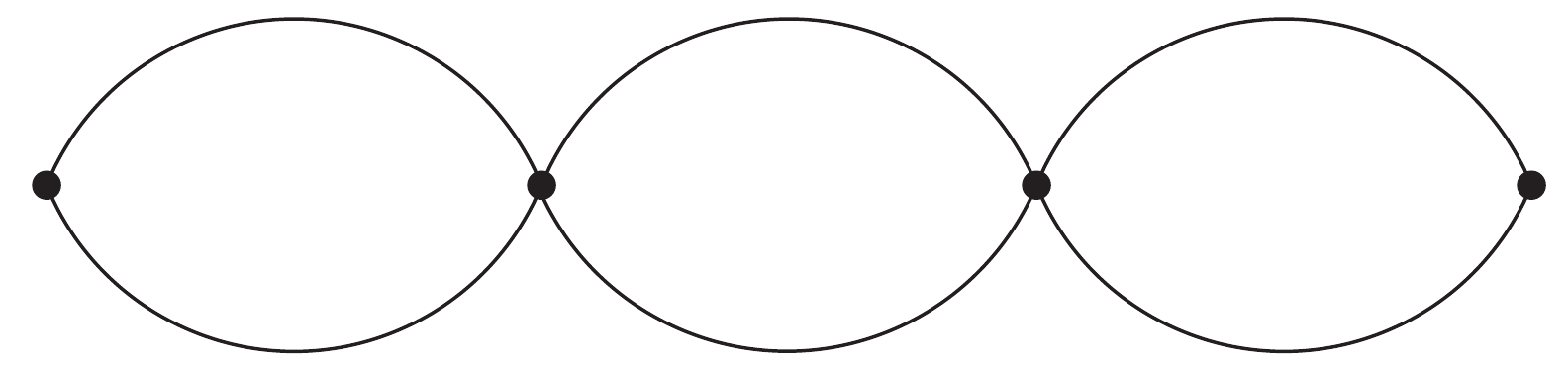}}}+\ldots\nonumber~,
\end{align}
where the ellipses denote higher order diagrams and we omitted all the combinatorial factors that multiply the various diagrams. The first diagram on the first line is just the sum of all the diagrams that do not connect the upper with the lower propagator, so in position space they just give $ G_1(\tau,\bar{\tau})^2\,({\delta_i}^k{\delta_j}^\ell+{\delta_i}^\ell{\delta_j}^k)\,|x|^{-2\Delta_X}$. Furthermore, the rightmost diagrams on the first and second lines both vanish for the chiral operators $\mathcal{Q}_I = {P_I}^{ij}\,X_i X_j$ defined in \eqref{eq:chirfirst}, since both diagrams are proportional to
\begin{equation}
{P_I}^{ij}\, \bar{h}_{ijp}\, h^{p m n} = 0~.
\end{equation}
Indeed, it is precisely these diagrams that give the descendant combinations $\mathcal{W}_I$ defined in \eqref{eq:descendants} the conformal dimension $\Delta_{\mathcal{W}} = \Delta_{X} + 1$ implied by \eqref{eq:descendantrelation}.
Lastly, the sum of the remaining two diagrams (with the appropriate combinatorial factors in front) is finite, and has been explicitly calculated in \cite{Baggio:2014ioa}.
Putting the various ingredients together, we obtain the results in \eqref{GRatio}.
\subsection*{Computation of $G_3(\tau,\bar{\tau})$}
The computation of the two-point function of the cubic chiral primary operator proceeds exactly in the same way as before, since there are no additional Feynman diagrams at this order---the contributing diagrams are identical to those in \eqref{eq:feynquadratic}, with one additional propagator connecting the left and right vertices, reflecting that the composite operator is now cubic in the fields. As a consequence, the answer can be immediately derived after some simple combinatorics, and leads to the result presented in \eqref{GRatio}.

\section{Explicit crossing equations}
\label{bootApp}

In this appendix we list the explicit expression used in the crossing equations for the four point function of four chiral operators in $\mathcal{N}=2$ theories with flavor symmetry $G=(\mathbb{Z}_3\times\mathbb{Z}_3)\rtimes S_3$, as well as $G\rtimes \mathbb{Z}_2$, $G\rtimes (\mathbb{Z}_2\times\mathbb{Z}_2)$, and $G\rtimes S_3$.

In addition to the superconformal block $\cG_{\Delta,\ell}$ defined in \eqref{SuperConf}, it will be useful to define $\tilde{\mathcal{G}}_{\Delta, \ell}$ by taking the expression for $\cG_{\Delta,\ell}$ and replacing $G_{\Delta',\ell'} \to(-1)^{\ell'}G_{\Delta',\ell'}$.

For $G$, we have
\es{Vs}{
&\vec V_{\bold1,\Delta,\ell}=
\begin{pmatrix}
\mathcal{G}_{+,\Delta,\ell}\\
0\\
0\\
0\\
\mathcal{G}_{-,\Delta,\ell}\\
\tilde{\mathcal{G}}_{+,\Delta,\ell}\\
\frac{1 }{2}\tilde{\mathcal{G}}_{+,\Delta,\ell}\\
-\frac{1 }{2}\tilde{\mathcal{G}}_{+,\Delta,\ell}\\
0\\
0\\
-\tilde{\mathcal{G}}_{-,\Delta,\ell}\\
-\frac{1 }{2}\tilde{\mathcal{G}}_{-,\Delta,\ell}\\
\frac{1 }{2}\tilde{\mathcal{G}}_{-,\Delta,\ell}\\
0\\
0\\
\end{pmatrix}\,,\qquad
\vec V_{\bold2,\Delta,\ell}=
\begin{pmatrix}
-\mathcal{G}_{+,\Delta,\ell}\\
0\\
0\\
\mathcal{G}_{-,\Delta,\ell}\\
0\\
0\\
\frac{1 }{2}\tilde{\mathcal{G}}_{+,\Delta,\ell}\\
\frac{1 }{2}\tilde{\mathcal{G}}_{+,\Delta,\ell}\\
-\tilde{\mathcal{G}}_{+,\Delta,\ell}\\
-\tilde{\mathcal{G}}_{+,\Delta,\ell}\\
0\\
-\frac{1 }{2}\tilde{\mathcal{G}}_{-,\Delta,\ell}\\
-\frac{1 }{2}\tilde{\mathcal{G}}_{-,\Delta,\ell}\\
\tilde{\mathcal{G}}_{-,\Delta,\ell}\\
\tilde{\mathcal{G}}_{-,\Delta,\ell}\\
\end{pmatrix}\,,\qquad
\vec V_{\bold2',\Delta,\ell}=
\begin{pmatrix}
-\mathcal{G}_{+,\Delta,\ell}\\
0\\
\mathcal{G}_{-,\Delta,\ell}\\
0\\
0\\
0\\
\frac{1 }{2}\tilde{\mathcal{G}}_{+,\Delta,\ell}\\
\frac{1 }{2}\tilde{\mathcal{G}}_{+,\Delta,\ell}\\
2\tilde{\mathcal{G}}_{+,\Delta,\ell}\\
0\\
0\\
-\frac{1 }{2}\tilde{\mathcal{G}}_{-,\Delta,\ell}\\
-\frac{1 }{2}\tilde{\mathcal{G}}_{-,\Delta,\ell}\\
-2\tilde{\mathcal{G}}_{-,\Delta,\ell}\\
0\\
\end{pmatrix}\,,
}
\es{Vs2}{
&\vec V_{\bold2'',\Delta,\ell}=
\begin{pmatrix}
-\frac12\mathcal{G}_{+,\Delta,\ell}\\
\mathcal{G}_{-,\Delta,\ell}\\
0\\
0\\
0\\
\tilde{\mathcal{G}}_{+,\Delta,\ell}\\
-\frac{1 }{4}\tilde{\mathcal{G}}_{+,\Delta,\ell}\\
\frac{1 }{4}\tilde{\mathcal{G}}_{+,\Delta,\ell}\\
0\\
0\\
-\tilde{\mathcal{G}}_{-,\Delta,\ell}\\
\frac{1 }{4}\tilde{\mathcal{G}}_{-,\Delta,\ell}\\
-\frac{1 }{4}\tilde{\mathcal{G}}_{-,\Delta,\ell}\\
0\\
0\\
\end{pmatrix}\,,\qquad
\vec V_{\bold2''',\Delta,\ell}=
\begin{pmatrix}
-\mathcal{G}_{+,\Delta,\ell}\\
-2\mathcal{G}_{-,\Delta,\ell}\\
-\mathcal{G}_{-,\Delta,\ell}\\
-\mathcal{G}_{-,\Delta,\ell}\\
2\mathcal{G}_{-,\Delta,\ell}\\
0\\
\frac{1 }{2}\tilde{\mathcal{G}}_{+,\Delta,\ell}\\
\frac{1 }{2}\tilde{\mathcal{G}}_{+,\Delta,\ell}\\
-\tilde{\mathcal{G}}_{+,\Delta,\ell}\\
\tilde{\mathcal{G}}_{+,\Delta,\ell}\\
0\\
-\frac{1 }{2}\tilde{\mathcal{G}}_{-,\Delta,\ell}\\
-\frac{1 }{2}\tilde{\mathcal{G}}_{-,\Delta,\ell}\\
\tilde{\mathcal{G}}_{-,\Delta,\ell}\\
-\tilde{\mathcal{G}}_{-,\Delta,\ell}\\
\end{pmatrix}\,,\qquad
\vec V_{\bold3',\Delta,\ell}=
\begin{pmatrix}
0\\
0\\
0\\
0\\
0\\
0\\
0\\
F_{+,\Delta,\ell}\\
0\\
0\\
0\\
0\\
F_{-,\Delta,\ell}\\
0\\
0\\
\end{pmatrix}\,,
}
\es{Vs30}{
&\vec { V}_{\bar{\bold3},2\frac{d-1}{3},0}=
\begin{pmatrix}
0\\
0\\
0\\
0\\
0\\
-F_{+,\Delta,\ell}\\
-\frac14|\tau|^2F_{+,\Delta,\ell}\\
0\\
\Re\tau F_{+,\Delta,\ell}\\
\Im\tau F_{+,\Delta,\ell}\\
-F_{-,\Delta,\ell}\\
-\frac14|\tau|^2F_{-,\Delta,\ell}\\
0\\
\Re\tau F_{-,\Delta,\ell}\\
\Im\tau F_{-,\Delta,\ell}\\
\end{pmatrix}\,,\qquad
\vec {\bold V}_{\bar{\bold3},\Delta,\ell}=
\begin{pmatrix}
0_4\\
0_4\\
0_4\\
0_4\\
0_4\\
-F^{11}_{+,\Delta,\ell}\\
-F^{22}_{+,\Delta,\ell}\\
0_4\\
-F^{12}_{+,\Delta,\ell}\\
-F^{21}_{+,\Delta,\ell}\\
-F^{11}_{-,\Delta,\ell}\\
-F^{22}_{-,\Delta,\ell}\\
0_4\\
-F^{12}_{-,\Delta,\ell}\\
-F^{21}_{-,\Delta,\ell}\\
\end{pmatrix}\,,\\
}
\es{Fdefs2}{
&F^{11}_{\pm,\Delta,\ell}=
\begin{pmatrix}
F_{\pm,\Delta,\ell}&0&0&0\\
0&F_{\pm,\Delta,\ell}&0&0\\
0&0&0&0\\
0&0&0&0\\
\end{pmatrix}\,,\quad
F^{12}_{\pm,\Delta,\ell}=
\begin{pmatrix}
0&0&F_{\pm,\Delta,\ell}&0\\
0&0&0&F_{\pm,\Delta,\ell}\\
F_{\pm,\Delta,\ell}&0&0&0\\
0&F_{\pm,\Delta,\ell}&0&0\\
\end{pmatrix}\,,\\
&F^{22}_{\pm,\Delta,\ell}=
\begin{pmatrix}
0&0&0&0\\
0&0&0&0\\
0&0&F_{\pm,\Delta,\ell}&0\\
0&0&0&F_{\pm,\Delta,\ell}\\
\end{pmatrix}\,,
\quad
F^{21}_{\pm,\Delta,\ell}=\frac{1}{\sqrt{3}}
\begin{pmatrix}
0&0&0&F_{\pm,\Delta,\ell}\\
0&0&-F_{\pm,\Delta,\ell}&0\\
0&-F_{\pm,\Delta,\ell}&0&0\\
F_{\pm,\Delta,\ell}&0&0&0\\
\end{pmatrix}\,,\\
}

where we define
 \es{Fdefine}{
&F_{\mp,\Delta,\ell}(u,v)=v^{\frac{d-1}{3}}G_{\Delta,\ell}(u,v)\mp u^{\frac{d-1}{3}}G_{\Delta,\ell}(v,u)\,,\\
&\mathcal{F}_{\mp,\Delta,\ell}(u,v)=v^{\frac{d-1}{3}}\mathcal{G}_{\Delta,\ell}(u,v)\mp u^{\frac{d-1}{3}}\mathcal{G}_{\Delta,\ell}(v,u)\,,\\
&\tilde{\mathcal{F}}_{\mp,\Delta,\ell}(u,v)=v^{\frac{d-1}{3}}\tilde{\mathcal{G}}_{\Delta,\ell}(u,v)\mp u^{\frac{d-1}{3}}\tilde{\mathcal{G}}_{\Delta,\ell}(v,u)\,.\\
}
For $G\rtimes \mathbb{Z}_2$, we have
\es{Vss}{
&\vec V_{\bold1^{E,O},\Delta,\ell}=
\begin{pmatrix}
\mathcal{G}_{+,\Delta,\ell}\\
0\\
0\\
\mathcal{G}_{-,\Delta,\ell}\\
\tilde{\mathcal{G}}_{+,\Delta,\ell}\\
\frac{1 }{2}\tilde{\mathcal{G}}_{+,\Delta,\ell}\\
-\frac{1 }{2}\tilde{\mathcal{G}}_{+,\Delta,\ell}\\
0\\
-\tilde{\mathcal{G}}_{-,\Delta,\ell}\\
-\frac{1 }{2}\tilde{\mathcal{G}}_{-,\Delta,\ell}\\
\frac{1 }{2}\tilde{\mathcal{G}}_{-,\Delta,\ell}\\
0\\
\end{pmatrix}\,,\qquad
\vec V_{\bold4,\Delta,\ell}=
\begin{pmatrix}
-2\mathcal{G}_{+,\Delta,\ell}\\
-2\mathcal{G}_{-,\Delta,\ell}\\
-\mathcal{G}_{-,\Delta,\ell}\\
2\mathcal{G}_{-,\Delta,\ell}\\
0\\
\tilde{\mathcal{G}}_{+,\Delta,\ell}\\
\tilde{\mathcal{G}}_{+,\Delta,\ell}\\
-2\tilde{\mathcal{G}}_{+,\Delta,\ell}\\
0\\
-\tilde{\mathcal{G}}_{-,\Delta,\ell}\\
-\tilde{\mathcal{G}}_{-,\Delta,\ell}\\
2\tilde{\mathcal{G}}_{-,\Delta,\ell}\\
\end{pmatrix}\,,\qquad
\vec V_{\bold2^{E,O},\Delta,\ell}=
\begin{pmatrix}
-\mathcal{G}_{+,\Delta,\ell}\\
0\\
\mathcal{G}_{-,\Delta,\ell}\\
0\\
0\\
\frac{1 }{2}\tilde{\mathcal{G}}_{+,\Delta,\ell}\\
\frac{1 }{2}\tilde{\mathcal{G}}_{+,\Delta,\ell}\\
2\tilde{\mathcal{G}}_{+,\Delta,\ell}\\
0\\
-\frac{1 }{2}\tilde{\mathcal{G}}_{-,\Delta,\ell}\\
-\frac{1 }{2}\tilde{\mathcal{G}}_{-,\Delta,\ell}\\
-2\tilde{\mathcal{G}}_{-,\Delta,\ell}\\
\end{pmatrix}\,,
}
\es{Vss2}{
&\vec V_{\bold2'^{E,O},\Delta,\ell}=
\begin{pmatrix}
-\frac12\mathcal{G}_{+,\Delta,\ell}\\
\mathcal{G}_{-,\Delta,\ell}\\
0\\
0\\
\tilde{\mathcal{G}}_{+,\Delta,\ell}\\
-\frac{1 }{4}\tilde{\mathcal{G}}_{+,\Delta,\ell}\\
\frac{1 }{4}\tilde{\mathcal{G}}_{+,\Delta,\ell}\\
0\\
-\tilde{\mathcal{G}}_{-,\Delta,\ell}\\
\frac{1 }{4}\tilde{\mathcal{G}}_{-,\Delta,\ell}\\
-\frac{1 }{4}\tilde{\mathcal{G}}_{-,\Delta,\ell}\\
0\\
\end{pmatrix}\,,\qquad
\vec V_{\bold6',\Delta,\ell}=
\begin{pmatrix}
0\\
0\\
0\\
0\\
0\\
0\\
F_{+,\Delta,\ell}\\
0\\
0\\
0\\
F_{-,\Delta,\ell}\\
0\\
\end{pmatrix}\,,
}
\es{Vss30}{
&\vec { V}_{{\bold6},2\frac{d-1}{3},0}=
\begin{pmatrix}
0\\
0\\
0\\
0\\
-F_{+,\Delta,\ell}\\
-\frac14\tau^2F_{+,\Delta,\ell}\\
0\\
\tau F_{+,\Delta,\ell}\\
-F_{-,\Delta,\ell}\\
-\frac14\tau^2F_{-,\Delta,\ell}\\
0\\
\tau F_{-,\Delta,\ell}\\
\end{pmatrix}\,,\qquad
\vec {\bold V}_{{\bold6},\Delta,\ell}=
\begin{pmatrix}
0_2\\
0_2\\
0_2\\
0_2\\
-F^{11}_{+,\Delta,\ell}\\
-F^{22}_{+,\Delta,\ell}\\
0_2\\
-F^{12}_{+,\Delta,\ell}\\
-F^{11}_{-,\Delta,\ell}\\
-F^{22}_{-,\Delta,\ell}\\
0_2\\
-F^{12}_{-,\Delta,\ell}\\
\end{pmatrix}\,,\\
}
where
\es{Fdefs}{
&F^{11}_{\pm,\Delta,\ell}=
\begin{pmatrix}
F_{\pm,\Delta,\ell}&0\\
0&0\\
\end{pmatrix}\,,\quad
F^{12}_{\pm,\Delta,\ell}=
\begin{pmatrix}
0&F_{\pm,\Delta,\ell}\\
F_{\pm,\Delta,\ell}&0\\
\end{pmatrix}\,,\quad F^{22}_{\pm,\Delta,\ell}=
\begin{pmatrix}
0&0\\
0&F_{\pm,\Delta,\ell}\\
\end{pmatrix}\,.
}
For $G\rtimes(\mathbb{Z}_2\times\mathbb{Z}_2)$ we have
\es{Vsss}{
&\vec V_{\bold1^{E,O},\Delta,\ell}=
\begin{pmatrix}
\mathcal{G}_{+,\Delta,\ell}\\
0\\
\mathcal{G}_{-,\Delta,\ell}\\
-\frac{3+\sqrt{3}}{24}\tilde{\mathcal{G}}_{+,\Delta,\ell}\\
\frac{-3+\sqrt{3}}{24}\tilde{\mathcal{G}}_{+,\Delta,\ell}\\
\frac{1 }{2}\tilde{\mathcal{G}}_{+,\Delta,\ell}\\
\frac{3+\sqrt{3}}{24}\tilde{\mathcal{G}}_{-,\Delta,\ell}\\
\frac{3-\sqrt{3}}{24}\tilde{\mathcal{G}}_{-,\Delta,\ell}\\
-\frac{1 }{2}\tilde{\mathcal{G}}_{-,\Delta,\ell}\\
\end{pmatrix}\,,\qquad
\vec V_{\bold4^{E,O},\Delta,\ell}=
\begin{pmatrix}
-2\mathcal{G}_{+,\Delta,\ell}\\
-\mathcal{G}_{-,\Delta,\ell}\\
2\mathcal{G}_{-,\Delta,\ell}\\
-\frac{3+2\sqrt{3} }{12}\tilde{\mathcal{G}}_{+,\Delta,\ell}\\
\frac{-3+2\sqrt{3} }{12}\tilde{\mathcal{G}}_{+,\Delta,\ell}\\
-\tilde{\mathcal{G}}_{+,\Delta,\ell}\\
\frac{3+2\sqrt{3} }{12}\tilde{\mathcal{G}}_{-,\Delta,\ell}\\
\frac{3-2\sqrt{3} }{12}\tilde{\mathcal{G}}_{-,\Delta,\ell}\\
\tilde{\mathcal{G}}_{-,\Delta,\ell}\\
\end{pmatrix}\,,\qquad
\vec V_{\bold4'^{E,O},\Delta,\ell}=
\begin{pmatrix}
-2\mathcal{G}_{+,\Delta,\ell}\\
\mathcal{G}_{-,\Delta,\ell}\\
0\\
\frac{\sqrt{3} }{12}\tilde{\mathcal{G}}_{+,\Delta,\ell}\\
-\frac{\sqrt{3} }{12}\tilde{\mathcal{G}}_{+,\Delta,\ell}\\
-\tilde{\mathcal{G}}_{+,\Delta,\ell}\\
-\frac{\sqrt{3} }{12}\tilde{\mathcal{G}}_{-,\Delta,\ell}\\
\frac{\sqrt{3} }{12}\tilde{\mathcal{G}}_{-,\Delta,\ell}\\
\tilde{\mathcal{G}}_{-,\Delta,\ell}\\
\end{pmatrix}\,,
}
\es{Vsss2}{
\vec V_{\bold6'^1,\Delta,\ell}&=
\begin{pmatrix}
0\\
0\\
0\\
0\\
0\\
-\frac29(9+5\sqrt{3})F_{+,\Delta,\ell}\\
0\\
0\\
-\frac29(9+5\sqrt{3})F_{-,\Delta,\ell}\\
\end{pmatrix}\,,\qquad 
\vec V_{\bold6^1,\Delta,\ell}=
\begin{pmatrix}
0\\
0\\
0\\
0\\
\frac29(9+5\sqrt{3})F_{+,\Delta,\ell}\\
0\\
0\\
\frac29(9+5\sqrt{3})F_{-,\Delta,\ell}\\
0\\
\end{pmatrix}
\,,\\
\vec V_{\bold6^2,\Delta,\ell}&=
\begin{pmatrix}
0\\
0\\
0\\
\frac29(9+5\sqrt{3})F_{+,\Delta,\ell}\\
0\\
0\\
\frac29(9+5\sqrt{3})F_{-,\Delta,\ell}\\
0\\
0\\
\end{pmatrix}\,.
}
For $G\rtimes S_3$ we have
\es{Vssss}{
&\vec V_{\bold1^{E,O},\Delta,\ell}=
\begin{pmatrix}
\mathcal{G}_{+,\Delta,\ell}\\
0\\
\mathcal{G}_{-,\Delta,\ell}\\
\tilde{\mathcal{G}}_{+,\Delta,\ell}\\
\frac12\tilde{\mathcal{G}}_{+,\Delta,\ell}\\
-\frac{1 }{2}\tilde{\mathcal{G}}_{+,\Delta,\ell}\\
-\tilde{\mathcal{G}}_{-,\Delta,\ell}\\
-\frac12\tilde{\mathcal{G}}_{-,\Delta,\ell}\\
\frac{1 }{2}\tilde{\mathcal{G}}_{-,\Delta,\ell}\\
\end{pmatrix}\,,\qquad
\vec V_{\bold6^{E,O},\Delta,\ell}=
\begin{pmatrix}
-3\mathcal{G}_{+,\Delta,\ell}\\
-2\mathcal{G}_{-,\Delta,\ell}\\
2\mathcal{G}_{-,\Delta,\ell}\\
0\\
\frac32\tilde{\mathcal{G}}_{+,\Delta,\ell}\\
\frac32\tilde{\mathcal{G}}_{+,\Delta,\ell}\\
0\\
-\frac32\tilde{\mathcal{G}}_{-,\Delta,\ell}\\
-\frac32\tilde{\mathcal{G}}_{-,\Delta,\ell}\\
\end{pmatrix}\,,\qquad
\vec V_{\bold2^{E,O},\Delta,\ell}=
\begin{pmatrix}
-\frac12\mathcal{G}_{+,\Delta,\ell}\\
\mathcal{G}_{-,\Delta,\ell}\\
0\\
\tilde{\mathcal{G}}_{+,\Delta,\ell}\\
-\frac14\tilde{\mathcal{G}}_{+,\Delta,\ell}\\
\frac14\tilde{\mathcal{G}}_{+,\Delta,\ell}\\
-\tilde{\mathcal{G}}_{-,\Delta,\ell}\\
\frac14\tilde{\mathcal{G}}_{-,\Delta,\ell}\\
-\frac14\tilde{\mathcal{G}}_{-,\Delta,\ell}\\
\end{pmatrix}\,,
}
\es{Vssss2}{
\vec V_{\bold6'^1,\Delta,\ell}=
\begin{pmatrix}
0\\
0\\
0\\
0\\
0\\
F_{+,\Delta,\ell}\\
0\\
0\\
F_{-,\Delta,\ell}\\
\end{pmatrix}\,,\qquad 
\vec V_{\bold6^1,\Delta,\ell}=
\begin{pmatrix}
0\\
0\\
0\\
0\\
-F_{+,\Delta,\ell}\\
0\\
0\\
-F_{-,\Delta,\ell}\\
0\\
\end{pmatrix}
\,,\qquad
\vec V_{\bold6^2,\Delta,\ell}=
\begin{pmatrix}
0\\
0\\
0\\
-F_{+,\Delta,\ell}\\
0\\
0\\
-F_{-,\Delta,\ell}\\
0\\
0\\
\end{pmatrix}\,.
}

\section{Exact results in $d=2$}
\label{2d}

In $d=2$, the family of Landau-Ginzburg CFTs with superpotential \eqref{superintro} is equivalent to a $T^2/\mathbb{Z}_3$ free orbifold CFT \cite{Greene:1988ut,Martinec:1989in}, and so can be solved exactly \cite{Dixon:1986qv}. In this appendix we summarize the results for the same CFT data that we have studied in $d=3$, for more details see \cite{Lin:2016gcl,Lerche:1989cs}.

The action for the orbifold theory is
\es{orb}{
S=\int d^2z\left( G_{\mu\nu}+B_{\mu\nu} \right)\partial \phi^{\mu}\bar\partial\phi^{\nu}+\text{fermions}\,,
}
where $z,\bar z$ are holomorphic spacetime coordinates and $\phi^{\mu}(z,\bar z)$ with $\mu,\nu=1,2$ are the target space coordinates on a torus $T^2$ with sides $2\pi R$, angle $2\pi/3$, metric
\es{torusMetric}{
ds^2=&(dx^1+\omega dx^2)(dx^1+\omega^2 dx^2)\,,\\
G_{\mu\nu}=& \begin{pmatrix} 1 & -\frac12 \\ -\frac12 & 1\end{pmatrix}\,,\qquad G^{\mu\nu}= \begin{pmatrix} \frac43 & \frac23\\ \frac23 &\frac43 \end{pmatrix}\,,
}
and $B$-field background
\es{B}{
B_{\mu\nu}=b\frac{2}{R^2}\begin{pmatrix} 0&1\\-1&0\end{pmatrix}\,.
}
The real parameters $R$ and $b$ parameterize the conformal manifold, and can be related to the complex parameter $\tau$ via the relation
\es{getTau}{
\left(\frac{4(-1+\tau^3)}{\tau(8+\tau^3)}\right)^3=J(y)\,,\qquad y\equiv b+{\rm i}\frac{\sqrt{3}}{4}R^2 \,,
}
where $J$ is the Klein invariant modular elliptic function, and the 12 roots of the polynomial in $\tau$ are permuted by the duality subgroup $A_4\subset S_4$. Note that unlike $d>2$, the XYZ theory in $d=2$ is free since $\tau  = 0$ corresponds to the decompactification limit $R \to \infty$.

\begin{figure}[t!]
\begin{center}
  \includegraphics[width=0.47\textwidth]{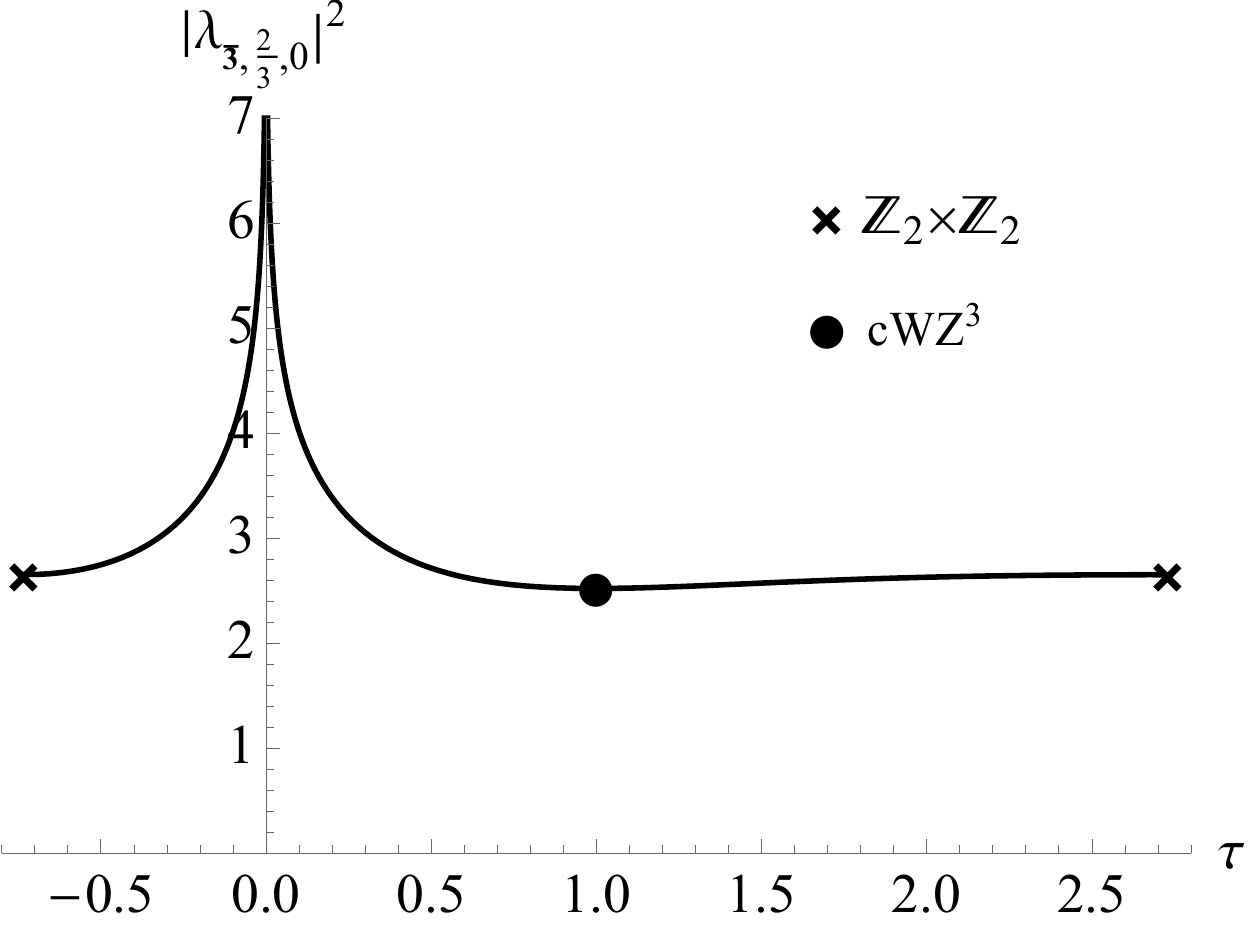}
   \includegraphics[width=0.47\textwidth]{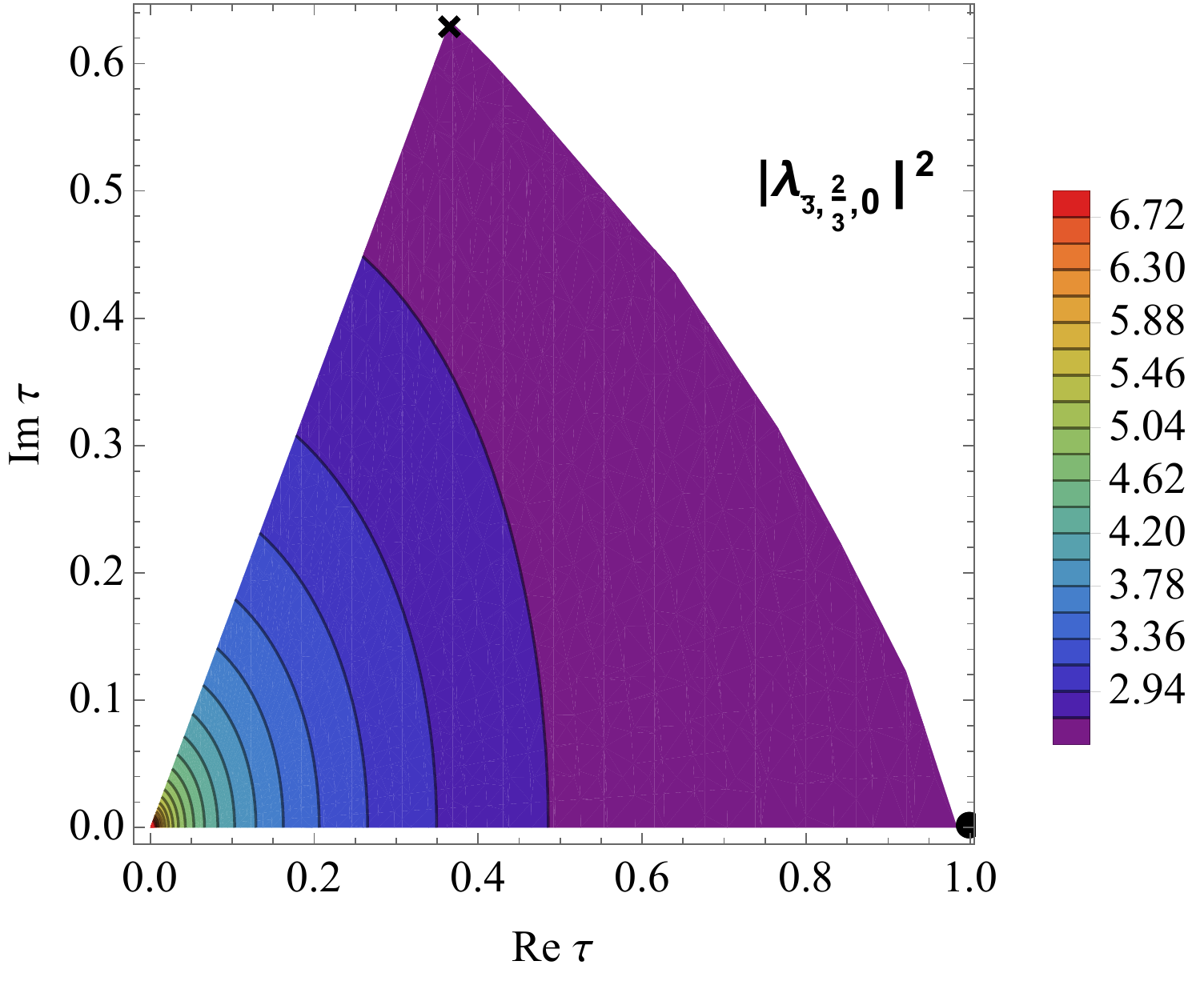}
 \caption{Exact chiral bilinear OPE coefficient squared $\lambda^2_{\bar{\bold3},\frac{2}{3},0}$ \eqref{OPE2d} for real $1-\sqrt{3}\leq\tau\leq1+\sqrt{3}$ ({\bf Left}), as well as for the fundamental domain $\mathbb{F}$ defined in Figure \ref{fund} ({\bf Right}). The cross and circle denote the enhanced symmetry points $\tau=1\pm \sqrt{3}, 1$ for the $\mathbb{Z}_2\times\mathbb{Z}_2$ and cWZ$^3$ models, respectively. Note that $\lambda^2_{\bar{\bold3},\frac{2}{3},0}$ diverges at the decompactification limit $\tau\to0$.}
\label{OPE2dB}
\end{center}
\end{figure}

The OPE coefficient of the chiral primary with scaling dimension $\frac23$ in our conventions\footnote{These relate to the definition in \cite{Lin:2016gcl} by a factor of $\frac{{|\tau|^2+2}}{2^{2/3}}$.} is then written as
\es{OPE2d}{
|\lambda_{\bar{\bold3},\frac23,0}|^2=\frac{{|\tau|^2+2}}{2^{2/3}}\abs{\sqrt{\frac{3}{2}}\frac{\Gamma\left(\frac23\right)^2}{\Gamma\left(\frac13\right)}R\sum_{v^1,v^2\in\mathbb{Z}}\exp\left[-\frac{\sqrt{3}\pi R^2}{2}\left(1-{\rm i}\frac{4}{\sqrt{3}R^2}b\right)|(v^1+\omega v^2)|^2\right]}^2\,,
}
where $R$ and $b$ can be written in terms of $\tau$ using \eqref{getTau}. In Figure \ref{OPE2dB} we plot $|\lambda_{\bar{\bold3},\frac23,0}|^2$ along the boundary of the manifold for real $1-\sqrt{3}\leq\tau\leq1+\sqrt{3}$, as well as for the entire fundamental domain $\mathbb{F}$ defined in Figure \ref{fund}.

\begin{figure}[t!]
\begin{center}
   \includegraphics[width=0.85\textwidth]{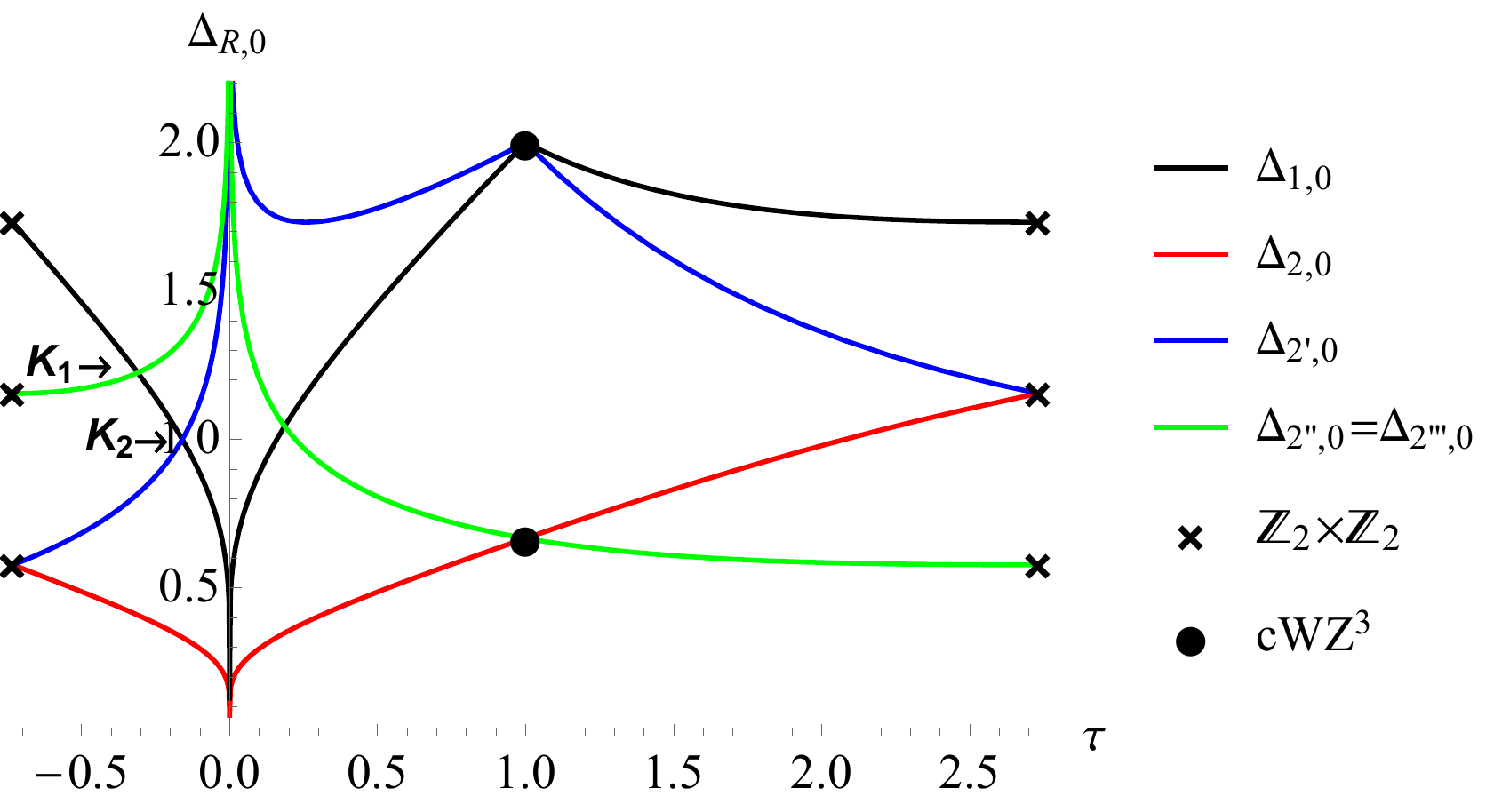}
 \caption{Exact scalar singlet and doublets scaling dimensions \eqref{del2d} for real $1-\sqrt{3}\leq\tau\leq1+\sqrt{3}$. The cross and circle denote the enhanced symmetry points $\tau=1\pm \sqrt{3}, 1$ for the $\mathbb{Z}_2\times\mathbb{Z}_2$ and cWZ$^3$ models, respectively. The points  $K_1$ and $K_2$ correspond to the kinks observed for this theory in \cite{Lin:2016gcl}. Note that some of the doublets diverge in the decompactification limit $\tau\to0$.}
\label{2DscalB}
\end{center}
\end{figure}

\begin{figure}[]
\begin{center}
   \includegraphics[width=.8\textwidth]{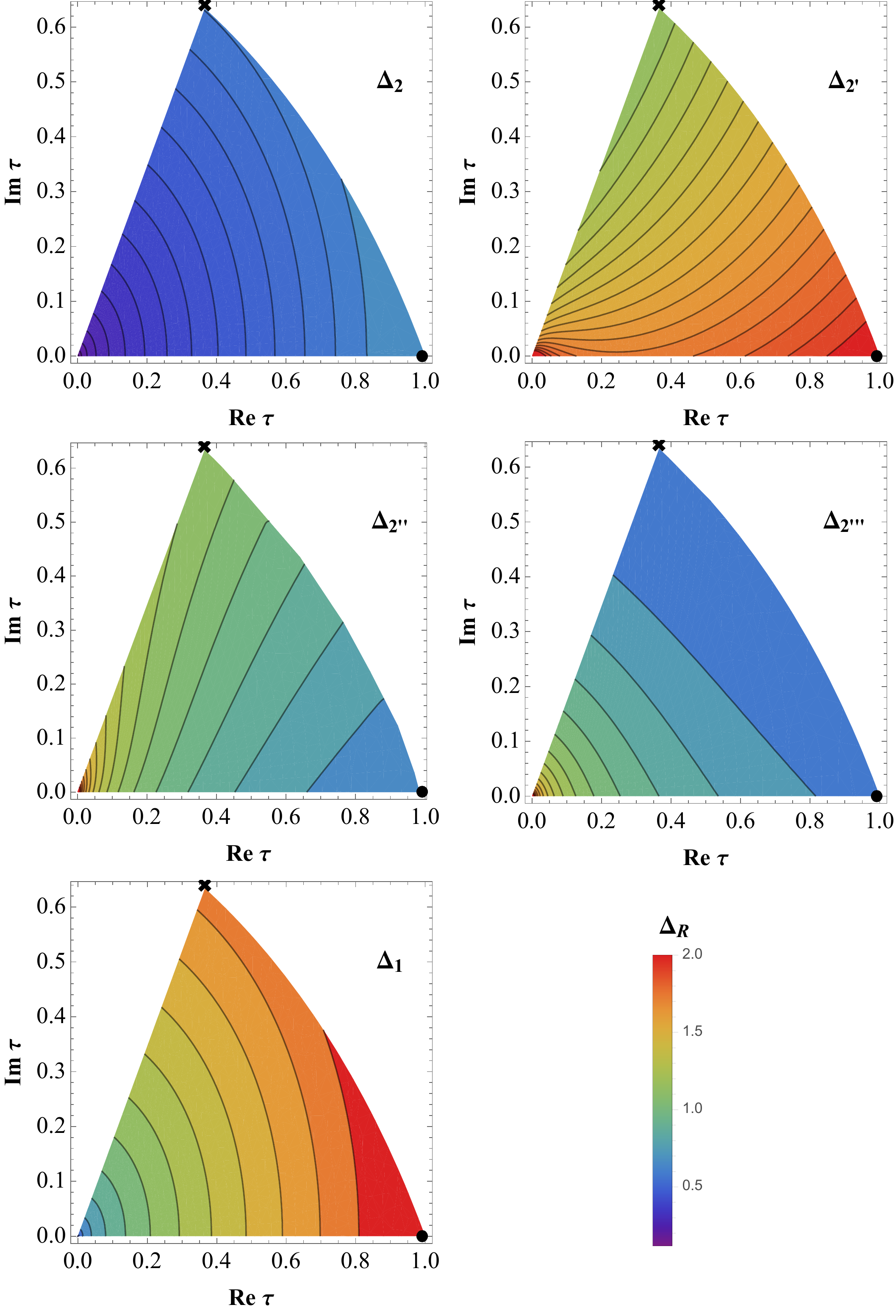}
 \caption{Exact scalar singlet and doublets scaling dimensions \eqref{del2d} in the fundamental domain $\mathbb{F}$ in Figure \ref{fund} for $d=2$. The cross and circle denote the $\mathbb{Z}_2\times\mathbb{Z}_2$ and cWZ$^3$ models, respectively.}
\label{2Dscal}
\end{center}
\end{figure}

The chiral scaling dimension of the singlet and doublets in the chiral-antichiral OPE can be written as
\es{del2d}{
h=&\frac12 G^{\mu\nu}\left[\frac{p_\mu}{R}+\frac12(G_{\mu\rho}+B_{\mu\rho})v^\rho R\right]\left[\frac{p_\nu}{R}+\frac12(G_{\nu\sigma}+B_{\nu\sigma})v^\sigma R\right]\,,\\
\bar h=&\frac12 G^{\mu\nu}\left[\frac{p_\mu}{R}-\frac12(G_{\mu\rho}-B_{\mu\rho})v^\rho R\right]\left[\frac{p_\nu}{R}-\frac12(G_{\nu\sigma}-B_{\nu\sigma})v^\sigma R\right]\,,\\
}
where the momentum $p^\mu$ and winding number $v^\mu$ are integers that must satisfy selection rules 
\es{selection}{
P(R,b)=&p^1-p^2\quad \text{mod 3}\,,\\
V(R,b)=&v^1+v^2\quad \text{mod 3}\,,
}
and $P$, $V$ are defined mod 3 and generically depend on the moduli $R$ and $b$. For instance, the singlet operator has $P(R,b)=V(R,b)=0$, while for the various doublets $P(R,b)$ and $V(R,b)$ are nontrivial functions of $R$ and $b$. For a given representation, there are many values of $p^\mu,v^\mu$ that satisfy the selection rules and $h=\bar h$. In practice, we scan over the possible values and extract the lowest possible scaling scaling dimension. In Figure \ref{2DscalB} we plot these scaling dimensions along the boundary of the manifold for real $1-\sqrt{3}\leq\tau\leq1+\sqrt{3}$. The points $K_1=(-.310,1.25)$ and $K_2=(-.160,1)$ for $(\tau,\Delta_\bold1)$ correspond 
to the kinks that were observed in the numerical bootstrap plot in \cite{Lin:2016gcl} of lowest singlet Virasoro primary scaling dimension. As discussed in \cite{Lin:2016gcl}, $K_2$ corresponds to a rational CFT with infinite higher spin currents, while $K_1$ has no enhanced symmetry. Here we observe that these kinks occur when the singlet scaling dimension coincides with one of the doublets. In Figure \ref{2Dscal} we plot the singlet and doublets scaling dimensions for the entire fundamental domain $\mathbb{F}$. We observe that these exact results in $d=2$ are in harmony with the general expectations based on dualities discussed in Section \ref{Morse}.

Lastly, we give the formula for the Zamolodchikov metric, which was computed in \cite{Cecotti:1990wz} to be
\es{Z2d}{
G(\tau,\bar\tau)=\frac{1}{4}\frac{1}{(\Im y)^2}\abs{\frac{\partial y}{\partial\tau}}^2\,,
}
where $y$ is defined implicitly in \eqref{getTau}. In terms of $y$, this metric is just the standard Weil-Peterson metric on $T^2$. 

\end{appendices}

\clearpage

\bibliography{3dConfMan}
\bibliographystyle{JHEP}

\end{document}